\documentclass[aps,pra,twocolumn,showpacs]{revtex4}

\usepackage{amsfonts}
\usepackage{amsmath}
\usepackage{amssymb}
\usepackage{graphicx}%
\begin{document}
\preprint{HEP/123-qed}
\title[Entanglement with Nilpotent Polynomials]
{Description of Quantum Entanglement with Nilpotent Polynomials}
\author{Aikaterini Mandilara}
\affiliation{Department of Physics, Washington University, Saint Louis, MO
63130, USA \footnote{Electronic address: admandil@wustl.edu}}
\author{Vladimir\ M.\ Akulin}
\affiliation{Laboratoire Aime Cotton, CNRS, Campus d'Orsay, 91405, Orsay, France}
\author{Andrei\ V.\ Smilga}
\affiliation{SUBATECH, Universit\'{e} de Nantes, 4, rue Alfred Kastler, BP\ 20722-44307,
Nantes-cedex 3, France\footnote{On leave of absence from ITEP, Moscow, Russia}}
\author{Lorenza\ Viola}
\affiliation{\mbox{Department of Physics and Astronomy, Dartmouth College, 6127 Wilder
Laboratory, Hanover, NH 03755, USA }\footnote{Electronic address: lorenza.viola@dartmouth.edu}}

\keywords{entanglement}
\pacs{03.67.Mn, 03.67.-a, 03.65.Ud, 03.65.Fd }

\begin{abstract}
We propose a general method for introducing {\em extensive}
characteristics of quantum entanglement.  The method relies on
polynomials of nilpotent raising operators that create entangled
states acting on a reference vacuum state. By introducing the notion
of \textit{tanglemeter}, the logarithm of the state vector represented
in a special canonical form and expressed via polynomials of nilpotent
variables, we show how this description provides a simple criterion
for entanglement as well as a universal method for constructing the
invariants characterizing entanglement.  We compare the existing
measures and classes of entanglement with those emerging from our
approach. We derive the equation of motion for the tanglemeter and, in
representative examples of up to four-qubit systems, show how the
known classes appear in a natural way within our framework. We extend
our approach to qutrits and higher-dimensional systems, and make
contact with the recently introduced idea of generalized entanglement.
Possible future developments and applications of the method are
discussed.

\end{abstract}

\volumeyear{year}
\volumenumber{number}
\issuenumber{number}
\eid{identifier}
\date[Date text ]{\today}




\startpage{1}

\maketitle
\begin{widetext}
\tableofcontents
\end{widetext}

\section{Introduction: How to characterize entanglement?}

Inseparability of quantum states of composite systems, discovered in the early
days of quantum mechanics by A. Einstein, B. Podolsky, and N. Rosen
\cite{Einstein} and named \textquotedblleft entanglement" (\textquotedblleft
Verschr\"{a}nkung\textquotedblright) by E. Schr\"{o}dinger \cite{Schrodinger},
became one of the central concepts of contemporary physics during the last
decade. Entanglement plays now a vital role within quantum information
science \cite{QI}, representing both the defining resource for quantum
communication -- where it enables, in particular, non-classical protocols such
as quantum teleportation \cite{Braunstein} and it leads to enhanced security
in cryptographic tasks \cite{Cryptography} -- and a key ingredient for
determining the efficiency of quantum algorithms and quantum computation
schemes \cite{Josza,Orus}. In addition, studies of entanglement have also
proved be relevant to fields as different as atomic physics \cite{Gershon},
quantum chaos \cite{Dima,Casati,Santos,Montangero}, quantum phase transitions
\cite{Arnesen,Osborne,Vidal,Eryigit,Stelmachovi,Amico,Somma,Chei}, and quantum
networks \cite{Cubitt}.

According to the original definition, the description of entanglement
relies on a specific partition of the composite physical system under
consideration. However, such a system can often be decomposed into a
number of subsystems in many different ways, each of the subsystems
possibly being a composite system by itself.  In order to avoid
ambiguity, given a partition of the composite system into $n$
subsystems, we call each of them an ``element" and characterize it by
a single, possibly collective, quantum number. Thus, the composite
system is a collection of the elements. We call this collection an
``assembly" in order to avoid confusion with an ``ensemble", which is
usually understood as a set of all possible realizations of a
many-body system with an associated probability distribution over these
realizations.  The $i$-th element is assumed to have Hilbert space of
dimension $d_{i}$.  Qubit, qutrit, and qudit are widely used names
for two-, three-, and $d$-level elements (with $d_{i}=2$, $d_{i}=3$,
and $d_{i}=d$, respectively).

The term ``entanglement'' has a transparent qualitative meaning: A pure
state of an assembly is entangled with respect to a chosen
partition when its state vector cannot be represented as a direct
product of state vectors of the elements.  This notion can also be
extended to generic mixed quantum states, whereby entanglement is
defined by the inability to express the assembly density operator as a
probabilistic combination of direct products of the density operators
of the elements. Intuitively, one expects that in the presence of
\textquotedblleft maximum\textquotedblright\ entanglement
\cite{Zeilinger}, the states of all subsystems comprising the overall
system are completely correlated in such a way that a measurement
performed on one part determines the states of all other parts.

The question is: how to quantitatively characterize entanglement for
an assembly of many elements? One would like to have a measure ranging
from zero for the product state to the maximum value for the maximally
entangled state. This can be easily accomplished for the bipartite
setting $n=2$ that is, for an assembly consisting just of two
distinguishable elements $A$ and $B$, each of arbitrary dimension.  In
this case, the von Neumann $\mathrm{Tr}[\rho_{A}\log\rho_{A}]$ or
linear $\mathrm{Tr}[\rho_{A}^{2}]$ entropies based on the reduced
density operator of either element, e.g.,
$\rho_{A}=\mathrm{Tr}_{B}[\left\vert \Psi\right\rangle \left\langle
\Psi\right\vert ]$, may be chosen as entanglement measures for a pure
state $\left\vert \Psi\right\rangle $ of the assembly. However,
already for the tripartite case $n=3$, characterizing and quantifying
entanglement becomes much harder. In fact, there are not one, but many
different characteristics of entanglement and, apart from the question
\textquotedblleft How much?", one has also to answer the question
\textquotedblleft In which way" \cite{Dur,Verstraete} are different
elements entangled? In other words, apart from separability criteria
\cite{Peres}, one also has to introduce inequivalent entanglement
measures \cite{Wootters,Vedral,Meyer,Wong,Rungta,Bennett,Heydari} and
entanglement classes \cite{Verstraete2,Dur,Acin,Akimasa}.

The identification of appropriate measures is not unique, and is
mostly dictated by convenience. The choice of classes may have a more
solid ground based on group theory, since their definition is related
to \textit{groups of local operations}. These are operations 
applied individually to each of the elements and forming a subgroup of
all possible transformations of the assembly state vector. One may
resort to group-theoretical methods, which allow one to construct a
set of invariants~\cite{Verstraete2,Grassl,Barnum,Leifer} under such
local operations. Under the action of local transformations, which can
be either unitary or, in the most general case, simply invertible, the
state vector of the system undergoes changes, still remaining within a
subset $\mathcal{O}$ (an \textit{orbit}) of the overall Hilbert space
$\mathcal{H}$. The dimension of the coset $\mathcal{H}/\mathcal{O}$,
that is the number of independent invariants identifying the orbit,
can be easily found for a generic quantum state. Still, there are
singular classes of orbits that require special consideration. Their
description depends critically on the number of elements and on the
detailed structure of the assembly.

The numerical values of the complete set of invariants may be chosen
as the \textquotedblleft orbit markers\textquotedblright\ that provide
one with an entanglement classification, although the choice of this set
is not, in general, unique. It turns out that generally accepted
measures of entanglement, such as concurrence \cite{Wootters} for two
qubits ($n=2$, $d_{i}=2$), and $3$-tangle \cite{Dur} for three qubits
($n=3$, $d_{i}=2$), are such invariants
\cite{Verstraete2,Gingrich}. Generalization of these measures
\cite{Verstraete2} to $n=4$, $d_{i}=2$ provides a connection between
measures and invariants characterizing different classes. Their
construction is not an easy task, and one needs to identify the
invariants that are able to distinguish inequivalent types of
entanglement \cite{Osterloh}. Moreover, the invariants are usually
high-order polynomials of the state amplitudes, with the maximum power
 growing linearly with the number of elements in the assembly. Therefore they
rapidly become rather awkward \cite{Leifer,MLV}.  A partial albeit not
unambiguous connections between the measures and classes are established
by the requirement \cite{Vidal2} for a measure to behave as a
so-called {\em monotone}. This means it should  be non-increasing on average
under the action of non-unitary invertible transformations of the
elements, also known as the family of {\em Local Operations and Classical
Communication}, LOCC \cite{Bennett2}.  The complete classification
problem remains unsolved even for relatively small assemblies (see
e.g. \cite{Luque} for recent results on the the five-qubit system).

Apart from polynomial-invariant constructions,
other schemes have been proposed to describe multipartite entanglement,
 including those 
 based on  generalization of Schmidt
decomposition \cite{Acin,Carteret2,Brun,Sudbery}, on invariant
eigenvalues \cite{Verstraete2}, on hyperdeterminants \cite{Akimasa},
 and on expectation values of
anti-linear operators \cite{Osterloh}. However, none of these
suggestion has been fully tested for $n>4$ qubits or more than three qutrits
($n>3$, $d=3$)~\cite{Briand}. Moreover, for the orbits of general
invertible local transformations, the complete sets of invariants are
unknown \cite{Kac} for assemblies consisting of $n$ qudits ($d_{i}=d$)
if $n>3$ and/or $d>4$. Still, a number of physically reasonable
suggestions \cite{Meyer,Heydari,Stockton,Linden} for entanglement
characterization have been attempted.

In this paper, we propose a different approach to entanglement
characterization. We focus exclusively on the case where an assembly
in a pure quantum state consists of {\em distinguishable} elements,
leaving generalizations to mixed states and to indistinguishable
elements for future studies. Our main aim is to construct
\textit{extensive} characteristics of entanglement. Thermodynamic
potentials linearly scaling with the number of particles in the system
offer examples of extensive characteristics widely employed in
statistical physics. The free energy given by the logarithm of the
partition function is a specific important example. We will introduce
similar characteristics for entangled states in such a way that their
values for a product state coincide with the sum of the 
corresponding values for unentangled groups of elements. This
technique is based on the notion of \textit{nilpotent} variables  and functions of these variables. 
An algebraic variable $x$ is called nilpotent if an integer $n$ exists
such that $x^n = 0$.
 In our case,
these variables are naturally provided by creation operators, whence the
logarithm function transforming products into sums plays the central
role in the construction.

Our approach is based on three main ideas: \textit{(i)} We express the
state vector of the assembly in terms of a polynomial of creation
operators for elements applied to a reference product
state. \textit{(ii)} Rather than working with the polynomial of
nilpotent variables describing the state, we consider its logarithm,
which is also a nilpotent polynomial. Due to the important role that
this quantity will play throughout the development, we call this quantity the
\textit{nilpotential} henceforth, by analogy to thermodynamic
potentials. \textit{(iii)} The nilpotential is not invariant under
local transformations, being different in general for different states
in the same orbit. We therefore specify a \textit{canonic}
\textit{form} of the nilpotential to which it can be reduced by means
 of local transformations. The nilpotential in  canonic form
is uniquely defined and contains complete information about the
entanglement in the assembly. We therefore call this quantity the
\textit{tanglemeter.}  The latter is, by construction, extremely
convenient as an \textit{extensive orbit marker}: the tanglemeter for
a system consists of several not interacting, unentangled groups of
elements equals the sum of tanglemeters of these groups.

Let us briefly explain these ideas, in the simplest example of $n$
qubits, which will be discussed in detail in Sect. II. An assembly of
$n$ qubits is subject to the $su(2)_{1}\oplus\ldots\oplus
su(2)_{i}\ldots\oplus su(2)_{n}$ Lie algebra of local transformations
\cite{note0}. As a reference state, we choose the 
Fock vacuum that is, the state $\left\vert
\mathrm{ O} \right\rangle =\left\vert
0,0,\ldots,0\right\rangle$  with all the qubits being in
the ground state. An arbitrary state of the assembly may be generated
via the action of a polynomial $F(\sigma_{i}^{+})$ in the nilpotent
operators $\sigma_{i}^{+}$ on the Fock vacuum. Here, the subscript $i$
enumerates the qubits, and the operator $\sigma_{i}^{+}$ creates the
state $|1\rangle$ out of the state $|0\rangle$. Evidently, $(\sigma
_{i}^{+})^{2}=0$, since the same quantum state cannot be created
twice. The family of all polynomials $F(\sigma_{i}^{+})$ forms a
ring. We note that \textit{anticommuting} nilpotent (Grassmann) variables
are widely employed in quantum field theory \cite{Berezin} and in
condensed matter physics \cite{Efetov}. However, the nilpotent
variables introduced here commute with one another.

In order to uniquely characterize entanglement, one must first 
select a convenient orbit marker. Following the idea of
Ref.~\cite{Carteret2}, we take as such a state $\left\vert
\Psi_{c}\right\rangle_{\mathcal{O}}$ lying in the orbit $\mathcal{O}$,
which is the ``closest'' to the reference state $\ \left\vert
\mathrm{ O}\right\rangle$ in the inner product sense, that is
$\left\vert \left\langle \mathrm{ O}\right.  \left\vert
\Psi_{c}\right\rangle_{\mathrm{ O}}\right\vert =\max$. 
 Once the state  $\left\vert
\Psi_{c}\right\rangle_{\mathrm{O}}$ is found, it is convenient 
to impose  a
non-standard normalization condition $\left\vert \left\langle \mathrm{O} \right.  \left\vert \Psi_{c}\right\rangle_{\mathrm{O}}\right\vert
=1$. We  call the resulting state $\left\vert
\Psi_{c}\right\rangle_{\mathrm{O}}$  \textit{canonic}. 
The canonic state is associated with the
canonic form of the polynomial $F_{c}(\sigma_{i}^{+})$, which begins
with a constant term equal to $1$.

We mainly work not with $F_{c}$ by itself, but with the tanglemeter --
a nilpotent polynomial $f_{c}=\ln F_{c}$ that can be explicitly
evaluated by casting the logarithm function in a Taylor series of the
nilpotent combination $F_{c}-1$. Since $(\sigma_{i}^{+})^{2}=0$, this
series is a polynomial  containing at most $2^{n}$ terms. Both the
tanglemeter and nilpotential ($f =\ln F$) resemble to the
eikonal, which is the logarithm of the regular \textit{semi-classical}
wave function in the position representation, multiplied by $-i$. The
difference is that in our case no approximation is made: $f$
represents the logarithm of the \textit{exact} state vector.

The nilpotential $f$ and the tanglemeter $f_{c}$ have several
remarkable properties: \textit{(i)} the tanglemeter provides a unique
and extensive characterization of entanglement; \textit{(ii)} a
straightforward entanglement criterion can be stated in terms of the cross
derivatives $\partial^{2}f$/$\partial
\sigma_{i}^{+}\partial\sigma_{j}^{+}$; \textit{(iii)} the dynamic
equation of motion for $f$ can be written explicitly and,
suggestively, in the rather general case has the same form as the well-known
\textit{classical} Hamilton-Jacobi equation for the eikonal.

The paper is organized as follows. In Sect. II, we analyze in detail
an assembly of $n$ qubits in terms of the nilpotent polynomials $F$
and $f$. We extend the notion of canonic forms to the group of
reversible local transformations $SL(2,\mathbb{C})$ and introduce the
idea of entanglement classes. We conclude the section by presenting
expressions relating the coefficients of $F_{C}$ and $f_{C}$ with
known measures of entanglement. 
To avoid confusion, we note that
 the subscript $c$ corresponds to  $su$-canonic forms 
in contrast to $C$,  which corresponds to  $sl$-canonic forms.
 Details of the calculations and some
proofs are given in Appendices A and B along with graphic
representations of the entanglement topology.

In Sect. III, we consider the evolution of the nilpotent polynomials
under the action of single-qubit and two-qubit Hamiltonians, and
derive an equation of motion for the nilpotential, which is distinct
from the Schr\"{o}dinger equation. For one important particular case
able to support \textit{universal quantum computation}~\cite{Barenco},
we show that this equation has a form of the classical $n$-dimensional
Hamilton-Jacobi equation. Describing quantum dynamics in terms of
nilpotentials suggests a computational algorithm for evaluating
tanglemeters, which can be performed by dynamically reducing the
polynomials to the forms canonic under either $SU(2)$ or
$SL(2,\mathbb{C})$ local transformations. In the example of a
four-qubit assembly, we explicitly illustrate how to identify the
resulting entanglement classes. This technique yields entanglement
classes consistent with the results of Ref.~\cite{Verstraete}. The
explicit analysis of these classes as well as details of derivation of
the equation of motion are given in Appendices B and C.

In Sect. IV, we extend our technique to assemblies of $d$-level
elements -- starting from qutrits~\cite{Cereceda,CHSH}. The
Cartan-Weyl decomposition of the $su(d)$ algebras suggests a natural
choice of nilpotent variables for qudits. For each element, we have
$d-1=r$ variables representing \textit{commuting} root vectors from
the corresponding Lie algebra, which has rank $r$.  For the
illustrative case of two and three qutrits, we discuss possible
choices of the canonic forms of the nilpotent polynomials. We further
extend the approach to the case where the assembly partition may change
as a result of the  merging of elements, such that the new assembly
consists of fewer number of elements with $d_{i}\neq d_{j}$, and consider
transformations of nilpotent polynomials associated with such a
change. Finally, we address a situation encountered in the framework
of \textit{generalized entanglement} \cite{Barnum1,Barnum2}, where the
rank $r$ of the algebras of allowed local transformations is less than
$d-1$. In other words, while the assembly is still assumed to be
composed of a number of distinct elements, the group of local
operations need not involve all possible local transformations. In
such a situation, the proper nilpotent variables are more complicated
than $\sigma^{+}$. In particular, they may have non-vanishing squares
\textit{etc}, with only $d_i$ powers vanishing.  In
addition, unlike in the conventional setting, entanglement relative to
the physical observables may exist not only among different
subsystems, but also within a {\em single} element.

We conclude by summarizing our results and discussing possible
developments and future applications of nilpotent polynomials and the
tanglemeter.

\vfill
\section{Entanglement characterization via nilpotent polynomials}

Consider $n$ qubits in a generic pure state $|\Psi\rangle$, 
\begin{widetext}
\begin{align}
\left\vert \Psi\right\rangle  &  =\sum_{\left\{  k_{i}\right\} ={0,1}}
\psi_{k_{n}k_{n-1}\ldots k_{1}}\left\vert k_{n},k_{n-1},\ldots , k_{1}
\right\rangle =\psi_{00\ldots0}\left\vert 0,0,\ldots0\right\rangle
\nonumber\\
& +   \psi_{10\ldots0}\left\vert 1,0,\ldots0\right\rangle +\psi_{01\ldots
0}\left\vert 0,1,\ldots0\right\rangle +\ldots+\psi_{11\ldots1}\left\vert
1,1,\ldots1\right\rangle\:, \label{EQ1}%
\end{align}
\end{widetext}
specified by $2^{n}$ complex amplitudes $\psi_{k_{n} k_{n-1}\ldots
k_{1}}$, i.e. by $2^{n+1}$ real numbers. The index $k_{i}=0,1$ 
corresponds to the ground or excited state of the
$i$-th qubit, respectively. When we take normalization into account
and disregard the global phase, 
 there are $2^{n+1}-2$ real parameters
characterizing the  assembly state.

It is natural to expect that any measure characterizing the intrinsic
entanglement in the assembly state remains invariant under unitary
transformations changing the state of each qubit.  A generic $SU(2)$
transformation is the exponential of an element of the $su(2)$ algebra,
\begin{equation}
U=\exp[\mathrm{i}(\sigma^{x}P^{x}+\sigma^{y}P^{y}+\sigma^{z}P^{z})]\ , 
\label{EQ1a}%
\end{equation}
where $\sigma^{x}$, $\sigma^{y}$, and $\sigma^{z}$ are Pauli matrices.
It depends on the three real parameters $P^{x}$, $P^{y}$, and
$P^{z}$. Such a transformation changes the amplitudes
$\psi_{k_{n}k_{n-1}\ldots k_{1}}$ in Eq.~(\ref{EQ1}), but preserves
some combinations of these amplitudes -- the invariants of local
transformations. Thus, local transformations move the state along an
orbit $\mathcal{O}$, while the values of the invariants serve as
markers of this orbit.

The first relevant question is: What is the maximum number of real
invariants required for the orbit identification, hence, for
entanglement characterization? A generic $SU(2)$ transformation represented by
Eq.~(\ref{EQ1a}) depends on three real parameters. Therefore, for
$n$ qubits the dimension of the coset $\mathcal{H}/\mathcal{O}$, that is the
number of different real parameters invariant under local unitary
transformations,  reads \cite{Linden,Carteret2}
\begin{equation}
D_{su}= 2^{n+1}-3n-2\ . \label{dimcoset}%
\end{equation}
 Mathematically, the counting in Eq.~(\ref{dimcoset}) corresponds to the number of the invariants of the group
$\otimes_i SU_i(2) \otimes U(1)$, where the factor $U(1)$ describes the multiplication of the wave function on the common phase
factor and the usual normalization condition $\langle \Psi | \Psi \rangle = 1$ 
for the wave function is imposed. 
It is more convenient for us to reformulate the same problem as 
 seeking  for the invariants of  
$\otimes_i SU_i(2) \otimes \mathbb{C}^*$, where $\mathbb{C}^*$ is the group of
multiplication by an arbitrary 
 nonzero complex number and no normalization condition on the wave function is imposed. This will allow us 
to choose a representative on the orbit of $\mathrm{O}$ with nonstandard normalization choice 
$\langle \Psi_C| \mathrm{O} \rangle = 1$.  

To be precise, the counting Eq.~(\ref{dimcoset}) is true for $n>2$ while the case $n=2$ is special: in
spite of the fact that $2^{3}-3\cdot2-2=0$, there \textit{is} a nontrivial
invariant of local transformations for two qubits. It has the form
\begin{equation}
I\ =\ \psi_{00}\psi_{11}-\psi_{01}\psi_{10}. \label{EQ2}%
\end{equation}
For a three-qubit system, five independent local invariants exist
\cite{Gingrich}, namely three real numbers
\begin{align}
I_{1}  &  =\psi_{kij}\psi^{\ast pij}\psi_{pmn}\psi^{\ast kmn}\ ,\nonumber\\
I_{2}  &  =\psi_{ikj}\psi^{\ast ipj}\psi_{mpn}\psi^{\ast mkn\ },\nonumber\\
I_{3}  &  =\psi_{ijk}\psi^{\ast ijp}\psi_{mnp}\psi^{\ast mnk\ }, \label{EQ3}%
\end{align}
and the real and the imaginary part of a complex number,
\begin{equation}
I_{4}+\mathrm{i}I_{5}=\psi_{ijk}\psi^{ijp}\psi_{mnp}\psi^{mnk}. \label{EQ3a}%
\end{equation}
Here, $\psi^{ijk}=\epsilon^{ii^{\prime}}\epsilon^{jj^{\prime}}\epsilon
^{kk^{\prime}}\psi_{i^{\prime}j^{\prime}k^{\prime}}$, with the summation
over repeated indexes taking values $0$ and $1$  implicit,
$\psi^{\ast}{}^{ijk}$ denotes the complex conjugate of $\psi_{ijk}$,
and $\epsilon^{ii^{\prime}}$ is the antisymmetric tensor of rank $2$.\
The quantity $2|I_4 + iI_5|$ is also known by the name \textit{residual entanglement} or 3-tangle $\tau$
Refs.~\cite{Wong,Akhtarshenas}.

Similar invariants can still be found for a four-qubit system.
However, with increasing $n$, the explicit form of the invariants
becomes less and less tractable and convenient for practical
use. Moreover, no explicit physical meaning can be attributed to such
invariants. We therefore suggest an alternative way to characterize
entanglement, which is based on: \textit{(i)} Specifying  the
canonic form of the state that unambiguously marks an orbit;
\textit{(ii)} Characterizing  this state with the help of
coefficients of a nilpotent polynomial; \textit{(iii)} Considering
 the logarithm of this polynomial, the tanglemeter. Thus, \textit{we
 construct  extensive invariants of local transformations as
the coefficients of the tanglemeter that is, nilpotential of the canonic
state.}

In this section, we proceed with illustrating the main technical
advantages of our description within the qubit setting, deriving an
entanglement criterion, and explaining how the invariants constructed by
our method are related to existing entanglement measures. We also
analyze a case important for certain applications involving indirect
measurements, where it is natural to consider a broader class of local
transformations constrained only by the requirement of unit
determinant. Specifically, we focus on the set of 
stochastic local operations assisted by classical communication 
 \cite{Bennett2}, which is widely employed in quantum
communication studies and protocols.
For qubits, such
transformations are known as SLOCC maps \cite{Bennett}.
 These operations do not necessarily preserve
the normalization of state vectors. However, as suggested by
\textit{Theorem 1} of Ref.~\cite{Verstraete2}, after a proper
renormalization they are described by the complexification
$sl(2,\mathbb{C})$ of the $su(2)$ algebra, such that the parameters
$(P_{i}^{x},P_{i}^{y},P_{i}^{z})$ specifying the transformation of
Eq.~(\ref{EQ1a}) on each qubit are now complex numbers. The
corresponding {\em real} positive invariants of local
$SL(2,\mathbb{C})$ transformations are monotones.

\subsection{Canonic form of entangled states\label{canonic}}

In order to unambiguously attribute a marker to each orbit, we specify
a canonic form of an entangled assembly state. To this end, we first
identify a reference state $\left\vert \mathrm{O}\right\rangle $ as a direct product of certain one-qubit
states. The latter can be chosen in an arbitrary way, but the choice
$|0 \rangle$,\ with the lowest energy level occupied, is the most
convenient to our scope. Thus, the reference state reads $\left\vert
\mathrm{O} \right\rangle \ =\ |0,\ldots0\rangle$. Drawing
parallels with quantum field theories and spin systems, we will call
$\left\vert \mathrm{O} \right\rangle $ the \textquotedblleft
ground\textquotedblright\ or \textquotedblleft
vacuum\textquotedblright  \ state. Then, following a suggestion of
Ref.~\cite{Carteret2}, by applying local unitary operations to a
generic quantum state $\left\vert \Psi \right\rangle $, we can bring
it into the \textquotedblleft canonic form\textquotedblright\
$\left\vert \Psi_{c}\right\rangle $ corresponding to the maximum
possible population of the reference state $\left\vert \mathrm{O} \right\rangle $. 
In other words, we apply a direct product
$U_{1}\otimes\ldots\otimes U_{n}$ of transformations as in
Eq.~(\ref{EQ1a}) to the state vector $\left\vert \Psi\right\rangle $,
and choose real parameters $(P_{i}^{x},P_{i}^{y},P_{i}^{z})$ to
maximize $\left\vert \left\langle \mathrm{O} \right\vert
U_{1}\otimes\ldots\otimes U_{n}\left\vert \Psi\right\rangle
\right\vert ^{2}$. The transformation $U_{1}\otimes\ldots\otimes
U_{n}$ satisfying this requirement can be seen to be unique up to
phase factors multiplying the upper states of each qubit. Modulo this
 uncertainty, the canonic state $\left\vert
\Psi_{C}\right\rangle =U_{1}\otimes\ldots\otimes U_{n}\left\vert
\Psi\right\rangle $ can serve as a valid orbit marker.

In fact, a generic unitary transformation of the $m$-th qubit, chosen
in the form $\exp[\mathrm{i}\sigma^{z}\phi_{m}]$
$\exp[\mathrm{i}\sigma^{x}g_{m}]$
$\exp[\mathrm{i}\sigma^{z}\varphi_{m}]$, $\phi_m,g_m,\varphi_m \in
\mathbb{R}$, equivalent to Eq.~(\ref{EQ1a}), results in
\begin{widetext}%
\begin{align}
\psi_{\ldots k_{m}=0\ldots} &  \rightarrow\psi_{\ldots k_{m}=0\ldots
}\mathrm{e}^{-\mathrm{i}\phi_{m}-\mathrm{i}\varphi_{m}}\cos g_{m}+\mathrm{i}\psi_{\ldots
k_{m}=1\ldots}\mathrm{e}^{\mathrm{i}\varphi_{m}-\mathrm{i}\phi_{m}}\sin
g_{m},\nonumber\\
\psi_{\ldots k_{m}=1\ldots} &  \rightarrow\psi_{\ldots k_{m}=1\ldots
}\mathrm{e}^{\mathrm{i}\phi_{m}+\mathrm{i}\varphi_{m}}\cos g_{m}+\mathrm{i}\psi_{\ldots
k_{m}=0\ldots}\mathrm{e}^{\mathrm{i}\phi_{m}-\mathrm{i}\varphi_{m}}\sin
g_{m}.\label{EQ4}%
\end{align}
\end{widetext}
Let us consider a generic infinitesimal local transformation of the
state in the canonic form. By expanding Eq.~(\ref{EQ4}) in series in
$g_{m}$ up to the second order, one obtains 
\begin{widetext}%
\begin{equation}\left\langle \mathrm{O} \Big| 
\otimes_i U_i \Big| \Psi\right\rangle \rightarrow{\bigg [}%
\Big(1-\sum\limits_{m}\tfrac{g_{m}^{2}}{2}\Big) \psi_{0\ldots0} + \mathrm{i}\sum_{m}g_{m}
\mathrm{e}^{2\mathrm{i}\varphi_{m}}
\psi_{0\ldots k_{m}=1\ldots0} - \sum
_{m>l}g_{m}g_{l} \mathrm{e}
^{2\mathrm{i}(\varphi_{m}+\varphi_{l})} \psi_{0\ldots k_{m}=1\ldots k_{l}=1\ldots0}{\bigg ]}\prod\limits_{r}%
\mathrm{e}^{-\mathrm{i}(\phi_{r}+\varphi_{r})}\,,\label{EQ5}%
\end{equation}
\end{widetext}
for the amplitude of the ground state. Since the parameters of the
transformation are arbitrary, the condition of maximum ground-state
population $\left\vert \left\langle \mathrm{O}|\Psi\right\rangle \right\vert ^{2}$ implies that the linear term
in Eq.~(\ref{EQ5}) vanishes,
\begin{equation}
\psi_{0\ldots k_{m}=1\ldots0}=0\:,\;\;\;\forall m\:. \label{EQ6}%
\end{equation}
This gives $n$ complex conditions and implicitly specifies $2n$ out of
$3n$ real parameters of the local transformation that maps a generic
state to the canonic form. The remaining $n$ parameters may be
identified in a generic case with the phase factors
$\mathrm{e}^{-\mathrm{i}(\phi_{r}+\varphi_{r})}$, where one can set
$\phi_{r}=0$ without loss of generality.  Two remarks are in order.

\noindent
\textit{(i) Special families of states} of measure zero in the assembly
Hilbert space may exist for which the system of equations (\ref{EQ6})
is degenerate and specifies less than $2n$ parameters. The simplest
example for $n=2$ is the Bell state, with
$\psi_{00}=\psi_{11}=1/\sqrt{2}$ and $\ \psi_{01}=\psi_{10}=0$. The
combination of two transformations of the form (\ref{EQ4}) gives a
state with the amplitudes
\begin{widetext}
\begin{align}
\label{n2trans}
& \psi'_{00} \ =\ \frac{ e^{-\mathrm{i}(\phi_1 + \phi_2)}}{\sqrt{2}}
\left[ e^{-\mathrm{i}(\varphi_1 + \varphi_2)} \cos g_1 \cos g_2 -
e^{\mathrm{i}(\varphi_1 + \varphi_2)} \sin g_1 \sin g_2 \right ]~~, \nonumber \\
& \psi'_{10} \ =\  \mathrm{i}\frac{ e^{\mathrm{i}(\phi_1 - \phi_2)}}{\sqrt{2}}
\left[ e^{-\mathrm{i}(\varphi_1 + \varphi_2)} \sin g_1 \cos g_2 +
e^{\mathrm{i}(\varphi_1 + \varphi_2)} \cos g_1 \sin g_2 \right ]~~, \nonumber  \\
&  \psi'_{01}  \ =\  \mathrm{i}\frac{e^{\mathrm{i}(\phi_2 - \phi_1)}}{\sqrt{2}}
\left[ e^{-\mathrm{i}(\varphi_1 + \varphi_2)} \cos g_1 \sin g_2 +
e^{\mathrm{i}(\varphi_1 + \varphi_2)} \sin g_1 \cos g_2 \right ]~~, \nonumber  \\
&\psi'_{11} \ =\  \frac{e^{\mathrm{i}(\phi_1 + \phi_2)}}{\sqrt{2}}
\left[ e^{\mathrm{i}(\varphi_1 + \varphi_2)} \cos g_1 \cos g_2 -
e^{-\mathrm{i}(\varphi_1 + \varphi_2)} \sin g_1 \sin g_2 \right ]\ .
\end{align}
\end{widetext}
One can see that the conditions $\psi_{01}^{\prime}=0$ and
$\psi_{10}^{\prime}=0$ are not independent: They both give
$g_{1}+g_{2}=0$ and $ \varphi_{1}+\varphi_{2}=0$ with arbitrary
$\phi_{1,2}$ (or, equivalently, $g_{1}-g_{2}=0,\ \varphi_{1}+\varphi_{2}= \pi/2$ with arbitrary
$\phi_{1,2}$).

 The orbit of this special state has $4$ parameters. On
the other hand, for a generic canonic state with
$\psi_{11}/\psi_{00}=\alpha$, $|\alpha|\neq1$, the conditions
$\psi_{01}^{\prime}=\psi_{10}^{\prime}=0$ imply $g_{1}=g_{2}=0$, and
only two phases $\phi_{1,2}+\varphi_{1,2}$ are arbitrary, whereas the
transformed state being independent of the differences
$\phi_{1,2}-\varphi_{1,2}$ in this case.

When $n$ grows, the pattern of such special classes of states becomes
more and more complicated. These families resemble ``catastrophe
manifolds'' where infinitesimal variation of the state amplitudes
$\psi$ result in a finite change of the local transformations reducing
the state to the canonic form. Here, we shall not discuss this further
and restrict ourselves to the generic case.

\noindent
{\it (ii)} As noticed in Ref.~\cite{Gingrich}, the conditions (\ref{EQ6}) are \textit{necessary but not
sufficient} in general for the state to have maximum ground-state
population $\left\vert \left\langle \mathrm{O}|\Psi
\right\rangle \right\vert ^{2}$.
For example, an $n=2$ state with $|\psi_{11}|>|\psi_{00}|$ and
$\psi_{01} =\psi_{10}=0$\ does \textit{not} have the maximum ground-state
 population, although it satisfies Eq.~(\ref{EQ6}): when
$|\psi_{11}|$ starts to exceed $|\psi_{00}|$, finite ``spin-flip"
operations must be applied to both qubits to reduce the state to
the canonic form.

For a generic $n$-qubit assembly state, the canonic form is unique up
to $n$ phase factors $\phi_{m}+\varphi_{m}$, and the state may be
characterized by $2^{n}-n-1$ complex ratios
$\alpha_{k_{n}k_{n-1}\ldots k_{1}}=\psi _{k_{n}k_{n-1}\ldots
k_{1}}/\psi_{0\ldots0}$ with $\sum_{m}k_{m}>1$, whereas the amplitude
of the vacuum state $\psi_{0\ldots0}$, after being factored out,
specifies the global phase and the normalization.  We disregard these
factors and normalize the wave function such that the amplitude of the
reference state $\psi_{0\ldots0}$ is set to unity. Then the parameters
$\alpha_{k_{n}k_{n-1}\ldots k_{1}}$ correspond to  the amplitudes
of the assembly states where at least two elements are excited. The
number of real parameters characterizing the canonic form equals
$2^{n+1}-2n-2$.   It is worth mentioning that all $\left\vert
\alpha _{k_{n}k_{n-1}\ldots k_{1}}\right\vert $ are invariant and,
moreover, in the case $n\geq3$, the ratios
$\alpha_{k_{n}k_{n-1}\ldots k_{1}} \alpha_{l_{n}l_{n-1}\ldots
l_{1}}/\alpha_{k_{n}^{\prime}k_{n-1}^{\prime}\ldots
k_{1}^{\prime}}\alpha_{l_{n}^{\prime}l_{n-1}^{\prime\prime}\ldots
l_{1}^{\prime}}$ are invariant if for each $m$ one of two conditions
$k_{m}^{\prime}=k_{m}$, $l_{m}^{\prime}=l_{m}$ or
$l_{m}^{\prime}=k_{m}$, $k_{m}^{\prime}=l_{m}$ are satisfied.
By specifying
 $n$  factors, we arrive at the 
bound Eq.~(\ref{dimcoset}) for the maximum number of invariants
characterizing entanglement.

\begin{figure}[ht]
{\centering{\includegraphics*[width=0.5\textwidth]{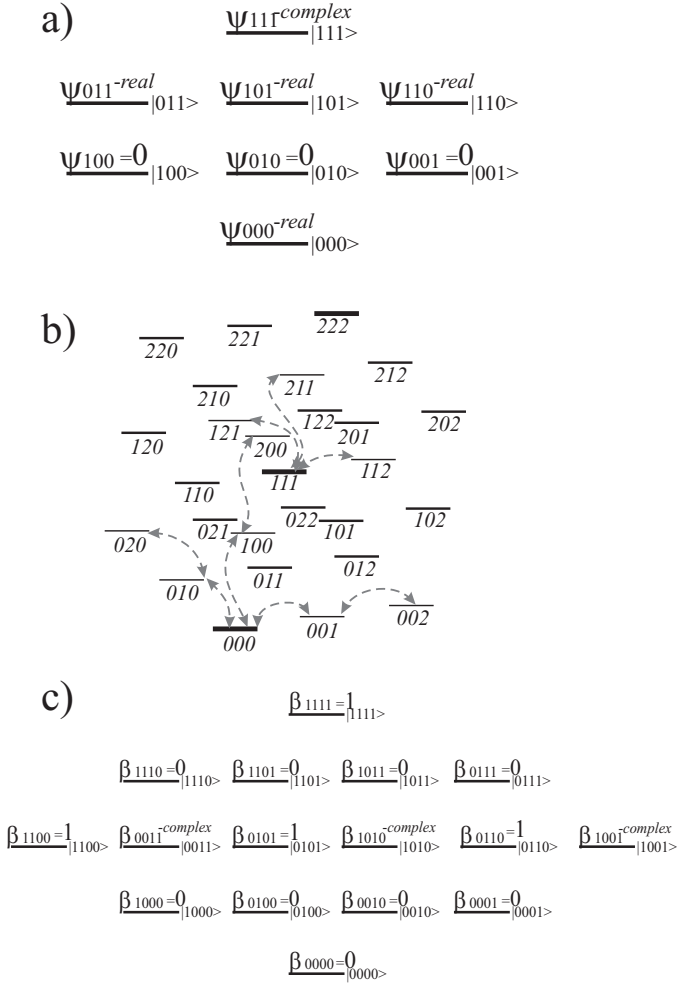}}} \vspace
{0.1cm}\caption{ a) Canonic form of the entangled state for three
qubits. By a local transformation the amplitudes of the lowest excited
states \textit{and} the phases of the second highest exited states are
set to zero.  The amplitude of the lowest state is taken as the common
factor determining the normalization and the global phase. b) In the
case of qudits, the group of local transformations is richer, and the
canonic form can be chosen such that it corresponds to maximum
population of the symmetric states $\left\vert k,k,k\right\rangle $
where $k=0,\ldots,d-1$. c) The structure of the tanglemeter for a
$4$-qubit system in the case of $SL(2,\mathbb{C})$ local
transformations. The scaling factors $q_{i}$ can be chosen in many
different ways, in particular such that some of the nonzero
coefficients equal one. }%
\label{FIG1}%
\end{figure}

Indeed, by an appropriate choice of the phase factors in Eq.~(\ref{EQ4}), one
can make  a set of $n$ non-zero amplitudes
$\alpha_{k_{n}k_{n-1}\ldots k_{1}}$ real and positive. For example, for a generic orbit
one can make $n$ amplitudes
$\alpha_{k_{n}k_{n-1}\ldots k_{1}}$ corresponding to the
next-to-highest excited states real and positive, with $\sum_{m}k_{m}=n-1$. In
Fig.~\ref{FIG1}a) we illustrate this for the simplest case $n=3$ where
the coefficients $\alpha_{011}$, $\alpha_{011}$, and $\alpha_{011}$
are chosen to be real and positive. As mentioned, the case $n=2$ is
special, since the next-to-highest excited state amplitudes coincide
with the first excited ones that vanish due to the requirement of
Eq.~(\ref{EQ6}).  Thus, a single parameter $\alpha_{11}$
characterizing entanglement can always be chosen real and positive, in
accordance with Eq.~(\ref{n2trans}), where we have only \textit{one}
free phase factor $\varphi_{1}+\varphi_{2}$. Some further discussion
is given in Appendix A.

Note that the determination of the canonic state for the orbit specified by
an arbitrary state vector $\left\vert \Psi\right\rangle $ of an
$n$-qubit assembly can be formulated as a standard quantum control
problem: the task is to find the global maximum of the  vacuum state population 
given by the functional
$\left\vert \left\langle \mathrm{O}\right\vert
U_{1}\otimes\ldots\otimes U_{n}\left\vert \Psi\right\rangle
\right\vert ^{2}$ starting from the initial
state $\left\vert \Psi\right\rangle $.  The 
 space of the control parameters $\left\{
P_{i}^{x},P_{i}^{y},P_{i}^{z}\right\}$ is $3n$-dimensional. Without taking advantage of
additional structure, the complexity of this procedure is in general
exponential in $n$ \cite{expn}.  A possibility to improve the efficiency of this
search is based on exploiting the solution of a set of differential
equations which is discussed in Sect.~\ref{u}.

We also note that,  the requirement of maximum vacuum-state population
alone {\em is}  insufficient  for  determination of the canonic state of assemblies consisting of
qudits with $d>2$.  Indeed, a local unitary transformation not
involving the vacuum state leaves this population intact, although it
changes the amplitudes of other states. To eliminate such ambiguity,
one needs to impose further constraints. As a possibility, one can
 maximize  by a sequence of step the populations of $d-1$ states $\left\vert
k,\ldots,k\right\rangle$, starting from $k=0$ and ending by
$k=d-2$. In this sequential procedure, maximization of the amplitude
of the state $\left\vert k,\ldots,k\right\rangle $ on the ($k+1$)-th step
is done by a restricted class of local transformations
belonging to the subgroup $SU(d-k)$ that acts non-trivially only on the
qudit states $\left\vert m\right\rangle$ with $m \geq k$.
This algorithm leads to a generalization of the condition of
Eq.~(\ref{EQ6}): now, the amplitudes of all states coupled to the
states $\left\vert k,\ldots,k\right\rangle $ by a single local
transformation $\in SU(d-k)$ such as $\psi_{0\ldots01}$,
$\psi_{0\ldots02}$, $\ldots$, $\psi_{1\ldots12}$, $\ldots$, $\psi_{k\ldots k, m>k}$
but not  $\psi_{k\ldots k, m<k}$
 vanish. The action of these local
transformations is indicated by dashed
arrows in Fig.~\ref{FIG1}~b). The remaining non-zero amplitudes normalized to  unit
vacuum amplitude characterize entanglement in the assembly of
qudits fairly unambiguously. One has still to fix $n(d-1)$ phase
factors of unitary transformations, but in analogy with the qubit
case, this can be done by setting real and positive some $n(d-1)$ of
$d^n - nd(d-1)/2 -1$ non-vanishing amplitudes. In Sect.~IV we  discuss 
this in more detail.

\subsection{Nilpotent polynomials for entanglement characterization}

The amplitudes $\psi$ or $\alpha$ are not the most convenient
quantities for characterizing entanglement, since they do not give an
immediate idea about the entanglement structure. For instance, for two
unentangled qubit pairs, each of which is in a Bell state, one finds
\begin{align}
\left\vert \Psi\right\rangle  &  =\frac{1}{2}\left(  \left\vert
0,0\right\rangle +\left\vert 1,1\right\rangle \right)  \otimes\left(  \left\vert
0,0\right\rangle +\left\vert 1,1\right\rangle \right) \nonumber\\
&  =\psi_{0000}\left\vert 0,0,0,0\right\rangle +\psi_{1100}\left\vert
1,1,0,0\right\rangle \nonumber\\
&  +\psi_{0011}\left\vert 0,0,1,1\right\rangle +\psi_{1111}\left\vert
1,1,1,1\right\rangle , \label{EQ8}%
\end{align}
where%
\begin{align}
\psi_{0000}  &  =\psi_{1100}=\psi_{0011}=\psi_{1111}=1/2\ ,\label{EQ9}\\
\alpha_{1100}  &  =\alpha_{0011}=\alpha_{1111}=1\:.\nonumber
\end{align}
In other words, though the system consists of two unentangled parts,
each of which is characterized by only one parameter, three non-zero
amplitudes $\alpha$ are present in the state vector. This is not
convenient and a better description of the entanglement is
desirable.

We now introduce a technique which serves this purpose. Consider a
standard raising operator
\[ \sigma_{i}^{+}=\left(
\begin{array}
[c]{cc}%
0 & 1\\
0 & 0
\end{array}
\right)\,,\]
acting in the $2$-dimensional Hilbert space of the $i$-th
qubit. Operators $\sigma _{i}^{+}$ acting on different qubits
commute. Since $\left( \sigma_{i}^{+}\right) ^{2}=0$, these operators
are nilpotent and they can be considered as
 nilpotent variables. Any quantum state $\left\vert
\Psi\right\rangle$ as in Eq.~(\ref{EQ1}) may be written in the form%
\begin{align}
\left\vert \Psi\right\rangle  &  =(\psi_{00\ldots0}+\psi_{00\ldots1}\sigma
_{1}^{+}+\ldots+\psi_{01\ldots0}\sigma_{n-1}^{+}+\psi_{10\ldots0}\sigma
_{n}^{+}\nonumber\\
&  +\psi_{00\ldots11}\sigma_{2}^{+}\sigma_{1}^{+}\ldots+\psi_{11\ldots0}%
\sigma_{n}^{+} \sigma_{n-1}^{+} +\ldots)\left\vert \mathrm{O}%
\right\rangle \nonumber\\
&  =\sum_{ \{k_{i} \}  = {0,1} }\psi_{k_{n}k_{n-1}\ldots k_{1}}%
\prod_{i=1}^{n}\left(  \sigma_{i}^{+}\right)  ^{k_{i}}\left\vert
\mathrm{O} \right\rangle , \label{EQ10}%
\end{align}
where each nilpotent monomial $\prod_{i=1}^{n}\left(  \sigma_{i}^{+}\right)
^{k_{i}}$ creates the basis state $\left\vert k_{n},\ldots,k_{1}\right\rangle
$ out of the vacuum state $\left\vert \mathrm{O}\right\rangle $. Let
$F(\{\sigma_{i}^{+}\})$ be the nilpotent polynomial%
\begin{align}
F(\{\sigma_{i}^{+}\})  &  =\sum_{\left\{  k_{i}\right\}  ={0,1}}\alpha
_{k_{n}k_{n-1}\ldots k_{1}}\prod_{i=1}^{n}\left(  \sigma_{i}^{+}\right)
^{k_{i}}\nonumber\\
&  =\sum_{\{k_{i}\}={0,1}}\frac{\psi_{k_{n}k_{n-1}\ldots k_{1}}}{\psi_{00\ldots0}%
}\prod_{i=1}^{n}\left(  \sigma_{i}^{+}\right)  ^{k_{i}}\ \label{EQ11}%
\end{align}
containing only the zeroth and first powers of each variable
$\sigma_{i}^{+}$.  A generic state $\left\vert \Psi\right\rangle $
normalized to unit vacuum amplitude $\psi_{0\ldots0}=1$ can thus be
written as $F(\{\sigma_{i}^{+}\})|\mathrm{O}\rangle$, with
$F(\{\sigma_{i}^{+}\})=1+\alpha _{00\ldots1}\sigma_{1}^{+}+\ldots$.

Next, define the nilpotential $f(\{\sigma_{i}^{+}\})$ given by the
logarithm of $F(\{\sigma_{i}^{+}\})$,
\begin{align}
f(\{\sigma_{i}^{+}\})  &  =\ln\left[  F(\{\sigma_{i}^{+}\})\right]
\label{EQ12}\\
& =\sum_{\{k_{i}\}={0,1}}\beta_{k_{n}k_{n-1}\ldots
k_{1}}\prod_{i=1}^{n}\left( \sigma_{i}^{+}\right) ^{k_{i}}\ .\nonumber
\end{align}
The coefficients $\beta_{k_{n}k_{n-1}\ldots k_{1}}$ and
$\alpha_{k_{n} k_{n-1}\ldots k_{1}}$ can be explicitly related to
each other by expanding $\ln F$ in  a Taylor series around $1$. This
calculation requires at most $n$ operations consisting of multiplications of the
polynomial $F-1$, which may generate an exponentially large ($\sim
2^{n}$) number of terms. Note that $\beta_{00\ldots0}=0$ since
$\alpha_{00\ldots0}=1$, so the nilpotential $f$ starts with the
first-order terms. Both $F$ and $f$ contain a finite number of
nilpotent terms, at most $2^{n}-1$, with the maximum-order term
proportional to the monomial $\prod_{i=1}^{n}\sigma_{i}^{+}$ given by the product of all
the nilpotent variables. The canonic form of the state vector corresponds to a
polynomial $F_{c}$, 
\begin{equation}
F_{c}=1+\alpha_{ij}\sigma_{i}^{+}\sigma_{j}^{+}\ldots\,,\label{EQ13}%
\end{equation}
which contains no linear monomials. The corresponding tanglemeter
$f_{c}$, also contains no linear terms and  reads
\begin{equation}
f_{c}(\{\sigma_{i}^{+}\})=\beta_{ij}\sigma_{i}^{+}\sigma_{j}^{+}%
+\ldots,\;\;\; \label{EQ13a}%
\end{equation}
with $\beta_{ij}=\alpha_{ij}$.

The discussion in Sect.~\ref{canonic} about the canonic form of the
state vector applies to the tanglemeter as well. For most purposes, it
suffices to employ the form which is unique up to phase changes of the
nilpotent variables given by the local transformations
$\sigma_{i}^{+}\mapsto\sigma_{i}^{+}\mathrm{e}^{2\mathrm{i}\phi_{i}}$. The
coefficients $\beta$ therefore remain invariant up to $n$ phase
factors, unless these factors are specified by additional
requirements. The phases $\phi_{i}$ may be chosen such that $n$ of
non-zero coefficients $\beta$ are set real and positive. Should the
tanglemeter of the generic state be defined unambiguously, we can
require this for the coefficients $\beta_{k_{n}\ldots k_{1}}$ with
$\sum_{i}k_{i}=n-1$, in the same way as it was done for $F$.  In
special cases, where one or several such coefficients equal zero, some
other conditions on the phases may be imposed.

The tanglemeter $f_{c}(\{\sigma_{i}^{+}\})$ immediately allows one to
check whether two groups $A$ and $B$ of qubits are entangled or
not. The following criterion holds:

\textbf{The entanglement criterion:} \textit{The parts }$A$\textit{ and }%
$B$\textit{ of a binary partition of an assembly of }$n$\textit{ qubits are
unentangled iff}
\begin{equation}
\frac{\partial^{2}f_{c}(\{x_{i}\})}{\partial x_{k}\partial x_{m}%
}=0\,, \;\;\; \forall k \in A,\:\forall m\in B\:.
\label{EQ14}%
\end{equation}
In other words, the subsystems $A$ and$~B$ of the partition are
disentangled iff $f_{A\cup B} =f_{A}(\left\{ x_{\in A}\right\} )+f_{B}
(\left\{ x_{\in B}\right\} )$, and no cross terms are present in the
tanglemeter.  Note that the criterion of Eq.~(\ref{EQ14}) holds not
only for the tanglemeter $f_{c}$, but for the nilpotential $f$ as
well. However, we 
formulate the criterion in terms of  $f_{c}$,  because the
coefficients $\beta$ of the tanglemeter are 
uniquely defined  by construction.

\subsection{Examples: Canonic forms for two, three, and four qubits}

For two qubits the result is immediate%
\begin{equation}
f_{c}=\beta_{11}\sigma_{2}^{+}\sigma_{2}^{+},\qquad F_{c}=1+\alpha_{11}%
\sigma_{2}^{+}\sigma_{2}^{+}=1+\beta_{11}\sigma_{2}^{+}\sigma_{1}^{+}\,,
\label{EQ14.1}%
\end{equation}
where the constant $\alpha_{11}=\beta_{11}$ can be chosen real. For
three qubits, the canonic forms of $F$ and $f$ also differ only by the
unity term,%
\begin{equation}
f_{c}=\beta_{3}\sigma_{2}^{+}\sigma_{1}^{+}+\beta_{5}\sigma_{3}^{+}\sigma
_{1}^{+}+\beta_{6}\sigma_{3}^{+}\sigma_{2}^{+}+\beta_{7}\sigma_{3}^{+}%
\sigma_{2}^{+}\sigma_{1}^{+}=F_{c}-1\,. \label{EQ14.2}%
\end{equation}
Here, we have introduced a shorter notation by considering the
indices of $\beta$ as binary representation of decimal numbers,
$011\rightarrow3$, \textit{etc}. One can make use of the fact that the
variables $\sigma_{i}^{+}$ are defined up to phase factors, and set
$\beta_{3}$, $\beta_{5}$, and $\beta_{6}$ real.
Expressing the
invariants of Eqs.~(\ref{EQ3})-(\ref{EQ3a}) via the parameters in the
canonic form, we obtain

\begin{widetext}%
\begin{align}
I_{1}  &  =1+2|\beta_{7}|^{2}\left(  \beta_{3}^{2}+\beta_{5}^{2}+\beta_{6}%
^{2}\right)  +|\beta_{7}|^{4}+2\beta_{3}^{2}+\beta_{3}^{4}+2\beta_{5}^{2}%
\beta_{6}^{2}\ ,\nonumber\\
\newline I_{2}  &  =1+2|\beta_{7}|^{2}\left(  \beta_{3}^{2}+\beta_{5}%
^{2}+\beta_{6}^{2}\right)  +|\beta_{7}|^{4}+2\beta_{5}^{2}+\beta_{5}%
^{4}+2\beta_{3}^{2}\beta_{6}^{2}\ ,\nonumber\\
\newline I_{3}  &  =1+2|\beta_{7}|^{2}\left(  \beta_{3}^{2}+\beta_{5}%
^{2}+\beta_{6}^{2}\right)  +|\beta_{7}|^{4}+2\beta_{6}^{2}+\beta_{6}%
^{4}+2\beta_{3}^{2}\beta_{5}^{2}\ ,\nonumber\\
\newline I_{4}+iI_{5}  &  =2\left(  \beta_{7}^{2}+4\beta_{3}\beta_{5}\beta
_{6}\right)  \ . \label{EQ14.3}%
\end{align}
This explicitly illustrates their linear independence.

The tanglemeter for four qubits reads%
\begin{align}
f_{c}  &  =\beta_{3}\sigma_{2}^{+}\sigma_{1}^{+}+\beta_{5}\sigma_{3}^{+}%
\sigma_{1}^{+}+\beta_{9}\sigma_{4}^{+}\sigma_{1}^{+}+\beta_{6}\sigma_{3}%
^{+}\sigma_{2}^{+}+\beta_{10}\sigma_{4}^{+}\sigma_{2}^{+}+\beta_{12}\sigma
_{4}^{+}\sigma_{3}^{+}\label{EQ14.4}\\
&  +\beta_{7}\sigma_{3}^{+}\sigma_{2}^{+}\sigma_{1}^{+}+\beta_{13}\sigma
_{4}^{+}\sigma_{3}^{+}\sigma_{1}^{+}+\beta_{11}\sigma_{4}^{+}\sigma_{2}%
^{+}\sigma_{1}^{+}+\beta_{14}\sigma_{4}^{+}\sigma_{3}^{+}\sigma_{2}^{+}%
+\beta_{15}\sigma_{4}^{+}\sigma_{3}^{+}\sigma_{2}^{+}\sigma_{1}^{+},\nonumber
\end{align}%
\end{widetext}%
while the coefficients $\alpha_{i}$ of the polynomial $F_c$ differ
from $\beta_{i}$ only at the last position%
\begin{equation}
\alpha_{15}=\beta_{15}+\beta_{3}\beta_{12}+\beta_{5}\beta_{10}+\beta_{9}%
\beta_{6}. \label{EQ14.5}%
\end{equation}
One may note that the sums of the indices of the factors in this
expression are equal. The latter is a general feature for the
relationship among the coefficients $\alpha$ and $\beta$: the
coefficients $\alpha$ are given by sums of terms, each of which
contains a product of the coefficients $\beta$ where the sum of the
indices equals the index of $\alpha$.
We also note that a proper 
set of invariants of $\otimes_{i=1}^4 SU_i(2) \otimes \mathbb{C}^*$
expressed via the components $\psi$ of the state vector can, in principle, 
be related to the tanglemeter coefficients, in analogy to the relation
between Eqs.~(\ref{EQ3}-\ref{EQ3a}) and Eq.~(\ref{EQ14.3}) for
  $\otimes_{i=1}^4 SU_i(2)$ invariants.

\subsection{Tanglemeter and entanglement classes for 
$SL(2,\mathbb{C})$ operations}

We now take a larger class of local operations 
and consider arbitrary invertible linear transformations  $GL$, 
instead of just the unitary transformations $SU$. 
Invertible transformations with non-zero determinant correspond
in general to indirect measurements, that are measurements performed
over an auxiliary system prepared  in a certain fixed quantum state
after it has interacted with the system under consideration.   Besides
allowing the realization of measurements more general than projective
Von Neumann measurements \cite{QI}, this procedure may serve as a tool
for quantum control and quantum state engineering \cite{Vogel}. In the
case where a \textit{single copy }\cite{purification} of a quantum
state is considered, the outcome of the measurements is not achieved
with certainty. Therefore, a stochastic factor allowing for the
outcome probability should be taken into account, whence the resulting
state vector has to be renormalized in accordance.  Since the normalization
factor in the latter require an information about the initial state vector, they do not 
strictly speaking form a group. In our approach, we do not impose a normalization condition 
on the wave function whatsoever and will be interested in finding the invariants
 \cite{Verstraete2}
of the transformations belonging to the group  
$ \mathcal{G} = \otimes_i SL_i(2, \mathbb{C}) \otimes  \mathbb{C}^*$,  where $ \mathbb{C}^*$
describes as before multiplication by an arbitrary nonzero complex number and  
the transformations  $\in SL_i(2,\mathbb{C})$ 
 multiply  the $i$th qubit state vector by a
$2\times2$ matrix of unit determinant. Another way to represent $\mathcal{G}$ is to express it as the
product  $\otimes_i GL_i(2,\mathbb{C})$ and factorize it over
$n-1$ redundant factors $\mathbb{C}^*$. The factors 
$\mathbb{C}^*$ in each $ GL_i(2,\mathbb{C})$ describe the same wave function transformations.

We  emphasize that the change of the wave function renormalization 
does not result exclusively from  the 
transformations $\in \mathbb{C}^*$, but from some   $ SL(2,\mathbb{C})$ 
transformations as well, and in particular, the transformations $\sim \exp {q\sigma_i^z}$ with complex $q$.
 Thus, considering just $\otimes_i SL_i(2,\mathbb{C})$
instead of the full group $\mathcal{G}$ may not have an explicit physical sense.
Still, we will  do it sometimes to better reveal  the mathematical structure of the results
obtained. Since the local $SL$ transformations comprise the key part of
  $\mathcal{G}$, and following the established usus we will  mainly refer to them and
talk about $sl$-entanglement in assemblies subject to indirect local measurements.
 
Though local, $SL$ operations can modify the 
 set of quantities which characterizes entanglement
 in assemblies subject to local $SU$ transformations.
 Since $SU\subset SL$, many different $su$-orbits  become equivalent
under local $SL$ transformations. In other words, the orbits of
 local $SL$ transformations contain the $su$-orbits as 
subsets. Classification of $sl$-orbits reveals the entanglement which
persists despite the indirect measurements. In order to distinguish this type
of entanglement from the invariants under local unitary
transformations, one may call it $sl$-entanglement.
In particular, classification for three-qubit assembly \cite{Dur}
shows that all generic quantum states  belong to one orbit of $\mathcal{G}$, 
which includes the canonic Greenberger-Horne-Zeilinger (GHZ) state $(\left\vert
000\right\rangle +\left\vert 111\right\rangle)/\sqrt{2}$ \cite{GHZ}.
The invariant Eq.~(\ref{EQ3a}) also known as $3$-tangle $\tau$ 
\cite{Wootters}, is different from zero only for this general orbit, and the value of $\tau$ calculated for the state amplitudes normalized to unit probability descriminates different $su$-orbits within this single general $sl$-orbit. The states with $\tau=1$ can be reduced to GHZ state by local unitary transformations, while for other states, with $\tau<1$ indirect local measurements are required for the purpose. Moreover, there are five singular orbits of $\mathcal{G}$ with $\tau=0$ that contain the states irreducible to GHZ-state. For  four-qubit assemblies, the classification \cite{Verstraete} becomes much more involved, but still it gives an idea about the types of entanglement and eventual measures.

 Each element of the $SL_{i}(2,\mathbb{C})$ group, 
 that is isomorphic
to the Lorentz group $SO(3,1)$, involves 6 parameters.
 In the general case, the number of invariants,
  \begin{equation}
 \label{invSL}
 D_{sl} \ =\  2^{n+1} - 6n - 2
 \end{equation}
is less than that for  unitary transformations  Eq.(~\ref{dimcoset}).
 This counting is valid and returns a positive value for $n \geq 4$ when the actions of different local operations are linearly independent. For $n=2,~3$ where the number of the parameters in the group is more than the number of the parameters in the wave function,
and the result of some local $SL$ are redundant, no invariants
exist. In particular, for two qubits, any generic state is equivalent under $\mathcal{G}$ to the
Bell state and for 3 qubits --- to the GHZ state. For four qubits, there are $6$ real  invariants, for
$n=5$, $  D_{sl}  = 32$, etc.

A smaller number of $sl$-invariants Eq.~(\ref{invSL}) as compared
to that of  $su$-invariants Eq.~(\ref{dimcoset}) implies that different $su$-orbits 
may belong to the same $sl$-orbit. In analogy
to the $su$-canonic state, one has to define a $sl$-canonic state as the marker of a $sl$-orbit.
In contrast to the unitary case where the canonic state has been defined by the
condition of maximum reference state population, for $SL$ transformations
we introduce directly canonic form of the tanglemeter. To this end, we
impose the following conditions: in addition to the requirement of the Eq.~(\ref{EQ6}), i.e.
all $n$ linear in $\sigma^+$ terms of the nilpotential equal to zero, we require that all
$n$ terms of $(n-1)$-th order vanish as well. In other words, the $sl$-tanglemeter
takes the form
\begin{equation}
f_{C}(\{\sigma_{i}^{+}\})=\sum_{\quad\Sigma_i k_{i}\neq \{1,n-1\}}\beta
_{k_{n} k_{n-1}\ldots k_{1}}\prod_{i=1}^{n}\left(  \sigma_{i}^{+}\right)
^{k_{i}}, \label{EQ14a}%
\end{equation}
and in this way we have specified $4n$ out of the $6n$ real
parameters of the local transformations that bring a given state to the
$sl$-canonic form. We thus left with $2n$ parameters that have to be specified. 

In contrast to the unitary case, where the nilpotent
variables $\sigma^+$ are defined up to arbitrary phase factors, for $SL$
transformations the
 variables in Eq.~(\ref{EQ14a}) are  defined up to a
\textit{complex-valued} scaling factor
$\sigma_{i}^{+}\mapsto\sigma_{i}^{+}q_{i}$.
One can further specify the $sl$-tanglemeter by choosing these factors such that
$n$ complex coefficients of the tanglemeter are set to unity. 
If convenient, one can impose another set of $n$ requirements.

As a first example, consider the three-qubit case. The $sl$-tanglemeter  Eq.~(\ref{EQ14a}) for a generic
three-qubit state  reads 
\begin{equation}
f_{C}=\sigma_{3}^{+}\sigma_{2}^{+}\sigma_{1}^{+}, \label{EQ14a1}%
\end{equation}
 where  the coefficient is set to unity by the scale freedom  in the definition
of the nilpotent variables.   
 The corresponding wave function $F_C$  is nothing but the GHZ state. This shows again that all generic
states belong to the same $sl$-orbit, which includes this state. There are, however, also three distinct
  singular classes of entangled
states of measure zero  
\cite{Dur} whose tanglemeters do not involve the product $\sigma_{1}^{+}\sigma_{2}^{+}\sigma_{3}^{+}$
 and  have one of the following forms,
\begin{align}
f_{C}=\sigma_{2}^{+}\sigma_{1}^{+}+\sigma_{3}^{+}\sigma_{1}^{+}\:,\nonumber \\%
f_{C}=\sigma_{3}^{+}\sigma_{1}^{+}+\sigma_{3}^{+} \sigma_{2}^{+}\:,\nonumber \\%
f_{C}=\sigma_{2}^{+}\sigma_{1}^{+}+\sigma_{3}^{+} \sigma_{2}^{+}\:.\nonumber %
\end{align}
 In this classification, we have only taken into account  the states whose 
tanglemeters involve  all three $\sigma_{i}^{+}$ such that no qubit is completely disentangled from the others. 

For a generic four-qubit state one finds the $sl$-tanglemeter
\begin{align}
f_{C}  &  =\beta_{3}\sigma_{2}^{+}\sigma_{1}^{+}+\beta_{5}\sigma_{3}^{+}%
\sigma_{1}^{+}+\beta_{9}\sigma_{4}^{+}\sigma_{1}^{+}+\beta_{6}\sigma_{3}%
^{+}\sigma_{2}^{+}\label{EQ14b}\\
&  +\beta_{10}\sigma_{4}^{+}\sigma_{2}^{+}+\beta_{12}\sigma_{4}^{+}\sigma
_{3}^{+}+\beta_{15}\sigma_{4}^{+}\sigma_{3}^{+}\sigma_{2}^{+}\sigma_{1}%
^{+}\,,\nonumber
\end{align}
where the scaling factors $q_{i}$ of the variables $\sigma_{i}^{+}$
can be specified such that this form becomes equivalent to the
expression given in \textit{Theorem 2} of Ref.~\cite{Verstraete}:
\begin{align}
f_{C}  &  =\beta_{3}\left(  \sigma_{1}^{+}\sigma_{2}^{+}+\sigma_{3}^{+}%
\sigma_{4}^{+}\right)  +\beta_{5}\left(  \sigma_{1}^{+}\sigma_{3}^{+}%
+\sigma_{2}^{+}\sigma_{4}^{+}\right) \label{EQ14bb}\\
&  +\beta_{6}\left(  \sigma_{1}^{+}\sigma_{4}^{+}+\sigma_{2}^{+}\sigma_{3}%
^{+}\right)  +\sigma_{1}^{+}\sigma_{2}^{+}\sigma_{3}^{+}\sigma_{4}%
^{+}.\nonumber
\end{align}
In Fig.~\ref{FIG1} c),  we illustrate the structure of $sl$-tanglemeter
for this case with an alternative choice of the scaling factors.

It is worth mentioning that, though any generic nilpotential can be
reduced to the canonic form of Eq.~(\ref{EQ14a}) this turns out to be impossible
for some sets of
states of measure zero, as it is
already the case for three qubits.  These sets may play an important
role for applications and can be grouped into special classes. Some of
these classes are shown in Sect.~\ref{u} at the example of four
qubits. There we also present an explicit algorithm for evaluation of
$sl$-tanglemeters based on the stationary solutions of dynamic
equations with feedbacks imposed on the parameters of local
transformations.  This yields the special entanglement classes in a
natural way  as singular stationary solutions.

We conclude this section by discussing the precise mathematical
meaning of the canonic states. The renormalization of the wave 
function that follows the maximization of the reference state
amplitudes by local $SU$ transformations belongs to the group $\mathbb{C}^*$
of multiplication by a complex number $\kappa$. Therefore, strictly speaking,
the applied transformations belong to the group $\otimes_iSU_i(2)\otimes \mathbb{C}^*$. However, the group $\otimes_iSU_i(2)$ does not affect the normalization of the state vector, while the requirement $\psi_{{\rm O}}=1$ imposed on the canonic state uniquely specifies the number $\kappa$ thus allowing to introduce the tanglemeter as a characteristic of $sl$-orbits. In other words, once the condition $\psi_{{\rm O}}=1$ is satisfied, the group $\otimes_iSU_i(2)\otimes \mathbb{C}^*$ becomes isomorphic to the group $\otimes_iSU_i(2)$.
 
This is no longer the case for indirect measurements. Neither the group $G$ nor its nontrivial part $\otimes_iSL_i(2,\mathbb{C})$ conserve the state normalization. By imposing the requirement $\psi_{{\rm O}}=1$, we mark an orbit of $G/\mathbb{C}^*$, and thereby specify the structure of the canonic state given by the state amplitude ratios $\psi_i/\psi_{{\rm O}}$ expressed in terms of the $sl$-tanglemeter coefficients. However, a state of same structure but with a different normalization can be physically achieved in many different ways, -- as a result of a single indirect measurement, or a sequence of two or more indirect measurements. The probability to obtain an outcome of the measurements that correspond to required $G/\mathbb{C}^*$ transformation thus depends on the particular choice of the measurement procedure. Therefore, the complex factor $\kappa$ can be an arbitrary number, irrelevent to the values of the $sl$-tanglemeter coefficients.
 
However, when we consider just the nontrivial part  $\otimes_iSL_i(2,\mathbb{C})$ of $G$, the factor $\kappa$ can bear certain physical significance. In fact, a transformation from $\otimes_iSL_i(2,\mathbb{C})$ may bring a state initially normalized to unit probability to another one, which differs from the canonic state only by a factor  $\kappa$. In this case the factor $\kappa$ is uniquely defined function of the initial state \cite{xx}. When the transformation is unitary $\kappa$ amounts to $1/\sqrt{\sum_{i}\left|\psi_i\right|^2}$ where the amplitudes $\psi_i$ of the canonic state are normalized to unity reference state amplitude, as required. For non-unitary $SL$ transformations this quantity is different. Therefore, $ln\left(\left|\kappa\sqrt{\sum_{i}\left|\psi_i\right|^2}\right|\right)$ can serve as a measure of non-unitarity of the transformation that discriminates different $su$-orbits that belong to the same $G$-orbit.

\subsection{How do the nilpotent polynomials relate to existing entanglement
measures}

In general, there is no universal and precise definition of proper measures
of entanglement \cite{MLV}, with the exception of bipartite entanglement: as
long as we are interested in entanglement between two parts $A$ and $B$ of a
quantum system in a pure state, natural measures of such entanglement do
exist. They are based on the reduced density operator $\rho _{A}$ of either
part, obtained by tracing over the quantum numbers corresponding to the
other part $B$. In particular, $S_{vN}=-\mathrm{Tr}[\rho _{A}\log \rho _{A}]$
and $S_{l}=1-\mathrm{Tr}[\rho _{A}^{2}]$, give the von Neumann and the
linear entropies, respectively \cite{QI}, as already mentioned in the
Introduction. Clearly, both characteristics can be directly related to the
tanglemeter parameters. However, the explicit formulae giving these
relations, which are simple for the case of two qubits%
\begin{align*}
S_{l}& =\frac{2\left\vert \beta _{11}\right\vert ^{2}}{\left( 1+\left\vert
\beta _{11}\right\vert ^{2}\right) ^{2}}, \\
S_{vN}& =\ln \left[ 1+\left\vert \beta _{11}\right\vert ^{2}\right] -\frac{%
\left\vert \beta _{11}\right\vert ^{2}}{1+\left\vert \beta _{11}\right\vert
^{2}}\ln \left[ \left\vert \beta _{11}\right\vert ^{2}\right] ~,
\end{align*}%
become awkward for larger numbers of qubits within the bipartition, as well
as for higher-dimensional elements. This reflects the fact that the
coefficients of nilpotent polynomials carry much more information about
entanglement than the simple bipartite correlations captured by the entropy
measures.

Another useful entanglement measure, \textit{concurrence} $C$, has been
introduced in Ref.~\cite{Wootters} in the context of mixed two-qubit states,
and has been employed for constructing the \textit{residual entanglement} $%
\tau $, as a measure characterizing three-qubit pure-state entanglement and
possibly beyond \cite{Wong,Akhtarshenas}. Both $C$ and $\tau $
may be expressed in terms of the amplitudes $\psi $ of the $su$-canonic
state and in terms of the tanglemeter coefficients $\beta $ Eq.~(\ref{EQ14.2}%
). The concurrence between the first and the second qubits reads%
\begin{align}
C_{12}& =2\Big||\psi _{000}\psi _{110}|-|\psi _{101}\psi _{011}|\Big|  \notag
\\
& =\frac{2||\beta _{6}|-|\beta _{5}\beta _{3}||}{1+\beta _{6}^{2}+\beta
_{5}^{2}+\beta _{3}^{2}+|\beta _{7}|^{2}}~.  \label{concurrence}
\end{align}%
The residual entanglement, or $3$-tangle, has the form of a fourth-order
polynomial in the amplitudes. For the canonic state, it reads 
\begin{align}
\tau _{3}& =4\Big|\left( \psi _{000}\psi _{111}\right) ^{2}+4\psi _{000}\psi
_{110}\psi _{101}\psi _{011}\Big|  \notag \\
& =\frac{4\left\vert \beta _{7}{}^{2}+4\beta _{6}\beta _{5}\beta
_{3}\right\vert }{\left( 1+\beta _{6}^{2}+\beta _{5}^{2}+\beta
_{3}^{2}+|\beta _{7}^{2}|\right) ^{2}}\,,  \label{threetangle}
\end{align}%
which up to a numerical factor is equal to the invariant $\left\vert
I_{4}+iI_{5}\right\vert $ of Eqs.~(\ref{EQ3a}, \ref{EQ14.3}) divided by the
normalization factor $\sum |\psi |^{2}=1+\beta _{6}^{2}+\beta _{5}^{2}+\beta
_{3}^{2}+|\beta _{7}^{2}|$. The presence of the normalization factor in the
denominators of Eqs.~(\ref{concurrence}), (\ref{threetangle}) is due to the
fact that these quantities are usually calculated for the state vector
normalized to unity while the coefficients $\beta $ refer to the tanglemeter
that is the logarithm of the canonical wave function with the normalization $%
\psi _{000}=1$.

What are convenient measures that can be introduced to characterize $sl$--entanglement?
 We have seen that all generic states of the assembly of three
qubits belong to the same orbit of $\mathcal{G}$ and strictly speaking there
are no invariant measures at all. However, the $su$-invariant $I_{4}+iI_{5}$ of
Eq.~(\ref{EQ3a}) remains invariant under the restricted class of
transformations $\otimes _{i=1}^{3}SL_{i}(2,\mathbb{C})$, while the other $%
SU $ invariants $I_{1,2,3}$ of Eq.~(\ref{EQ3}) depending on \ both $\psi $ and $%
\psi ^{\ast }$ change under $SL$ transformations. Hence, in this restricted
sense it may serve as a measure for $sl$-entanglement.

The measures characterizing the $sl$-entanglement for a 4-qubit assembly can
be constructed in a similar way. We take products of several factors $\sim
\psi $ (but not the factors $\sim \psi ^{\ast }$) and convolute it over $%
SU(2)$ indices with invariant tensors $\epsilon ^{ii^{\prime }}$ \cite{footnote0}.
 The simplest combination 
\begin{equation}
I^{(2)}\ =\ \psi _{ijkl}\psi ^{ijkl}\,  \label{J2}
\end{equation}%
is $sl$-invariant and can be taken as a characteristic of $sl$-entanglement,
remaining not invariant only with respect to the transformations $\in 
\mathbb{C}^{\ast }$. There are three different $sl$-invariants $\sim \psi
^{4}$, 
\begin{align}
I_{12}^{(4)}& =I_{34}^{(4)}\ =\ \psi _{ijkl}\psi ^{ijmn}\psi _{opmn}\psi
^{opkl}\,,  \notag \\
I_{13}^{(4)}& =I_{24}^{(4)}\ =\ \psi _{ikjl}\psi ^{imjn}\psi _{ompn}\psi
^{okpl}\,,  \notag \\
I_{14}^{(4)}& =I_{23}^{(4)}\ =\ \psi _{iklj}\psi ^{imnj}\psi _{omnp}\psi
^{oklp}\,.  \label{4-th}
\end{align}%
The ratios $I_{12}^{(4)}/(I^{(2)})^{2}$, $I_{13}^{(4)}/(I^{(2)})^{2}$, and $%
I_{14}^{(4)}/(I^{(2)})^{2}$ are in addition invariant with respect to
multiplication of the state vector by an arbitrary complex constant and
thereby they are invariants of $\mathcal{G}$. Were these ratio linearly
independent, they would give us a complete characterization of $4$-qubit
entanglement, since the $4$-qubit $sl$-tanglemeter Eq.~(\ref{EQ14bb}) involves $3
$ complex parameters. However, they are not. The following identity 
\begin{equation}
I_{12}^{(4)}+I_{13}^{(4)}+I_{14}^{(4)}\ =\ \frac{3}{2}\left( I^{(2)}\right)
^{2}\,  \label{identJ}
\end{equation}%
makes these quantities inconvenient for the entanglement characterization.

We therefore turn to the $6$-th order invariants and consider following
three functionally independent combinations
\begin{widetext}
\begin{align}
I_{12}^{(6)}\ & =\frac{1}{6}\ \left( \psi _{ingd}\psi _{mrko}\psi
_{sjph}-\psi _{ingo}\psi _{mrkh}\psi _{sjpd}\right) \psi ^{mrgd}\psi
^{inph}\psi ^{sjko}\,,  \notag \\
I_{23}^{(6)}\ & =\frac{1}{6}\ \left( \psi _{ijpo}\psi _{mngh}\psi
_{srkd}-\psi _{ijpd}\psi _{mngo}\psi _{srkh}\right) \psi ^{mrgd}\psi
^{inph}\psi ^{sjko}\,,  \notag \\
I_{13}^{(6)}\ & =\frac{1}{6}\ \left( \psi _{ijkh}\psi _{mnpd}\psi
_{srgo}-\psi _{ijgh}\psi _{mnkd}\psi _{srpo}\right) \psi ^{mrgd}\psi
^{inph}\psi ^{sjko}\,,  \label{6-th}
\end{align}
\end{widetext}
whose differences give the invariants Eq.~(\ref{4-th}) multiplied by $I^{(2)}$%
. Explicit form of these invariants for a generic state is awkward. However
they take a simple form 
\begin{align}
I^{(2)}& =\psi _{0000}^{2}(t+x+y+z)  \notag \\
I_{12}^{(6)}\ & =4\ \psi _{0000}^{6}\left( t+x-y-z\right) (tx-yz)\,,  \notag
\\
I_{23}^{(6)}\ & =4\ \psi _{0000}^{6}\left( t-x+y-z\right) (ty-xz)\,,  \notag
\\
I_{13}^{(6)}\ & =4\ \psi _{0000}^{6}\left( t-x-y+z\right) (tz-xy)\,,
\label{I6can}
\end{align}%
for the canonic state, where the $sl$-tanglemeter Eq.~(\ref{EQ14bb}) suggests%
\begin{eqnarray*}
\psi _{1100} &=&\psi _{0011}=\psi _{0000}\sqrt{x} \\
\psi _{1001} &=&\psi _{1001}=\psi _{0000}\sqrt{y} \\
\psi _{0101} &=&\psi _{1010}=\psi _{0000}\sqrt{z} \\
\psi _{1111} &=&\psi _{0000}t
\end{eqnarray*}%
for the nonvanishing amplitudes, and where the notations $x=\beta _{3}^{2}$, 
$y=\beta _{5}^{2}$, $z=\beta _{6}^{2}$, and $t=\psi _{1111}=1+x+y+z$ are
employed. We introduce new variables%
\begin{align}
X\ & =\psi _{0000}^{2}\left( t+x-y-z\right) ,  \notag \\
Y\ & =\ \psi _{0000}^{2}\left( t-x+y-z\right) \,,  \notag \\
Z\ & =\psi _{0000}^{2}\left( t-x-y+z\right) \,,  \label{REP}
\end{align}%
and find%
\begin{align*}
I_{12}^{(6)}\ & =X(I^{(2)}X-YZ)\,, \\
I_{23}^{(6)}\ & =Y(I^{(2)}Y-XZ)\,, \\
I_{13}^{(6)}\ & =Z(I^{(2)}Z-XY)\,.
\end{align*}%
Solving this system of equations yields%
\begin{align}
X& =\sqrt{(I_{12}^{(6)}+P)/I^{(2)}}\ \,,  \notag \\
\ Y& =\sqrt{(I_{23}^{(6)}+P)/I^{(2)}}\,,  \notag \\
\ Z& =\sqrt{(I_{13}^{(6)}+P)/I^{(2)}}\,,  \label{SYS}
\end{align}%
where $P$ is a root \ of a cubic equation 
\begin{equation*}
(I_{13}^{(6)}+P)(I_{23}^{(6)}+P)(I_{12}^{(6)}+P)=\left( I^{(2)}\right)
^{2}P^{2}.
\end{equation*}%
Different roots of these equations  and different signs of the square roots in
Eq.~(\ref{SYS}) yield different $sl$-canonic states, that either coincide or
are related by $SL$ transformations. The amplitudes of these states can be
written explicitly%
\begin{widetext}
\begin{eqnarray}
\psi _{0000} &=&\frac{\sqrt{\sqrt{I_{13}^{(6)}+P}+\sqrt{I_{23}^{(6)}+P}+%
\sqrt{I_{12}^{(6)}+P}-\left( I^{(2)}\right) ^{3/2}}}{\sqrt{2}\left(
I^{(2)}\right) ^{1/4}},  \notag \\
\psi _{1100} &=&\psi _{0011}=\frac{\sqrt{\sqrt{I_{13}^{(6)}+P}-\sqrt{%
I_{23}^{(6)}+P}-\sqrt{I_{12}^{(6)}+P}+\left( I^{(2)}\right) ^{3/2}}}{2\left(
I^{(2)}\right) ^{1/4}},  \notag \\
\psi _{1001} &=&\psi _{1001}=\frac{\sqrt{\sqrt{I_{23}^{(6)}+P}-\sqrt{%
I_{13}^{(6)}+P}-\sqrt{I_{12}^{(6)}+P}+\left( I^{(2)}\right) ^{3/2}}}{2\left(
I^{(2)}\right) ^{1/4}},  \notag \\
\psi _{0101} &=&\psi _{1010}=\frac{\sqrt{\sqrt{I_{12}^{(6)}+P}-\sqrt{%
I_{23}^{(6)}+P}-\sqrt{I_{13}^{(6)}+P}+\left( I^{(2)}\right) ^{3/2}}}{2\left(
I^{(2)}\right) ^{1/4}},  \notag \\
\psi _{1111} &=&\frac{\sqrt{I_{13}^{(6)}+P}+\sqrt{I_{23}^{(6)}+P}+\sqrt{%
I_{12}^{(6)}+P}+\left( I^{(2)}\right) ^{3/2}}{2\sqrt{2}\left( I^{(2)}\right)
^{1/4}\sqrt{\sqrt{I_{13}^{(6)}+P}+\sqrt{I_{23}^{(6)}+P}+\sqrt{I_{12}^{(6)}+P}%
-\left( I^{(2)}\right) ^{3/2}}},  \label{AMPL}
\end{eqnarray}
\end{widetext}
while the ratios $\psi _{1100}/\psi _{0000}$, $\psi _{1001}/\psi _{0000}$,
and $\psi _{0101}/\psi _{0000}$ yield the $sl$-tanglemeter coefficients $%
\beta _{3}$, $\beta _{5}$, and $\beta _{6}$, respectively. Thus, the $sl$-entanglement
 in the $4$-qubit assembly can be completely characterized by
three independent scale--invariant complex ratios 
\begin{widetext}
\begin{eqnarray*}
\beta _{3} &=&\frac{\sqrt{\sqrt{I_{13}^{(6)}+P}-\sqrt{I_{23}^{(6)}+P}-\sqrt{%
I_{12}^{(6)}+P}+\left( I^{(2)}\right) ^{3/2}}}{\sqrt{2}\sqrt{\sqrt{%
I_{13}^{(6)}+P}+\sqrt{I_{23}^{(6)}+P}+\sqrt{I_{12}^{(6)}+P}-\left(
I^{(2)}\right) ^{3/2}}}, \\
\beta _{5} &=&\frac{\sqrt{\sqrt{I_{23}^{(6)}+P}-\sqrt{I_{13}^{(6)}+P}-\sqrt{%
I_{12}^{(6)}+P}+\left( I^{(2)}\right) ^{3/2}}}{\sqrt{2}\sqrt{\sqrt{%
I_{13}^{(6)}+P}+\sqrt{I_{23}^{(6)}+P}+\sqrt{I_{12}^{(6)}+P}-\left(
I^{(2)}\right) ^{3/2}}}, \\
\beta _{6} &=&\frac{\sqrt{\sqrt{I_{12}^{(6)}+P}-\sqrt{I_{23}^{(6)}+P}-\sqrt{%
I_{13}^{(6)}+P}+\left( I^{(2)}\right) ^{3/2}}}{\sqrt{2}\sqrt{\sqrt{%
I_{13}^{(6)}+P}+\sqrt{I_{23}^{(6)}+P}+\sqrt{I_{12}^{(6)}+P}-\left(
I^{(2)}\right) ^{3/2}}},
\end{eqnarray*}
\end{widetext}
coming from the invariants Eqs.~(\ref{6-th},\ref{J2}).

As for a measure characterizing entanglement for $4$ qubits, one has to
consider at least two quantities. The first is the sum $\sum \left\vert \psi
^{2}\right\vert $ over the amplitudes Eq.~(\ref{AMPL}), which gives the
regular normalization of the canonic-like state. Once the invariants of Eqs.~(%
\ref{J2}), (\ref{6-th}) are calculated for a state normalized to $1$, this sum
shows how the $SL$ transformation required for setting the state to the
canonic form is different from a unitary\ transformation, whence $\left\vert
\ln \sum \left\vert \psi ^{2}\right\vert \right\vert $ provides us with a
measure of this nonunitary. The root $P$ and the signs of the square roots
Eq.~(\ref{SYS}) have to be chosen such that $\left\vert \ln \sum \left\vert
\psi ^{2}\right\vert \right\vert $ is minimum. This quantity discriminates
different $su$-orbits that belong to the same $sl$-orbit in analogy to the $3
$-tangle, which discriminates different $su$-orbits within a single generic $%
SL$ orbit of $3$-qubit assembly. It may serve as a measure of $su$-entanglement
 within a $sl$-orbit. The second quantity has to discriminate
different $sl$-orbits and serve as a measure of $sl$-entanglement. A natural
candidate for that is the sum of moduli squared of the $sl$-tanglemeter
coefficients $\beta $, which takes value $0$ for $GHZ$ canonic state,
remaining larger for all other states. The choice of $P$ and the signs in
Eq.~(\ref{SYS}) has to be done such that this quantity is minimum. This
measure shows us how close is the orbit to the $GHZ$ orbit.

\begin{figure}[h]
{\centering{\includegraphics*[width=0.5\textwidth]{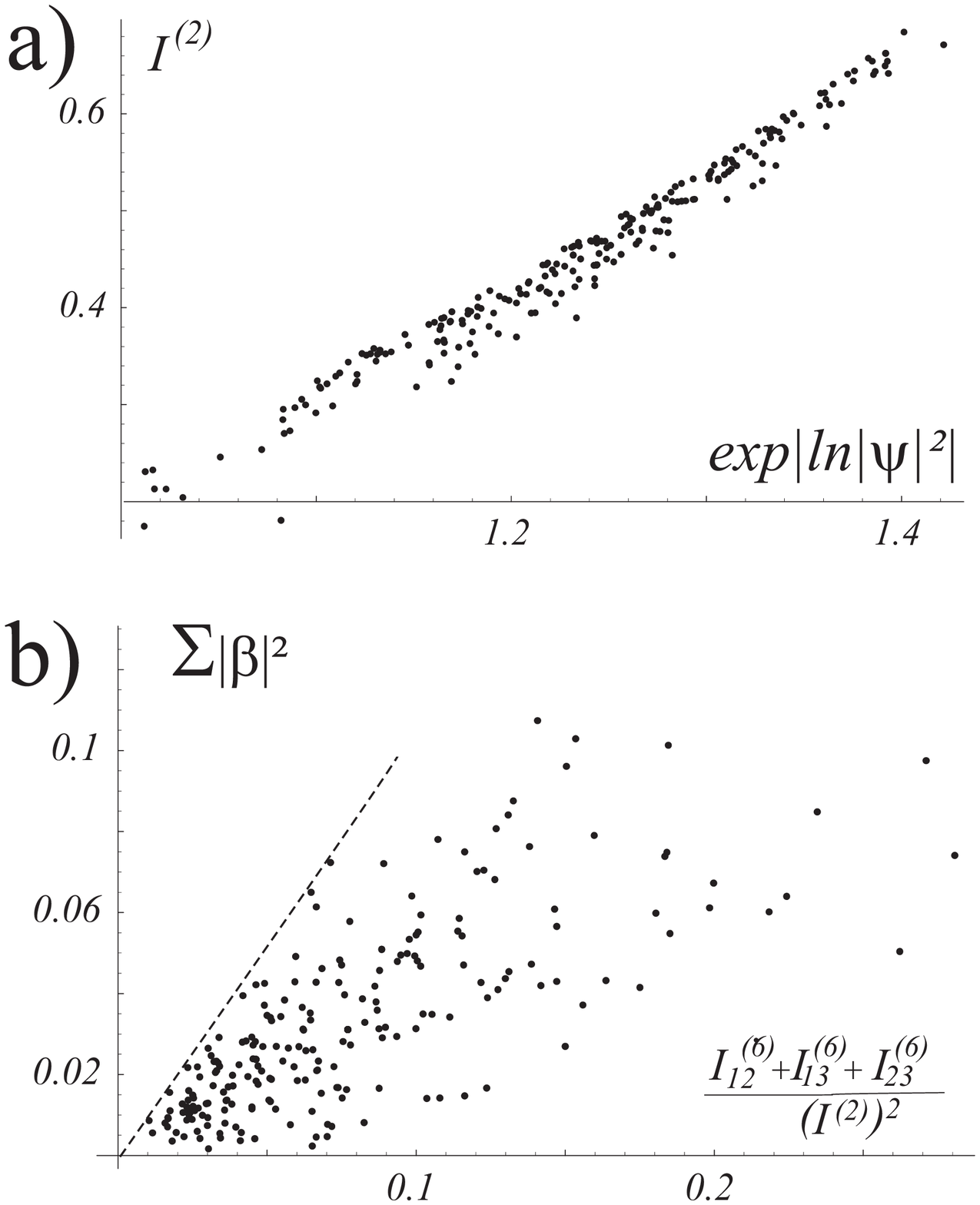}}} \vspace{0.1cm}
\caption{a) The polynomial invariant $\left\vert I^{(2)}\right\vert $
plotted versus the non-unitarity measure $\exp \left\vert \ln \sum\left\vert\psi ^{2}\right\vert \right\vert $ for a set of $\sim 10^{2}$ randomly chosen assembly states. b) Same for the combination $\left(
\left\vert I_{13}^{(6)}\right\vert +\left\vert I_{23}^{(6)}\right\vert
+\left\vert I_{12}^{(6)}\right\vert \right) /$ $\left\vert
I^{(2)}\right\vert ^{2}$ and the $\sum \left\vert \protect\beta %
^{2}\right\vert $given by the $sl$-tanglemeter.  }
\label{FIGpoly}
\end{figure}

One may also ask for simpler measures that would be polynomials on the state
amplitudes $\psi $. It turns out that two such characteristics, $\left\vert
I^{(2)}\right\vert $ and $\left( \left\vert I_{13}^{(6)}\right\vert
+\left\vert I_{23}^{(6)}\right\vert +\left\vert I_{12}^{(6)}\right\vert
\right) /$ $\left\vert I^{(2)}\right\vert ^{2}$, can be directly associated
with the $su$ and $sl$-measures. In Fig.~\ref{FIGpoly} we show these
characteristics plotted versus $\exp \left\vert \ln \sum \left\vert \psi
^{2}\right\vert \right\vert $ and $\sum \left\vert \beta ^{2}\right\vert $,
respectively, for a variety of $\sim 10^{2}$ randomly chosen assembly states
normalized to unity. One sees that $\left\vert I^{(2)}\right\vert $ strongly
correlates with the measure of non-unitarity, while the sum of the moduli of 
$6$-th order invariants majorates the $sl$-entanglement measure based on the 
$sl$-tanglemeter coefficients. Other combinations of invariants do not correlate with the tanglemeter coefficients.

\section{Quantum state operations and dynamics in terms of the nilpotent
polynomials}

In this section, we first describe the effects of local and gate
transformations on the assembly state vector as algebraic
manipulations of the corresponding polynomials $F$ and $f$. In
principle, by applying a properly chosen sequence of finite local
transformations, one can reduce a nilpotent polynomial to the canonic
form, thereby specifying the tanglemeter. However, straightforwardly
applying these transformations is not a very practical way to proceed,
since it usually requires lengthy calculations.

We therefore turn  to infinitesimal transformations second, and derive
the equations of motion describing the dynamics of the nilpotential under
continuous local and gate operations. We show that, for an important
class of Hamiltonians supporting universal quantum computation, the
dynamic equation for the nilpotential acquires a well-known
Hamilton-Jacobi form.

We thirdly demonstrate how to determine the tanglemeter with the help
of such equation. To this end, a proper feedback is required, ensuring
that the parameters of infinitesimal $SU(2)$ or $SL(2,\mathbb{C})$
transformations are adjusted to track current values of the
nilpotential coefficients. The tanglemeter appears as a stable
stationary solution, that is a focus of the resulting equation.  We
illustrate this method in the example of four qubits. We show how to
find the $sl$-tanglemeter for a generic $4$-qubit state and how to
explicitly identify a number of special classes that cannot be reduced
to this form. For these classes we suggest alternative natural
tanglemeters.

\subsection{Local operations}

A general local unitary transformation
Eq. (\ref{EQ1a}) applied to $i$-th qubit
can also be expressed in the equivalent form%
\begin{equation}
U_{i}=\mathrm{e}^{A_{i}\sigma_{i}^{-}}\mathrm{e}^{B_{i}\sigma_{i}^{z}%
}\mathrm{e}^{C_{i}\sigma_{i}^{+}},~~ \label{EQ16}%
\end{equation}
which better suits the consideration in terms of nilpotential polynomials,
since each step of the local transformation can be expressed as
an operation linear in $\sigma$, that is
\begin{align}
\mathrm{e}^{C_{i}\sigma_{i}^{+}}&=1+C_i\sigma_{i}^{+}\,,\nonumber\\
\mathrm{e}^{B_{i}\sigma_{i}^{z}}&=\mathrm{cos}\left(B_i\right)+\mathrm{sin}\left(B_i\right)\sigma_{i}^{z}\,,\nonumber\\
\mathrm{e}^{A_{i}\sigma_{i}^{-}}&=1+A_i\sigma_{i}^{-}\,.\nonumber
\end{align}
 The
explicit expressions
\begin{align}
A_{i} & =\frac{\left( \mathrm{i}P_{i}^{x}-P_{i}^{y}\right) \sin
P}{P\cos P+\mathrm{i}P_{i}^{z}\sin P};\ B_{i}=\log\left[ \cos
P+\frac{\mathrm{i}
P_{i}^{z}\sin P}{P}\right]; \nonumber\\ C_{i} &
=\frac{\left( \mathrm{i}P_{i}^{x}+P_{i}^{y}\right) \sin P}{P\cos
P+\mathrm{i}P_{i}^{z}\sin P};\ P=\sqrt{%
{\textstyle\sum\limits_{\kappa=x,y,z}}
\left(  P_{i}^{\kappa}\right)  ^{2}} \label{EQ17}%
\end{align}
relate the parameters in Eq.~(\ref{EQ1a}) and Eq.~(\ref{EQ16}).

The transformations Eq.~(\ref{EQ16}) act on the state vector $|\Psi\rangle
= F(\{\sigma_i^+\}) |\mathrm{ O} \rangle $ and yield a transformed
state $F'(\{\sigma_i^+\}) |\mathrm{ O} \rangle $. One can formalize the
rules allowing one to obtain $F'$ from $F$. Bearing in mind that
$\sigma^- |0\rangle = 0$ and $\sigma^z |0\rangle = - |0\rangle$, one
can represent the action of $\sigma_i^+, \sigma_i^z, \sigma_i^-$ as
appropriate differential operations for the nilpotent variable $\sigma^+_i$.  The application of the operator
$\sigma_{i}^{+}$ is straightforward -- it is a direct multiplication:
this operation eliminates the terms that were proportional to
$\sigma_{i}^{+}$ prior to the multiplication.  The application of
$\sigma_{i}^{-}$ is a kind of inverse: it can be considered as a
derivative with respect to the variable $\sigma_{i}^{+}$, which
eliminates the terms independent of $\sigma_{i}^{+}$ and makes the
terms linear in $\sigma_{i}^{+}$ independent of this variable
\cite{integration}.  Finally, the application of $\sigma_{i}^{z}$
changes the signs of the terms independent of $\sigma_{i}^{+}$, and
leaves intact  terms linear in $\sigma^+_i$. These actions are summarized by the following 
formulae%
\begin{align}
\sigma_{i}^{+}F  &  =\sigma_{i}^{+}F\,,\nonumber\\
\sigma_{i}^{-}F  &  =\frac{\partial F}{\partial\sigma_{i}^{+}}\,,\label{EQ21aa}\\
\sigma_{i}^{z}F  &  =-F+2\sigma_{i}^{+}\frac{\partial F}{\partial\sigma
_{i}^{+}},\nonumber
\end{align}
while each unitary operation $U\left(\sigma_i^x,\sigma_i^y,\sigma_i^z\right)$
can be represented by a differential operator $U\left(\sigma_i^+,\frac{\partial }{\partial\sigma_{i}^{+}},
2\sigma_{i}^{+}\frac{\partial}{\partial\sigma_{i}^{+}}-1\right)$.

By sequentially applying the three transformations of Eq.~(\ref{EQ16})
to $F$, a local transformation can be interpreted as multiplication by
an exponential function of $\sigma_{i}^{+}$, followed by a linear
transformation $\sigma_{i}^{+}\mapsto \mathrm{e}^{2B_{i}}(A_{i} +
\sigma_{i}^{+})$ of the variable $\sigma_{i}^{+}$ and multiplication
by $e^{-B_i}$, leading to
\begin{widetext}%
\begin{align}
U_{i}F(\sigma_{1}^{+},\ldots,\sigma_{i}^{+},\ldots,\sigma_{n}^{+})  
=  & \, \mathrm{e}^{A_{i}\sigma_{i}^{-}}\mathrm{e}^{B_{i}\sigma_{i}^{z}}F(\sigma
_{1}^{+},\ldots,\sigma_{i}^{+},\ldots,\sigma_{n}^{+})\mathrm{e}^{C_{i}
\sigma_{i}^{+}} 
 = \mathrm{e}^{A_{i}\sigma_{i}^{-}}\mathrm{e}^{B_{i}\sigma
_{i}^{z}}G(\sigma_{1}^{+},\ldots,\sigma_{i}^{+},\ldots,\sigma_{n}
^{+})\nonumber\\
= & \, \mathrm{e}^{-B_{i}}G(\sigma_{1}^{+},\ldots,\mathrm{e}^{2B_{i}}
A_{i}+\mathrm{e}^{2B_{i}}\sigma_{i}^{+},\ldots,\sigma_{n}^{+})~.
\label{EQ18a}%
\end{align}
\end{widetext}
Since local operations on different qubits commute, this single-qubit
transformation may be straightforwardly generalized to $n$ qubits.

Note that in order to cast $F$ into the canonic form,%
\begin{widetext}
\begin{equation}
F(\{\sigma_{i}^{+}\})  
=\sum_{\left\{  k_{i}\right\}  ={0,1}}\alpha
_{k_{n} k_{n-1}\ldots k_{1}}\prod_{i=1}^{n}\left(  \sigma_{i}^{+}\right)
^{k_{i}} \: \mapsto \label{EQ19}
F_{c}(\{\sigma_{i}^{+}\})  
=\sum_{\left\{  k_{i}\right\}  ={0,1};\,k_{1}%
+\ldots+k_{n}\neq 1}\alpha_{k_{n} k_{n-1}\ldots k_{1}}^{\prime}\prod
_{i=1}^{n}\left(  \sigma_{i}^{+}\right)  ^{k_{i}} \ , 
\nonumber
\end{equation}
\end{widetext}
one has to solve a set of nonlinear equations for the parameters
$A_{i}$, $B_{i}$, and $C_{i}$. This can be done explicitly only for at most
 four qubits, while for a larger system an efficient
numerical technique is required.  This task can be accomplished by an
iterative procedure in the spirit of the Newton algorithm, that is, by
consecutively applying a series $U_{n}\ldots U_{2}U_{1}$ of linear
transformations $U_{i}$, each of which eliminates the terms linear in
$\sigma_{i}^{+}$. However, this procedure may require infinitely many
iterations, since a linear transformation applied to one of the
$\sigma_{i}^{+}$ may (and usually does) generate terms linear in other
$\sigma_{j\neq i}^{+}$ .
In Sect.~\ref{u} we show how dynamic equations describing the
evolution of the nilpotential $f$ under local transformations offer a
better tool to solve this problem.

\subsection{Two-qubit gate operations}

Quantum gates are unitary transformations acting on finite subsets of
qubits in the assembly.  In particular, two-qubit gates $U_{ij}$
operate non-trivially on the  pair $\{i,j\}$.  Thanks to general {\em
universality} results \cite{QI,Barenco}, an arbitrary non-local
transformation on $n$ qubits may be expressed as a finite sequence of
arbitrary single-qubit operations and two-qubits operations drawn from
a standard set, applied to both individual and pairs of qubits
according to a certain quantum network.  Thus, starting from an
initial computational state, {\em any} state may be reached through
the application of a quantum circuit built from gates in the set.  We
consider here the simplest choice for the standard two-qubit gate
operation,
\begin{align}
U_{ij}  &  \hspace*{-.5mm}=\hspace*{-.5mm}
\exp\left[  \mathrm{i}t(\sigma_{i}^{+}\sigma_{j}^{-}+\sigma
_{i}^{-}\sigma_{j}^{+})\right]  =\exp\left[  \mathrm{i}\frac{t(\sigma_{i}%
^{x}\sigma_{j}^{x}+\sigma_{i}^{y}\sigma_{j}^{y})}{2}\right]  ~\nonumber\\
& \hspace*{-.5mm} =\hspace*{-.5mm}\cos^{2}\frac{t}{2}+
\sigma_{i}^{z}\sigma_{j}^{z}\sin^{2}\frac{t}%
{2}+\mathrm{i}\frac{\sigma_{i}^{x}\sigma_{j}^{x}+\sigma_{i}^{y}\sigma_{j}^{y}%
}{2}\sin t\,, \label{EQ20}%
\end{align}
depending on the single parameter $t \in {\mathbb R}$, where the
tensor product symbol, $\sigma_{i}^{\kappa}\sigma
_{j}^{\varkappa}=\sigma_{i}^{\kappa}\otimes\sigma_{j}^{\varkappa}$, is
implicit.

Only the terms of $F$ that contain $\sigma_{i}^{+}$ or
$\sigma_{j}^{+}$ are affected by the transformation Eq.~(\ref{EQ20}). The
terms that either do not contain these variables or are proportional
to their product are left intact.

The nilpotent polynomials $A_{i}=A_{i}\left( \left\{ \sigma_{k\neq
i,j}^{+}\right\} \right)$ and $A_{j}=A_{j}\left( \left\{ \sigma_{k\neq
i,j}^{+}\right\} \right)$ which are the coefficients in front of the
variables $\sigma_{i}^{+}$ and $\sigma_{j}^{+}$, respectively, undergo
a unitary rotation
\begin{align}
A_{i}\sigma_{i}^{+}  &  \mapsto A_{i}\sigma_{i}^{+}\cos t+\mathrm{i}%
A_{j}\sigma_{j}^{+}\sin t\nonumber\\
A_{j}\sigma_{j}^{+}  &  \mapsto A_{j}\sigma_{j}^{+}\cos t+\mathrm{i}%
A_{i}\sigma_{i}^{+}\sin t\,. \label{EQ21}%
\end{align}
in the same way as the components of a qubit state vector do under an
$SU(2)$ transformation.

\vfill
\subsection{Local and gate operations in terms of the nilpotential}

Equations (\ref{EQ18a})-(\ref{EQ21}) are particular cases of general
expressions for transformations of nilpotent polynomials $F$ under the
action of unitary operations. We now consider this in terms of the
nilpotential.

The  rules of  Eq.~(\ref{EQ16}), allows one  to express
  the   action   of   a  unitary   operation   $U\left(   \left\{
\sigma_{i}^{+},\sigma_{i}^{-},\sigma_{i}^{z}\right\}  \right)  $
as a differential operator acting  on  the
nilpotent polynomial $F$ 
\begin{equation}
F^{\prime}=U\left(  \left\{  \sigma_{i}^{+},\frac{\partial}{\partial\sigma
_{i}^{+}},2\sigma_{i}^{+}\frac{\partial}{\partial\sigma_{i}^{+}}-1\right\}
\right)  F\,, \label{EQ21ab}%
\end{equation}
while for the nilpotential one finds%
\begin{equation}
f^{\prime}=\log\left(  U\left(  \left\{  \sigma_{i}^{+},\frac{\partial
}{\partial\sigma_{i}^{+}},2\sigma_{i}^{+}\frac{\partial}{\partial\sigma
_{i}^{+}}-1\right\}  \right)  \mathrm{e}^{f}\right)\,. \label{EQ21ac}%
\end{equation}
Note that a generic transformation Eq.~(\ref{EQ21ac}) of an initially
canonic polynomial does {\em not} necessarily results in another
canonic polynomial.

Let us consider two particular cases of the general transformation of
Eq.~(\ref{EQ21ac}): \textit{(i)} A local unitary operation 
Eq.~(\ref{EQ1a}) with $P_{i}^{x} =P\cos\phi$, $P_{i}^{y}=P\sin\phi$,
$P^{z}=0$; \textit{(ii)} The two-qubit gate  Eq.~(\ref{EQ20}). They
 transform the nilpotential according to
\begin{widetext}%
\begin{equation}
f^{\prime}=f+\ln\left(  \cos P +\mathrm{ie}^{\mathrm{i}\phi}\frac{\partial
f}{\partial\sigma_{i}^{+}}\sin P\right)  -i\sigma_{i}^{+}\frac{\mathrm{e}%
^{\mathrm{i}\phi}\left(  \frac{\partial f}{\partial\sigma_{i}^{+}}\right)
^{2}-\mathrm{e}^{-\mathrm{i}\phi}}{1+\mathrm{ie}^{\mathrm{i}\phi}%
\frac{\partial f}{\partial\sigma_{i}^{+}}\tan P}\tan P\,, \label{EQ21bc}%
\end{equation}
and
\begin{align}
f^{\prime}=f+2\,\sigma_{i}^{+}\,\sigma_{j}^{+}\,\left(  1-\cos t\right)
\,\frac{\partial^{2}f}{\partial\sigma_{i}^{+}\partial\sigma_{j}^{+}}  &
-\left(  \sigma_{j}^{+}-\sigma_{j}^{+}\,\cos t-\mathrm{{i}\,}\sigma_{i}%
^{+}\,\sin t\right)  \,\frac{\partial f}{\partial\sigma_{j}^{+}}-\left(
\sigma_{i}^{+}-\sigma_{i}^{+}\,\cos t-\mathrm{{i}\,}\sigma_{j}^{+}\,\sin
t\right)  \frac{\partial f}{\partial\sigma_{i}^{+}}\nonumber\\
&  -\mathrm{{i}\,}\sigma_{i}^{+}\,\sigma_{j}^{+}\,\bigg(  \frac{\sin2t\,\,}%
{2}\left(  \frac{\partial f}{\partial\sigma_{i}^{+}}\right)  ^{2}%
+2\mathrm{i}\,\sin^{2}t\,\frac{\partial f}{\partial\sigma_{i}^{+}}%
\,\frac{\partial f}{\partial\sigma_{j}^{+}}+\frac{\sin2t\,\,}{2}\bigg(
\frac{\partial f}{\partial\sigma_{j}^{+}}\bigg)^{2}\bigg)\,,
\label{EQ18bd}%
\end{align}
\end{widetext}
respectively.

\subsection{Equations of motion for the nilpotential}

Consider now an infinitesimal unitary  transformation $U=1-\mathrm{i\,d}t\,H$  which is not necessarily local.
The increment $\Delta f $ of the nilpotential $f$ suggested by the Eq.~(\ref{EQ21bc}) reads 
\begin{equation}
\Delta f=\log\left(  U\mathrm{e}^{f}\right)  -\log\left(  \mathrm{e}%
^{f}\right)  = \log\left(  1-\mathrm{i\;d}t\;\mathrm{e}^{-f}H\mathrm{e}%
^{f}\right)  \ . \label{EQ21ad}%
\end{equation}

This yields the following dynamic equation for $f$, 
\begin{equation}
\mathrm{i}\frac{\partial f}{\partial t}=\mathrm{e}^{-f}H\mathrm{e}^{f}\:, 
\label{EQ21af}%
\end{equation}
which we discuss in detail in the rest of this section.

\subsubsection{Local Hamiltonians}

We begin with the case of a local Hamiltonian
\begin{align}
H  &  =\sum_{i}P_{i}^{x}(t)\sigma_{i}^{x}+P_{i}^{y}(t)\sigma_{i}^{y}+P_{i}%
^{z}(t)\sigma_{i}^{z}\nonumber\\
&  =\sum_{i}P_{i}^{-}(t)\sigma_{i}^{+}+P_{i}^{+}(t)\sigma_{i}^{-}+P_{i}%
^{z}(t)\sigma_{i}^{z}\ , \label{EQ48}%
\end{align}
where $P_{i}^{\pm}=P_{i}^{x}\pm iP_{i}^{y}$, 
and we first separately consider only the  term
$H_i=P_{i}^{-}(t)\sigma_{i}^{+}+P_{i}^{+}(t)\sigma_{i}^{-}+P_{i}%
^{z}(t)\sigma_{i}^{z}$ in the sum. Upon substituting it in
 Eq.~(\ref{EQ21af}) and splitting the nilpotential $f$ on the right hand side
 in two parts, the part
$f_{0}=f-\sigma_{i}^{+}\partial f/\partial\sigma_{i}^{+}$ independent
of $\sigma_{i}^{+}$, and the part $f_{1}=\sigma_{i}^{+}\partial
f/\partial\sigma_{i}^{+}$ linear in $\sigma_{i}^{+}$, we obtain
\begin{equation}
\mathrm{i}\frac{\partial f}{\partial t}=\mathrm{e}^{-\sigma_{i}^{+}\partial
f/\partial\sigma_{i}^{+}}H_i\mathrm{e}^{\sigma_{i}^{+}\partial f/\partial
\sigma_{i}^{+}}\,. \label{EQ48a}%
\end{equation}
The part $f_{0}$ commutes with the derivatives entering the
Hamiltonian and therefore cancels. Substitution of Eq.~(\ref{EQ21aa})
into the Hamiltonian of Eq.~(\ref{EQ48}) followed by expansion  over the
nilpotent variable $\sigma_{i}^{+}$ results in 
\begin{widetext}%
\begin{equation}
\mathrm{i}\frac{\partial f}{\partial t}=-P_{i}^{z}+P_{i}^{-}\sigma_{i}%
^{+}+\left(  2P_{i}^{z}\sigma_{i}^{+}+P_{i}^{+}\right)  \frac{\partial
f}{\partial\sigma_{i}^{+}}-P_{i}^{+}\sigma_{i}^{+}\left(  \frac{\partial
f}{\partial\sigma_{i}^{+}}\right)^{2}\,. \label{EQ48b}%
\end{equation}
Straightforward generalization of this equation to the
case of the Hamiltonian  Eq.~(\ref{EQ48}) yields
\begin{equation}
\mathrm{i}\frac{\partial f}{\partial t}=\sum_{i=1}^{n}\bigg[  -P_{i}^{z}%
+P_{i}^{-}\sigma_{i}^{+}+\left(  2P_{i}^{z}\sigma_{i}^{+}+P_{i}^{+}\right)
\frac{\partial f}{\partial\sigma_{i}^{+}}-P_{i}^{+}\sigma_{i}^{+}\left(
\frac{\partial f}{\partial\sigma_{i}^{+}}\right)  ^{2}\bigg]  .
\label{EQ49.11}%
\end{equation}
Another equivalent form of the same equation reads
\begin{equation}
\mathrm{i}\frac{\partial f}{\partial t}=\sum_{i=1}^{n}\left[  P_{i}^{z}\left(
2f_{i}\sigma_{i}^{+}-1\right)  +P_{i}^{x}\left(  f_{i}+\sigma_{i}^{+}%
-f_{i}^{2}\sigma_{i}^{+}\right)  +\mathrm{i}P_{i}^{y}\left(  f_{i}-\sigma
_{i}^{+}-f_{i}^{2}\sigma_{i}^{+}\right)  \right] \, , \label{EQ49.10}%
\end{equation}
\end{widetext}
where we denote $f_{i}={\partial f}/{\partial\sigma_{i}^{+}}$. Note
that the coefficients $P_{i}^{x,y,z}$ can be  functions of
time.  Also, note that the right hand-side of
Eqs.~(\ref{EQ48b})-(\ref{EQ49.10}) does not depend on the constant term
in $f$. In other words, the latter, though evolving with time by
itself, does not affect the evolution of the ``essential''
coefficients in the nilpotential in front  of the nilpotent
variables and their products.

\subsubsection{Binary interactions}

We now consider the binary interaction
\begin{equation}
H=\sum_{i,j,\kappa,\varkappa}G_{ij}^{\kappa\varkappa}(t)\sigma_{i}^{\kappa
}\sigma_{j}^{\varkappa}\,,\qquad\kappa,\varkappa=+,-,z \,,\label{EQ50}%
\end{equation}
among the qubits. Note that the local transformations
Eq.~(\ref{EQ49.10}) can be absorbed into the time dependence of the
coupling coefficients $G_{ij}^{\kappa\varkappa}(t)$ by simply passing
to the interaction representation. In order to achieve universal
evolution in this representation, one needs to consider all nine
coefficients $G_{ij}^{\kappa\varkappa}$ characterizing the
interaction of Eq.~(\ref{EQ50}) between a  pair $\left\{i,j\right\}$ of qubits 
 as being different from zero. An alternative way is to chose such a representation
  that tensor $G_{ij}^{\kappa\varkappa}(t)$ in
Eq.~(\ref{EQ50}) takes the form of a diagonal spherical tensor with
respect to the upper
indices. In this representation, the Hamiltonian
\begin{align}
H  &  =\sum_{i,\kappa}P_{i}^{\kappa}(t)\sigma_{i}^{\kappa}+\sum_{i,j}%
G_{ij}^{+-}(t)\left(  \sigma_{i}^{+}\sigma_{j}^{-}+\sigma_{i}^{-}\sigma
_{j}^{+}\right) \label{EQ51}\\
&  +\sum_{i,j}G_{ij}^{zz}(t)\sigma_{i}^{z}\sigma_{j}^{z}+\sum_{i,j}G_{ij}%
^{++}(t)\left(  \sigma_{i}^{+}\sigma_{j}^{+}+\sigma_{i}^{-}\sigma_{j}%
^{-}\right)  \,,\nonumber
\end{align}
apart of the local operations Eq.~(\ref{EQ49.11})
involves also the binary interactions determined by only three real
coupling parameters $G_{ij}^{zz}(t)$, $G_{ij}^{+-}(t)=G_{ij}^{-+}(t)$,
and $G_{ij}^{++} (t)=G_{ij}^{--}(t)$. 

The explicit forms of the equations of motion Eq.~(\ref{EQ21af}) for the
Hamiltonians Eqs.~(\ref{EQ50}),(\ref{EQ51}) are rather awkward.  We note,
however, that universal evolution is achieved \cite{Barenco} with
an even simpler Hamiltonian
\begin{align}
H  &  =\sum_{i}P_{i}^{+}(t)\sigma_{i}^{+}+\sum_{i}P_{i}^{-}(t)\sigma_{i}%
^{-}\label{EQ52}\\
&  +\sum_{i<j}G_{ij}(t)\left(  \sigma_{i}^{+}\sigma_{j}^{-}+\sigma_{i}%
^{-}\sigma_{j}^{+}\right) ,\nonumber
\end{align}
with $P_{i}^{+}(t)=P_{i}^{-}(t)^{\ast}$, which depends on a smaller
set of operators, $\sigma_{i}^{+}$, $\sigma_{i}^{-}$, and $\left(
\sigma_{i}^{+}\sigma_{j}^{-}+\sigma_{i}^{-}\sigma_{j}^{+}\right)
$. Repeated commutators of these operators satisfy the Lie-algebraic
bracket generation condition for complete controllability, that is,
all-order commutators span the full space of Hermitian operators for
the assembly, and thus ensures universal evolution. It therefore
suffices to specify the form of Eq.~(\ref{EQ21af}) for the Hamiltonian
of Eq.~(\ref{EQ52}).

In Appendix C, we derive the corresponding equation of motion for $f$.
It reads%
\begin{align}
\mathrm{i}\frac{\partial f}{\partial t}  &  =\sum_{i}\left[  P_{i}%
^{-}(t)\sigma_{i}^{+}+P_{i}^{+}(t)\frac{\partial f}{\partial\sigma_{i}^{+}%
}\left(  1-\sigma_{i}^{+}\frac{\partial f}{\partial\sigma_{i}^{+}}\right)
\right] \label{EQ56}\\
&  +\sum_{i \neq j}G_{ij}(t)\sigma_{j}^{+}\frac{\partial f}{\partial\sigma_{i}^{+}
}\left(  1-\sigma_{i}^{+}\frac{\partial f}{\partial\sigma_{i}^{+}}\right)
\,.\nonumber
\end{align}
Note that Eq.~(\ref{EQ56}) formally resembles the Hamilton-Jacobi equation for
the mechanical action of classical systems with the Hamiltonian
\begin{align}
H  &  =\sum_{i}\left[  P_{i}^{-}(t)x_{i}+P_{i}^{+}(t)p_{i}\left(  1-x_{i}%
p_{i}\right)  \right] \nonumber\\
&  + \sum_{i \neq j}G_{ij}(t)\left[  x_{j}p_{i}\left(  1-x_{i}p_{i}\right)
\right] \,. \label{EQ56a}%
\end{align}
where $p_{i}={\partial f}/{\partial\sigma_{i}^{+}}$ plays a role of
the momentum, while $x_{i}=\sigma_{i}^{+}$ are the
coordinates. Comparing with the conventional classical Hamilton-Jacobi
equation, the only essential difference is the factor $\mathbf{i}$ multiplying
the time derivative and the presence of complex parameters that can be
interpreted as time-dependent forces and masses.  After cumbersome
calculations taking into account the fact that the constants of
motions entering the action function are nilpotent variables, one can
reproduce the finite transformations of Eqs.~(\ref{EQ21bc}-\ref{EQ18bd}).

\begin{figure}[hb]
{\centering{\includegraphics*[width=0.5\textwidth]{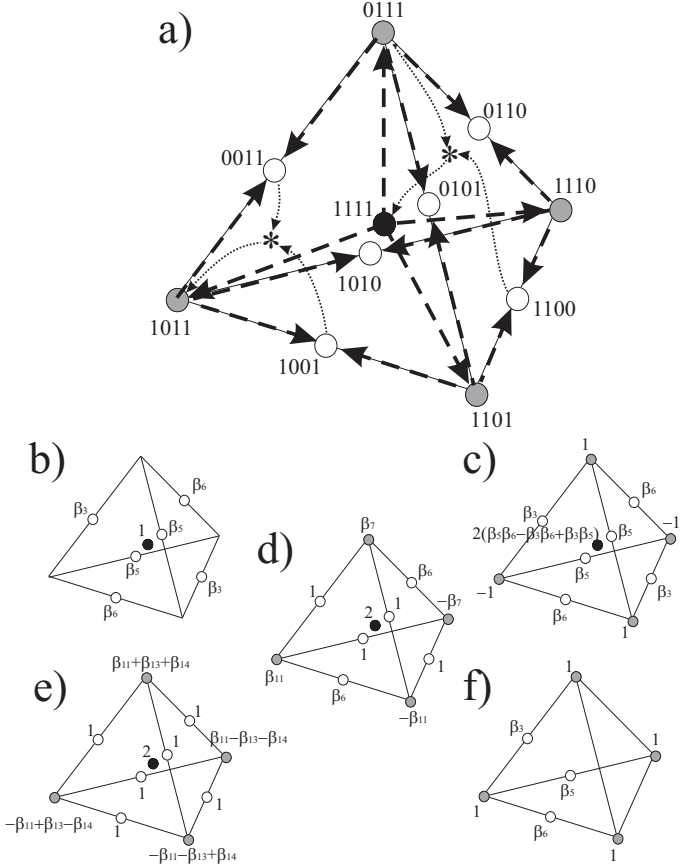}}} \vspace
{0.1cm}\caption{Diagram illustrating the dynamics of the coefficients
$\beta_{i}$ of the nilpotential $f$ for a $4$-qubit system subject to
$SL(2,\mathbb{C})$ transformations. The coefficients corresponding to
quadric, cubic, and bilinear terms are shown by black, gray, and white
circles, respectively. a) Flux directions. The $SL(2,\mathbb{C})$
transformations are applied in such a way, that the polynomials remain
in the canonic form with respect to $su(2)$ transformations, with all
linear terms $\beta_{i}\sigma _{i}^{+}=0$. The dashed arrows show the
linear contributions of different states to the time derivatives of
the neighboring states, and the dotted arrows jointed by asterisk
depict two types of bilinear contributions. The other, similar terms
may be constructed by symmetry. b) Canonic form for the generic state
and for singular classes corresponding to one c), two d), three e),
and four f) vanishing eigenvalues of the determinant
Eq.~(\ref{EQ62e}).}%
\label{FIG2}%
\end{figure}

\subsubsection{\label{u} Dynamic equations for the nilpotential and 
construction of $SU(2)$ and  $SL(2,\mathbb{C})$ tanglemeters}

The dynamic equation Eq.~(\ref{EQ56}) suggests an algorithm for evaluation
of the tanglemeter. This is based on the idea of feedback
by adjusting the parameters $P_i$ of the local transformations in
Eq.~(\ref{EQ56}) in function of the current values of the tanglemeter's 
coefficients.
 To this end, we fix the terms linear in $\sigma_{i}^{+}$ in the
local Hamiltonian of Eq.~(\ref{EQ48}) as
\begin{equation}
P_{i}^{-}=\left(  P_{i}^{+}\right)^{\ast}=-\mathrm{i}\beta_{i}\,,
\label{EQ60}%
\end{equation} 
where 
\begin{equation}
\beta_{i}=\left. \frac{\partial f}{\partial\sigma_{i}^{+}}
\right |_{\sigma \longrightarrow 0}\,, \label{EQ58}%
\end{equation}
are the coefficients of the linear terms in the nilpotential at a
given time.  From Eqs.~(\ref{EQ56},\ref{EQ60}) we find the evolution of these coefficients
under local transformations,%
\begin{equation}
\frac{\partial\beta_{i}}{\partial t}=-\beta_{i}+\sum_{i=1}^{n}\bigg[
\beta_{j}^{\ast} \left. \frac{\partial^{2}f}{\partial\sigma_{j}^{+}\partial\sigma
_{i}^{+}}\right |_{\sigma\longrightarrow0}\hspace*{-3mm}-\beta_{i}\left\vert \beta
_{i}\right\vert ^{2}\bigg] \,. \label{EQ61}%
\end{equation}

When $f$ is close to $f_c$, the matrix of the second derivatives
$M_{ij}=(\partial^{2}f/\partial\sigma_{j}^{+}\partial\sigma_{i}^{+})
|_{\sigma\longrightarrow0}$ can be explicitly expressed in terms of
the ``second excited state amplitudes", $\psi_{0\ldots
k_{i}=1,k_{j}=1,\ldots0}$, which enter Eq.~(\ref{EQ5}) and were
introduced when discussing the $su$-canonic form of states. According
to Eq.~(\ref{EQ13a}),
\begin{align}
M_{ij}^C = & \left. \frac{\partial^2 f_{c}} 
{\partial\sigma_{j}^{+}\partial\sigma_{i}^{+}} \right 
|_{\sigma\longrightarrow 0} \nonumber \\
= & \beta_{ij} =\alpha_{ij} = \psi_{0\ldots k_{i}
=1,k_{j}=1,\ldots0}/\psi_{0,\ldots0}\,. \nonumber
\end{align} 
The condition of the maximum reference state population for a state in
the canonic form suggested by Eqs.~(\ref{EQ5}-\ref{EQ6}) implies
that the population increment,
\begin{align*}
\delta\rho_{\left\vert \mathrm{O}\right\rangle }  & \hspace*{-.8mm} = 
\hspace*{-.5mm}\left\vert
\psi_{0\ldots0}\right\vert^2 \hspace*{-.5mm} \bigg[ \bigg \vert 1\hspace*{-.5mm} - 
\hspace*{-.5mm}\frac{1}{2}\sum_{i,j}g_{i}\left(  
\mathrm{e}^{2\mathrm{i}(\varphi_{i}+\varphi_{j})}
M_{ij}+\delta_{ij}\right)  g_{j} \bigg\vert^{2} \hspace*{-2mm}- 1 
\hspace*{-.5mm}\bigg] \\
&  = - {\rm Re} \left\vert \psi_{0\ldots0}\right\vert^{2} 
\sum_{i,j}g_{i}\left(e^{2i(\varphi_{i}+\varphi_{j})}
M_{ij}+\delta_{ij}\right) g_{j} \,,%
\end{align*}
is always negative. As the phases $\varphi_i$ are arbitrary, it
follows that the eigenvalues of $M_{ij}$ lie within the unit circle
and hence their real parts lie in the interval $(-1,1)$. Therefore the equation
(\ref{EQ61}) linearized in the vicinity of the canonic state
\begin{align}
\frac{\partial\beta_{i}}{\partial t}  &  =-\beta_{i}+\sum_{j}M_{ij}\beta
_{j}^{\ast}\,,\nonumber\\
\frac{\partial\beta_{i}^{\ast}}{\partial t}  &  =-\beta_{i}^{\ast}+\sum
_{j}M_{ij}^{\ast}\beta_{j}\,, \label{EQ62}%
\end{align}
has a stable stationary point at $\left\{ \beta_{i}\right\} =0$, which
implies that the coefficients $\beta_{i}$ standing in front of the
linear terms $\beta_{i}\sigma_{i}^{+}$ in the nilpotential tend to
zero exponentially. The presence of the nonlinear terms $\beta_{i}\left\vert
\beta_{i}\right\vert ^{2}$ yet accelerates this trend.  
Therefore, an arbitrary nilpotential $f$ subject to the local transformation  with
the parameters of the Hamiltonian Eq.~(\ref{EQ48}) chosen according to the feedback conditions
Eqs.~(\ref{EQ60},\ref{EQ58}), rapidly converges to the tanglemeter $f_c$.
The problem of finding an efficient numerical algorithm for determining the tanglemeter for
large assemblies is thereby solved.   Verification that the outcome
indeed corresponds to the global maximum of the reference state
population should finalize the procedure. Note however, that the maximum vacuum state population obtained with local transformations corresponds to the maximum population of the ground state for each qubit. On the other hand, for a given set of single-qubit density matrices, the local operations maximizing the ground state population of each qubits are uniquely defined. Therefore, the only maximum of the reference state population is the global one, and hence, no matter what the initial state is, the procedure indeed converges to the canonic state and no verification is required.

A  procedure of reducing the nilpotential to the canonic form  can be carried out also for 
$SL$ transformations. At the first stage of this procedure,
 we  reduce it to the $su$-canonic form so that
the terms linear in $\sigma_i^+$ are absent. 
 Then we apply $SL$ operations. 
An element of the $SL(2, \mathbb{C})$ group can be
represented as $\exp\{ it (P_i^- \sigma_i^+ + P_i^+ \sigma_i^- + P_i^z \sigma_i^z ) \}$,
where  $P_{i}^{-}$ and $P_{i}^{+}$ are no longer complex conjugates  and 
$P_i^z$ is also a complex number.

Finding the $sl$-canonic state can also be formulated as a 
control problem, based on the feedback. 
We choose the parameters $P_i$ in the Hamiltonian Eq.~(\ref{EQ49.11})
in such a way that the terms in the nilpotential involving
the monomials of order one and of order $n-1$ in $\sigma_i^+$ would decrease exponentially with time.
To this end, we may choose at this stage $P_i^z = 0$ and impose two conditions, ({\it i}) the condition 
\begin{equation}
P_{j}^{-}=-\sum_{i=1}^{n}\left.  P_{i}^{+}\frac{\partial^{2}f}{\partial
\sigma_{i}^{+}\partial\sigma_{j}^{+}}\right\vert _{\sigma\rightarrow0}%
=-\sum_{i=1}^{n}P_{i}^{+}\beta_{i,j}\,, \label{EQ62a}%
\end{equation}
expressing $P^-_i$ via $P^+_i$ 
which is keeping the nilpotential in the form of tanglemeter, and ({\it ii}) the condition
\begin{align}
&  \left.  \mathrm{i}\frac{\partial^{n-1}f}{%
{\textstyle\prod_{i\neq j}}
\partial\sigma_{i}^{+}}\right\vert _{\sigma\rightarrow0}=-P_{j}^{+}%
\frac{\partial^{n}f}{%
{\textstyle\prod_i}
\partial\sigma_{i}^{+}}\label{EQ62b}\\
&  +\sum_{m=1}^{n}P_{k}^{+}\left.  \frac{\partial^{n-1}}{%
{\textstyle\prod_{i\neq m}}
\partial\sigma_{i}^{+}}\left[  \sigma_{i}^{+}\left(  \frac{\partial
f}{\partial\sigma_{i}^{+}}\right)  ^{2}\right]  \right\vert _{\sigma
\rightarrow0}\,,\nonumber
\end{align}
which ensures the exponential decrease of all $n$ coefficients in
front of the second-highest order terms.

After having eliminated the monomials of orders $1$ and $n-1$, we can specify the
scaling parameters $P_i^z$ such that $n$ additional conditions are imposed on
the tangemeter coefficients. For example, one can set to unity the
coefficients in front of the highest order term and set $(n-1)$ coefficients
in front of certain monomials equal to $(n-1)$ coefficients of other monomials.
Within the group $G$, multiplication by a complex number $\kappa\in \mathbb{C}^*$ allows one
to normalize the canonic state to unit vacuum-state amplitude and thereby to get
rid of the constant term in the $sl$-tanglemeter $f_C$ Eq.~(\ref{EQ14a}).

  The  condition Eq.~(\ref{EQ62b}) on $P_{j}^{+}$ is written
implicitly as  a
set of $n$ linear equations. These equations can be resolved for generic states
   as we will show in the
next section  in the four-qubit example Eq.~(\ref{EQ62d}).  However,
they have no solution  when the
 determinant of the system vanishes.  These singularities correspond to
 singular classes of entangled states  and require  special
consideration.

\subsubsection{Example: Classes and $sl$-tanglemeters for $4$ qubits}

Now with the examble of $4$-qubit assembly we illustrate the procedure of evaluation
of the $sl$-tanglemeter with the help of dynamic equations supplimented by the feedback conditions.
 The $su$-tanglemeter Eq.~(\ref{EQ14.4}) has 11 complex coefficients
$\beta_{i}$ and one has to solve a system of eleven first-order nonlinear differential  equations.
Instead of presenting this awkward
system explicitly, in Fig.~\ref{FIG2} we schematically depict
contributions to the time derivatives of the $\beta_{i}$ which are
either linear or bilinear in the tanglemeter coefficients, and can be interpreted
as a sort of ``entanglement fluxes'' \cite{Cubitt}. One notices that
the coupling of the second-order terms
$\beta_{ij}\sigma_{i}^{+}\sigma_{j}^{+}$ to the fourth-order term
$\beta_{15}\sigma_{4}^{+}\sigma_{3}^{+} \sigma_{2}^{+}\sigma_{1}^{+}$
occur via the third-order terms $\beta_{7}
\sigma_{3}^{+}\sigma_{2}^{+}\sigma_{1}^{+}$, $\beta_{13}\sigma_{4}^{+}
\sigma_{3}^{+}\sigma_{1}^{+}$, $\beta_{11}\sigma_{4}^{+}\sigma_{2}^{+}
\sigma_{2}^{+}$,
$\beta_{14}\sigma_{2}^{+}\sigma_{3}^{+}\sigma_{4}^{+}$, and thus the
time evolution of all $\beta_{i}$ stops when this third-order
coefficients $\beta_{7}$, $\beta_{13}$, $\beta$, and $\beta_{14}$
vanish.  Therefore, by setting the time dependence of the parameters
$P_{1}^{+}$, $P_{2}^{+}$, $P_{3}^{+}$ and $P_{4}^{+}$ such that they
drive all four third-order coefficients to zero, we gradually reduce
the $SU(2)$ canonic form of Eq.~(\ref{EQ14.4}) to the
$SL(2,\mathbb{C})$ form of Eq.~(\ref{EQ14b}). This control process results in an exponentially
fast vanishing of the coefficients $\beta_{7}$, $\beta_{13}$,
$\beta_{11}$, and $\beta_{14}$. 

We now write down explicitly the differential equations for these
coefficients,%
\begin{align}
\mathrm{i}\overset{.}{\beta}_{14}  &  =\!-P_{1}^{+}\beta_{15}+2P_{2}^{+}%
\beta_{6}\beta_{10}+2P_{3}^{+}\beta_{6}\beta_{12}+2P_{4}^{+}\beta_{10}%
\beta_{12},\nonumber\\
\mathrm{i}\overset{.}{\beta}_{13}  &  =2P_{1}^{+}\beta_{5}\beta_{9}-P_{2}%
^{+}\beta_{15}+2P_{3}^{+}\beta_{5}\beta_{12}+2P_{4}^{+}\beta_{9}\beta
_{12},\nonumber\\
\mathrm{i}\overset{.}{\beta}_{11}  &  =2P_{1}^{+}\beta_{3}\beta_{9}+2P_{2}%
^{+}\beta_{3}\beta_{10}-P_{3}^{+}\beta_{15}+2P_{4}^{+}\beta_{9}\beta
_{10},\nonumber\\
\mathrm{i}\overset{.}{\beta}_{7}  &  =2P_{1}^{+}\beta_{3}\beta_{5}+2P_{2}%
^{+}\beta_{3}\beta_{6}+2P_{3}^{+}\beta_{5}\beta_{6}-P_{4}^{+}\beta_{15}\,,
\label{EQ62d}%
\end{align}
and see that, in the general case, by a proper choice of the
paremeters $P_i^+$ one can impose the feedback conditions such that
these equations take the form
\begin{equation}
\overset{.}{\beta}_{7}=-\beta_{7};\quad\overset{.}{\beta}_{11}=-\beta
_{11};\quad\overset{.}{\beta}_{13}=-\beta_{13};\quad\overset{.}{\beta}%
_{14}=-\beta_{14}\,.\nonumber
\end{equation}
The evolution implied by these equations brings the nilpotential in the form
of $sl$-tanglemeter $f_C$.
We note that the coefficients $P_i^-$ should satisfy the requirement
Eq.~(\ref{EQ62a}) which ensures that the nilpotential always
remains in the form of the tanglemeter $f_c$ during this evolution even if
the state does not remain in the same $su$-orbit.
  We thus arrive at the
tanglemeter $f_{C}$ of Eq.~(\ref{EQ14b}), defined up to the scaling
factors. Now we can invoke the scaling of the nilpotent variables
$\sigma_{i}^{+}$and reduce the tanglemeter $f_{c}$ to the $sl$-canonic
form $f_{C}$ of Eq.~(\ref{EQ14bb}), unless one of the bilinear
coefficients vanish. The latter case  corresponds to a
measure-zero manifold and the canonic form may be chosen as
\begin{align}
f_{C}  &  =\sigma_{2}^{+}\sigma_{1}^{+}+\beta_{5}\left(  \sigma_{3}^{+}%
\sigma_{1}^{+}+\sigma_{4}^{+}\sigma_{2}^{+}\right) \label{EQ14bbb}\\
&  +\beta_{6}\left(  \sigma_{4}^{+}\sigma_{1}^{+}+\sigma_{3}^{+}\sigma_{2}%
^{+}\right)  +\sigma_{4}^{+}\sigma_{3}^{+}\sigma_{2}^{+}\sigma_{1}%
^{+},\nonumber
\end{align}
or in any equivalent form resulting from a the permutation of the
indices. More singular classes are discussed in Appendix B.

Reducing $f$ to the canonic forms of
Eqs.~(\ref{EQ14bb},\ref{EQ14bbb}) is unattainable when the determinant
\begin{equation}
\left\vert
\begin{array}
[c]{cccc}%
-\beta_{15} & 2\beta_{6}\beta_{10} & 2\beta_{6}\beta_{12} & 2\beta_{10}%
\beta_{12}\\
2\beta_{5}\beta_{9} & -\beta_{15} & 2\beta_{5}\beta_{12} & 2\beta_{9}%
\beta_{12}\\
2\beta_{3}\beta_{9} & 2\beta_{3}\beta_{10} & -\beta_{15} & 2\beta_{9}%
\beta_{10}\\
2\beta_{3}\beta_{5} & 2\beta_{3}\beta_{6} & 2\beta_{5}\beta_{6} & -\beta_{15}%
\end{array}
\right\vert \label{EQ62e}%
\end{equation}
vanishes, and it becomes impossible to impose the required
feedback conditions. In that event, we loose control over the dynamics of
$\beta_{7}$, $\beta_{13}$, $\beta_{11},$, and $\beta_{14}$, and some
linear combinations of these coefficients cannot be set to zero by a
proper choice of $P_{i}^{+}$. Consider this singular case in more
detail.  The determinant Eq.~(\ref{EQ62e}) is equal to zero when one
or more of its eigenvalues, 
\begin{widetext}%
\begin{align}
\gamma_{1}  &  =\beta_{15}-2\sqrt{\beta_{5}\beta_{6}\beta_{9}\beta_{10}%
}+2\sqrt{\beta_{3}\beta_{6}\beta_{9}\beta_{12}}-2\sqrt{\beta_{3}\beta_{5}%
\beta_{10}\beta_{12}}\,,\nonumber\\
\gamma_{2}  &  =\beta_{15}+2\sqrt{\beta_{5}\beta_{6}\beta_{9}\beta_{10}%
}-2\sqrt{\beta_{3}\beta_{6}\beta_{9}\beta_{12}}-2\sqrt{\beta_{3}\beta_{5}%
\beta_{10}\beta_{12}}\,,\nonumber\\
\gamma_{3}  &  =\beta_{15}-2\sqrt{\beta_{5}\beta_{6}\beta_{9}\beta_{10}%
}-2\sqrt{\beta_{3}\beta_{6}\beta_{9}\beta_{12}}+2\sqrt{\beta_{3}\beta_{5}%
\beta_{10}\beta_{12}}\,,\nonumber\\
\gamma_{4}  &  =\beta_{15}+2\sqrt{\beta_{5}\beta_{6}\beta_{9}\beta_{10}%
}+2\sqrt{\beta_{3}\beta_{6}\beta_{9}\beta_{12}}+2\sqrt{\beta_{3}\beta_{5}%
\beta_{10}\beta_{12}}\,, \label{EQ62.1}%
\end{align}
\end{widetext}
vanish. Let us first focus on the case where only the first eigenvalue
is zero. This implies that six coefficients in front of the bilinear
terms and the coefficient $\beta_{15}$ in front of the $4$-order term
are no longer independent parameters -- the last one being the
function of the first ones explicitly given by $\gamma_{1}=0$. The
eigenvector
\[
\left(  -\sqrt{\beta_{6}\beta_{10}\beta_{12}},\sqrt{\beta_{5}\beta_{9}%
\beta_{12}},-\sqrt{\beta_{3}\beta_{9}\beta_{10}},\sqrt{\beta_{3}\beta_{5}%
\beta_{6}}\right)\,,
\]
corresponding to $\gamma_{1}$ gives the combination of the cubic
terms that cannot be eliminated, %
\begin{align*}
&  -\lambda\sqrt{\beta_{6}\beta_{10}\beta_{12}}\sigma_{4}^{+}\sigma_{3}%
^{+}\sigma_{2}^{+}+\lambda\sqrt{\beta_{5}\beta_{9}\beta_{12}}\sigma_{4}%
^{+}\sigma_{3}^{+}\sigma_{1}^{+}\\
&  -\lambda\sqrt{\beta_{3}\beta_{9}\beta_{10}}\sigma_{4}^{+}\sigma_{2}%
^{+}\sigma_{1}^{+}+\lambda\sqrt{\beta_{3}\beta_{5}\beta_{6}}\sigma_{3}%
^{+}\sigma_{2}^{+}\sigma_{1}^{+}\,.%
\end{align*}
Clearly, this combination is determined up to a scaling factor
$\lambda$. The nilpotential of Eq.~(\ref{EQ14.4}) thus takes the form
\begin{widetext}%
\begin{align}
f  &  =\beta_{3}\sigma_{2}^{+}\sigma_{1}^{+}+\beta_{5}\sigma_{3}^{+}\sigma
_{1}^{+}+\beta_{9}\sigma_{4}^{+}\sigma_{1}^{+}+\beta_{6}\sigma_{3}^{+}%
\sigma_{2}^{+}+\beta_{10}\sigma_{4}^{+}\sigma_{2}^{+}+\beta_{12}\sigma_{4}%
^{+}\sigma_{3}^{+}\nonumber\\
&  +\lambda\left(  \sqrt{\beta_{3}\beta_{5}\beta_{6}}\sigma_{3}^{+}\sigma
_{2}^{+}\sigma_{1}^{+}-\sqrt{\beta_{3}\beta_{9}\beta_{10}}\sigma_{4}^{+}%
\sigma_{2}^{+}\sigma_{1}^{+}+\sqrt{\beta_{5}\beta_{9}\beta_{12}}\sigma_{4}%
^{+}\sigma_{3}^{+}\sigma_{1}^{+}-\sqrt{\beta_{6}\beta_{10}\beta_{12}}%
\sigma_{4}^{+}\sigma_{3}^{+}\sigma_{2}^{+}\right) \nonumber\\
&  +2\left(  \sqrt{\beta_{5}\beta_{6}\beta_{9}\beta_{10}}-\sqrt{\beta_{3}%
\beta_{6}\beta_{9}\beta_{12}}+\sqrt{\beta_{3}\beta_{5}\beta_{10}\beta_{12}%
}\right)  \sigma_{4}^{+}\sigma_{3}^{+}\sigma_{2}^{+}\sigma_{1}^{+}\,,
\label{EQ14.4a}%
\end{align}
\end{widetext}
specified in terms of seven complex parameters $\left( \lambda,\beta
_{i}\right) $. We can eliminate four of these complex parameters by
setting the scaling factors of $\sigma_{i}^{+}$, and arrive at the
form%
\begin{align}
f_{C}  &  =\beta_{3}\left(  \sigma_{2}^{+}\sigma_{1}^{+}+\sigma_{4}^{+}%
\sigma_{3}^{+}\right)  +\beta_{5}\left(  \sigma_{3}^{+}\sigma_{1}^{+}%
+\sigma_{4}^{+}\sigma_{2}^{+}\right) \nonumber\\
&  +\beta_{6}\left(  \sigma_{4}^{+}\sigma_{1}^{+}+\sigma_{3}^{+}\sigma_{2}%
^{+}\right) \nonumber\\
&  +\sigma_{3}^{+}\sigma_{2}^{+}\sigma_{1}^{+}-\sigma_{4}^{+}\sigma_{2}%
^{+}\sigma_{1}^{+}+\sigma_{4}^{+}\sigma_{3}^{+}\sigma_{1}^{+}-\sigma_{4}%
^{+}\sigma_{3}^{+}\sigma_{2}^{+}\nonumber\\
&  +2\left(  \beta_{5}\beta_{6}-\beta_{3}\beta_{6}+\beta_{3}\beta_{5}\right)
\sigma_{4}^{+}\sigma_{3}^{+}\sigma_{2}^{+}\sigma_{1}^{+}\,, \label{EQ14.4bc}%
\end{align}
which depends only on three complex parameters. This combination can
be considered as a class of the polynomials that cannot be reduced to
the canonic forms of Eqs.~(\ref{EQ14bb},\ref{EQ14bbb}) by the
sequential application of infinitesimal transformations preserving the
canonic $SU(2)$ form. Permutation of indices of $\sigma_{i}^{+}$ give
equivalent classes.

Next, we consider the case where two of the eigenvalues, say
$\gamma_{1}$ and $\gamma_{2}$, of Eq.~(\ref{EQ62.1}) are zero, that
is,%
\begin{align}
\beta_{15}  =2\beta_{3}\beta_{12}\,,
\;\;\;\beta_{5}\beta_{10}  =\beta_{3}\beta_{12}\,.\nonumber
\end{align}
The corresponding eigenvectors%
\begin{align*}
&  \left(  -\sqrt{\beta_{6}\beta_{10}\beta_{12}},\sqrt{\beta_{5}\beta_{9}%
\beta_{12}},-\sqrt{\beta_{3}\beta_{9}\beta_{10}},\sqrt{\beta_{3}\beta_{5}%
\beta_{6}}\right)\,, \\
&  \left(  -\sqrt{\beta_{6}\beta_{10}\beta_{12}},-\sqrt{\beta_{5}\beta
_{9}\beta_{12}},\sqrt{\beta_{3}\beta_{9}\beta_{10}},\sqrt{\beta_{3}\beta
_{5}\beta_{6}}\right)\,,
\end{align*}
suggest the form of the nilpotential 
\begin{widetext}%
\begin{align*}
f  &  =\beta_{3}\sigma_{2}^{+}\sigma_{1}^{+}+\beta_{5}\sigma_{3}^{+}\sigma
_{1}^{+}+\beta_{9}\sigma_{4}^{+}\sigma_{1}^{+}+\beta_{6}\sigma_{3}^{+}%
\sigma_{2}^{+}+\beta_{10}\sigma_{4}^{+}\sigma_{2}^{+}+\beta_{12}\sigma_{4}%
^{+}\sigma_{3}^{+}\\
&  +\lambda\left(  \sqrt{\beta_{3}\beta_{5}\beta_{6}}\sigma_{3}^{+}\sigma
_{2}^{+}\sigma_{1}^{+}-\sqrt{\beta_{3}\beta_{9}\beta_{10}}\sigma_{4}^{+}%
\sigma_{2}^{+}\sigma_{1}^{+}+\sqrt{\beta_{5}\beta_{9}\beta_{12}}\sigma_{4}%
^{+}\sigma_{3}^{+}\sigma_{1}^{+}-\sqrt{\beta_{6}\beta_{10}\beta_{12}}%
\sigma_{4}^{+}\sigma_{3}^{+}\sigma_{2}^{+}\right) \\
&  +\mu\left(  \sqrt{\beta_{3}\beta_{5}\beta_{6}}\sigma_{3}^{+}\sigma_{2}%
^{+}\sigma_{1}^{+}+\sqrt{\beta_{3}\beta_{9}\beta_{10}}\sigma_{4}^{+}\sigma
_{2}^{+}\sigma_{1}^{+}-\sqrt{\beta_{5}\beta_{9}\beta_{12}}\sigma_{4}^{+}%
\sigma_{3}^{+}\sigma_{1}^{+}-\sqrt{\beta_{6}\beta_{10}\beta_{12}}\sigma
_{4}^{+}\sigma_{3}^{+}\sigma_{2}^{+}\right) \\
&  +2\beta_{3}\beta_{12}\sigma_{4}^{+}\sigma_{3}^{+}\sigma_{2}^{+}\sigma
_{1}^{+},
\end{align*}
\end{widetext}
which after a proper scaling can be simplified%
\begin{align}
f_{C}  &  =\sigma_{2}^{+}\sigma_{1}^{+}+\sigma_{3}^{+}\sigma_{1}^{+}%
+\sigma_{4}^{+}\sigma_{2}^{+}+\sigma_{4}^{+}\sigma_{3}^{+}+\beta_{6}\left(
\sigma_{4}^{+}\sigma_{1}^{+}+\sigma_{3}^{+}\sigma_{2}^{+}\right) \nonumber\\
&  +\beta_{7}\left(  \sigma_{3}^{+}\sigma_{2}^{+}\sigma_{1}^{+}-\sigma_{4}%
^{+}\sigma_{3}^{+}\sigma_{2}^{+}\right)  +\beta_{11}\left(  \sigma_{4}%
^{+}\sigma_{2}^{+}\sigma_{1}^{+}-\sigma_{4}^{+}\sigma_{3}^{+}\sigma_{1}%
^{+}\right) \nonumber\\
&  +2\sigma_{4}^{+}\sigma_{3}^{+}\sigma_{2}^{+}\sigma_{1}^{+}\,.\label{EQ14.4c}%
\end{align}
Again, this depends on three complex parameters and leads to
equivalent classes under indices permutations.

Similarly, the case where $\gamma_{1}=\gamma_{2}=\gamma_{3}=0$ yields%
\[
\beta_{15}/2=\sqrt{\beta_{5}\beta_{10}}=\sqrt{\beta_{3}\beta_{12}}=\sqrt
{\beta_{6}\beta_{9}}\,, \]
which after scaling results in%
\begin{align}
f_{C} & =\sigma_{2}^{+}\sigma_{1}^{+}+\sigma_{3}^{+}\sigma_{1}^{+}%
+\sigma_{4}^{+}\sigma_{2}^{+}+\sigma_{4}^{+}\sigma_{3}^{+}+\sigma_{4}%
^{+}\sigma_{1}^{+}+\sigma_{3}^{+}\sigma_{2}^{+}\nonumber\\ &
+\beta_{14}\left( \sigma_{3}^{+}\sigma_{2}^{+}\sigma_{1}^{+}-\sigma
_{4}^{+}\sigma_{2}^{+}\sigma_{1}^{+}+\sigma_{4}^{+}\sigma_{3}^{+}\sigma
_{1}^{+}-\sigma_{4}^{+}\sigma_{3}^{+}\sigma_{2}^{+}\right) \nonumber\\
& +\beta_{13}\left( \sigma_{3}^{+}\sigma_{2}^{+}\sigma_{1}^{+}+\sigma
_{4}^{+}\sigma_{2}^{+}\sigma_{1}^{+}-\sigma_{4}^{+}\sigma_{3}^{+}\sigma
_{1}^{+}-\sigma_{4}^{+}\sigma_{3}^{+}\sigma_{2}^{+}\right) \nonumber\\
& +\beta_{11}\left( \sigma_{3}^{+}\sigma_{2}^{+}\sigma_{1}^{+}-\sigma
_{4}^{+}\sigma_{2}^{+}\sigma_{1}^{+}-\sigma_{4}^{+}\sigma_{3}^{+}\sigma
_{1}^{+}+\sigma_{4}^{+}\sigma_{3}^{+}\sigma_{2}^{+}\right) \nonumber\\
&
+2\sigma_{4}^{+}\sigma_{3}^{+}\sigma_{2}^{+}\sigma_{1}^{+}\,.\label{EQ14.4d}%
\end{align}

The last case, where all four $\gamma_{i}=0$, may be realized by
setting to zero just three parameters, say
$\beta_{15}=\beta_{10}=\beta_{12}=0$. This enables us to dynamically
eliminate one more of the bilinear coefficients, say $\beta_{9}$, and
to set all the third-order coefficients equal to one by scaling.  This
yields a singular canonic form,
\begin{align}
f_{C}  &  =\sigma_{3}^{+}\sigma_{2}^{+}\sigma_{1}^{+}+\sigma_{4}^{+}\sigma
_{3}^{+}\sigma_{1}^{+}+\sigma_{4}^{+}\sigma_{2}^{+}\sigma_{1}^{+}+\sigma
_{4}^{+}\sigma_{3}^{+}\sigma_{2}^{+}\nonumber\\
&  +\beta_{3}\sigma_{2}^{+}\sigma_{1}^{+}+\beta_{5}\sigma_{3}^{+}\sigma
_{1}^{+}+\beta_{6}\sigma_{3}^{+}\sigma_{2}^{+}, \label{EQ14bbbb}%
\end{align}
which still depends on three parameters and allows permutations. All
five $sl$-tanglemeters obtained from the dynamic equations and
dependent on three complex parameters are depicted in Fig.~\ref{FIG2}.

\section{Entanglement beyond qubits}

In this section we show how the nilpotent polynomials approach may be
extended to describe situations more general than assemblies of
qubits. In particular, we discuss in detail the case of qutrits, each
qutrit being transformed by $SU(3)$ or $SL(3,\mathbb{C})$ groups, and
construct appropriate nilpotent polynomials $F$ and $f=\ln F$ for
these systems. We define the canonic form $F_{c}$ for $F$ and the
tanglemeter $f_{c}$. Next, we generalize this technique to qudits
$d$-level elements--qudits.

Remarkably, the nilpotent polynomials formalism allows us to make
contact with the framework of generalized entanglement, introduced in
Refs.~\cite{Barnum1,Barnum2}. While the latter provides a notion of
entanglement which relies directly on \textit{physical observables}
and, as such, is meaningful even in the absence of an underlying
system partition (see also~\cite{GE}), an important special case
arises in the situations where an element structure is specified, but, due
to some operational or fundamental constraint, the rank $r$ of the
algebra of local transformations is smaller than $d-1$, where $d$ is the element
dimension. In this context, special attention is devoted to
spin-$1$ systems, namely three-level systems restricted to evolve
under the action of spin operators living in the $so(3) \equiv su(2)$ subalgebra of
$su(3)$. In particular, we show how to introduce nilpotent polynomials
for characterizing generalized entanglement within a single element of
an assembly. In such a case, one encounters a new kind of  the nilpotent
variables whose squares do not vanish and only some higher powers  do. 
 We also consider entanglement among
different elements of such an assembly. 

We conclude by extending the
nilpotent polynomials formalism to the case of multipartite
entanglement among groups of elements comprising the assembly,
that is, to the case where different elements of an assembly merge,
thereby creating a new assembly with elements of higher dimensions.

\subsection{Qutrits and qudits}

In order to describe entanglement among $d$-level elements of an
assembly, one needs to invoke the Lie algebras of higher rank $su(d)$
and their complex versions $sl(d,\mathbb{C})$. The construction of
nilpotent polynomials for such systems is based on the so-called
\textit{Cartan-Weyl decomposition}.  We illustrate this for qutrits,
$d=3$, and then generalize to arbitrary $d$.

\subsubsection{Nilpotent variables for qutrits}

Let us start by reminding some basic facts about group theory and Lie
algebra representation theory \cite{Cornwell,Humphreys}. The $d^{2}-1$
generators of the algebra $sl(d)$, the complexified $su(d)$, may be
decomposed into 3 sets:

\begin{description}
\item[(i)] a set $\mathcal{H}$ of $r=d-1$ linearly independent, mutually commuting generators of a
Cartan subalgebra {\sl span}($\mathcal{H}$) ($r$ is the rank of the algebra and {\sl span}($\mathcal{H}$) 
is the vector space spanned by $\mathcal{H}$).  In a
faithful matrix representation, the most natural choice for Cartan
generators are traceless $d\times d$ diagonal matrices.

\item[(ii)] a set $\{\mathcal{E}\}$ of $d(d-1)/2$ \textquotedblleft raising\textquotedblright\
generators spanning a nilpotent subalgebra {\sl span}($\mathcal{E}) \subset su(d)$.  The elements
of $\{\mathcal{E}\}$ and of  {\sl span}($\mathcal{E})$ can be
represented by the matrices with nonzero elements only above the main diagonal.

\item[(iii)] $d(d-1)/2$ Hermitian conjugate \textquotedblleft
lowering\textquotedblright\ generators spanning a nilpotent subalgebra
{\sl span}$\{\mathcal{F}\}$ represented by matrices with nonzero elements only below the
diagonal.
\end{description}

In the case of $su(2)$, each of the above sets contains a single
generator: $\mathcal{H} = \{\sigma^{3}=\sigma_z \},\
\mathcal{E}= \{\sigma^{+}\}$ and $\mathcal{F}%
= \{\sigma^{-}\}$. For $su(3)$ having eight
generators represented by the eight Gell-Mann $\lambda^{a}$ matrices
\cite{Cornwell}, the Cartan subalgebra involves two generators usually
chosen as%
\begin{equation}%
\begin{array}
[c]{cc}%
\lambda^{3}=\left(
\begin{array}
[c]{ccc}%
1 & 0 & 0\\
0 & -1 & 0\\
0 & 0 & 0
\end{array}
\right)  , & \lambda^{8}=\dfrac{1}{\sqrt{3}}\left(
\begin{array}
[c]{ccc}%
1 & 0 & 0\\
0 & 1 & 0\\
0 & 0 & -2
\end{array}
\right)  .
\end{array}
\label{EQ22}%
\end{equation}
The basis $\mathcal{E}$ for the raising nilpotent subalgebra  is comprised of $3$ elements,%
\begin{equation}
s^{+}=\frac{\lambda^{1}+i\lambda^{2}}{2}=\left(
\begin{array}
[c]{ccc}%
0 & 1 & 0\\
0 & 0 & 0\\
0 & 0 & 0
\end{array}
\right)  , \label{EQ24}%
\end{equation}%
\begin{equation}
u^{+}=\frac{\lambda^{4}+i\lambda^{5}}{2}=\left(
\begin{array}
[c]{ccc}%
0 & 0 & 1\\
0 & 0 & 0\\
0 & 0 & 0
\end{array}
\right)  , \label{EQ25}%
\end{equation}%
\begin{equation}
t^{+}=\frac{\lambda^{6}+i\lambda^{7}}{2}=\left(
\begin{array}
[c]{ccc}%
0 & 0 & 0\\
0 & 0 & 1\\
0 & 0 & 0
\end{array}
\right)  , \label{EQ26}%
\end{equation}
while $\mathcal{F}$ includes their Hermitian conjugates.

The elements of $\mathcal{E}$ and $\mathcal{F}$ are the root vectors
of $su(3)$. It means that for $e\in\mathcal{E}$, $f\in\mathcal{F}$,
and $h\in {\sl span} (\mathcal{H})$, the commutator $[e,h]$ is proportional to $e$,
and the commutator $[f,h]$ is proportional to $f$. The subalgebra
$\mathcal{E}$ is nilpotent, meaning that the multiple commutators
$[e_{1},\ldots,[e_{p-1} ,e_{p}]\ldots]$ vanish starting at some
level $p-1$. For $su(3)$ with only one nontrivial commutator
$[s^{+},t^{+}]=u^{+}$,  double commutators already vanish. For $su(d)$,
they vanish at the level $d-1$, coinciding with the rank of the
algebra $r$.

A generic pure state of a qutrit may be represented as%
\begin{equation}
|\psi\rangle=\psi_{0}|0\rangle+\psi_{1}|1\rangle+\psi_{2}|2\rangle=\left(
\psi_{0}+\psi_{1}t^{+}+\psi_{2}u^{+}\right)  |0\rangle\,, 
\label{psiqutrits}%
\end{equation}
with
\begin{equation}
|0\rangle\equiv\left(
\begin{array}
{c}%
0\\
0\\
1
\end{array}
\right)  ,\;|1\rangle\equiv\left(
\begin{array}
{c}%
0\\
1\\
0
\end{array}
\right)  ,\;|2\rangle\equiv\left(
\begin{array}
{c}%
1\\
0\\
0
\end{array}
\right)  .
\label{012}
\end{equation}
Equation (\ref{psiqutrits}) generalizes a similar representation for the qubit
pure states extensively discussed so far. The operators $t^{+},u^{+}$
belong to $\mathcal{E}$ and \textit{commute}. Note that even for
larger $d$, such a set of $d-1=r$ commuting nilpotent operators always
exists \cite{note2}, allowing one  to express a generic qudit state
as a first-order polynomial of commuting nilpotent variables.

The state $|0\rangle$, which by definition is annihilated by all ``lowering''
generators of ${\cal F}$, plays the special role of the {\it Fock
vacuum} or {\it reference} state.  But the choice Eq.~(\ref{012}) is not
unique. Actually, {\it any} qutrit state can serve as a valid
reference state.  For each $|\tilde 0 \rangle = U|0 \rangle,\ U \in
SU(3)$, the analysis above applies if one merely changes (conjugates)
accordingly the Cartan-Weyl decomposition and defines
$$
\tilde {\cal H} = U{\cal H}U^{-1}\,, \;\;\tilde {\cal E} = U{\cal
E}U^{-1}\,, \;\; \tilde {\cal F} = U{\cal F}U^{-1}\,.
$$ 
Again, {\sl span}($\tilde {\mathcal E}$) and {\sl span}($\tilde {\mathcal F}$) are nilpotent; again, 
$|\tilde 0 \rangle$ are annihilated by the elements of $\tilde {\cal F}$, 
etc. For example, if the choice $|\tilde 0 \rangle = |1\rangle$ is made, the 
elements of ${\cal E}$ are $\tilde s^+ = - u^+, \tilde t^+ = - t^-, \tilde u^+
= s^+$.  In this case, a generic qutrit state would be expressed as
\begin{equation}
|\psi\rangle \hspace{-.5mm}= \hspace{-.5mm}\left( \hspace{-.5mm}
- \psi_{0}\tilde t^{+}+\psi_{1}+\psi_{2}\tilde u^{+}\right)\hspace{-.5mm}
|1\rangle\ \hspace{-.8mm}= \hspace{-.5mm}
\left(  \psi_{0}t^{-}+\psi_{1}+\psi_{2}s^{+}\right)\hspace{-.5mm}
|1\rangle\, , \label{vector}%
\end{equation}
The choice of Eq.~(\ref{psiqutrits}) is more natural for discussing
qutrits, whereas the choice made in Eq.~(\ref{vector}) allows one to
associate $\lambda^{3}$ with the spin projection operator $S_{z}$ for
a spin-$1$ system, with the vacuum state $|\tilde 0 \rangle =|1\rangle$
naturally corresponding to the {\em lowest} eigenvalue $-1$ of this
operator.  The state $|1\rangle$ is so-called {\em extremal weight} state for the
representation and is discussed in Sect. IVC in more details.  In any case, this
choice is mainly a matter of state labeling, which in most cases is
dictated by convenience and can then be done accordingly for each
particular physical problem.

In analogy with Eq.~(\ref{EQ10}), the state of $n$ qutrits can then be
written as
\begin{align}
|\Psi\rangle &  =\ \hspace{-5mm}
\sum_{\{\nu_{i},\eta_{i}=0,1\,/\nu_{i}\eta_{i}=0\}}%
\psi_{2\nu_n +\eta_n \ldots 2\nu_1 + \eta_1}
\prod_{i=1}^{n}(u_{i}^{+})^{\nu_{i}}
(t_{i}^{+})^{\eta_{i}}|\mathrm{O}\rangle\nonumber\\
&  =F(u_{i}^{+},t_{i}^{+})|\mathrm{O}\rangle\ , \label{qutvec}%
\end{align}
where $|\mathrm{O}\rangle = |0,\ldots,0 \rangle$ denotes the
reference state. Our next steps are: {\it (i)} to consider the
nilpotential $f=\ln F$; {\it (ii)} to bring it into the canonic form
specified by the requirement of maximum population of the state
$|\mathrm{O}\rangle$ followed by the requirement of maximum
population of the state $\left|1\right\rangle=\left\vert 1\ldots1\right\rangle $, as discussed at
the end of Sect.~\ref{canonic}; and {\it (iii)} to normalize
$|\Psi_C\rangle$ by the condition $\langle \mathrm{ O}|\Psi_C \rangle
 = 1$. The tanglemeter $f_{C}$ thereby obtained provides one with simple
entanglement characteristics and relevant insights. Let us explain how
this construction works for an assembly of two qutrits.

\subsubsection{Entanglement of two qutrits}

We select $|\mathrm{O}\rangle=|0,0\rangle$ as the 
reference state.  The relevant nilpotent variables are $u_{i}^{+}$ and
$t_{i}^{+}$, where the index $i=1,2$ labels the qutrits. An arbitrary assembly
state can then be written as
\begin{align}
|\Psi\rangle &  =F(u_{i}^{+},t_{i}^{+})|\mathrm{O}\rangle=
(\psi_{00}+\psi_{01}t_{1}^{+}\nonumber\label{EQ27}\\
&  +\psi_{02}u_{1}^{+}+\psi_{10}t_{2}^{+}+\psi_{20}u_{2}^{+}+\psi_{11}%
t_{1}^{+}t_{2}^{+}\nonumber\\
&  +\psi_{12}t_{2}^{+} u_{1}^{+} +\psi_{21}u_{2}^{+} t_{1}^{+} +\psi_{22} u_{2}^{+}
 u_{1}^{+})|\mathrm{ O}\rangle\:.
\end{align}
The requirement of maximum population of $|\mathrm{ O}\rangle$
dictates that the linear terms in $F(u_{i}^{+},t_{i}^{+})$
vanish. This is proved by inspecting the variation of the population
under local unitary transformations, similarly to what was done for
qubits (see Eq.~(\ref{EQ5})).  Imposing also our normalization
constraint dictating that the expansion of $F$ starts with $1$, we bring
$F_c$ into the form%
\begin{equation}
F_{c}=1+\alpha_{11}t_{2}^{+}t_{1}^{+}+\alpha_{12} t_{2}^{+} u_{1}^{+}%
+\alpha_{21}u_{2}^{+} t_{1}^{+}+\alpha_{22}u_{2}^{+}u_{1}^{+}\,. \label{EQ28}%
\end{equation}
Thus, we are left with the reference state and four  excited states.
The form in Eq.~(\ref{EQ28}) is invariant with respect to the subgroup
$SU_2(2) \otimes SU_1(2)$ of local transformations which can mix the
levels $|1\rangle$ and $|2\rangle$ for each qutrit, but preserve the
population of the state $|\mathrm{ O}\rangle$.  Using this
freedom, we further restrict the canonic form by acting on
$|\Psi\rangle$ with the transformations of $SU_2(2) \otimes SU_1(2)$
that maximizes the population of the state $|11\rangle=t_{2}^{+}
t_{1}^{+}|0\rangle$, while the population of the reference state
$|\mathrm{ O}\rangle$ is already maximized. This
eliminates the terms $\propto t_{2}^{+} u_{1}^{+}$ and $\propto u_{2}^{+}
t_{1}^{+}$ in $F_c$.  Thus, we finally obtain
\begin{equation}
F_{c}(u_{2}^{+},t_{2}^{+},u_{1}^{+},t_{1}^{+})=1+\alpha_{11}t_{2}^{+}t_{1}%
^{+}+\alpha_{22}u_{2}^{+}u_{1}^{+}\:. \label{EQ28a}%
\end{equation}
The polynomial $F_{c}$ depends only on two complex parameters
$\alpha_{ii}$, which can be set real and positive by a local phase
transformation,%
\begin{equation}
\exp\left(  i\gamma_{2}\lambda_{2}^{3}+i\delta_{2}\lambda_{2}^{8}+
i\gamma_{1}\lambda_{1}^{3}+i\delta_{1}\lambda_{1}^{8}\right)  . \label{EQ29}%
\end{equation}
Thus, two real parameters are sufficient for characterizing entanglement
between two qutrits, as illustrated in Fig.~\ref{FIG1} b). This is
consistent with the result obtained by a straightforward application
of the bipartite Schmidt decomposition.

When counting the number of invariants $N_{I}$ \ for a two-qutrit
state with the help of the expression
\begin{equation}
D_{su}=2\cdot3^{n}-8n-2 \:, \label{EQ29aa}%
\end{equation}
we find $N_{I}=0$ for $n=2$. As for the two-qubit case, this
number differs from the actual number of independent parameters
because the phase transformation of Eq.~(\ref{EQ29}) has more
parameters than the number of the coefficients in Eq.~(\ref{EQ28}),
and some of them act on the coefficients in the same way
(cf. Eq.~(\ref{n2trans}) and the discussion thereabout). The counting
given by Eq.~(\ref{EQ29aa}) holds for $n\geq 3$.

\subsubsection{Entanglement in a generic qudit assembly}

The above analysis suggests the following generalization to an
assembly of $n$ qudits. Let $d_{i}$ denote the local dimension of the
$i$-th element with the associated $su(d_{i})$ algebra. For each
element $i$, we choose a reference state $|0\rangle_i$ and perform the
corresponding Cartan-Weyl decomposition ${\cal H}_i \oplus {\cal
E}_{i} \oplus {\cal F}_{i}$ for the generators of this algebra, in such a way that $f|0
\rangle_i = 0$ for all $f \in {\cal F}_i$). The most convenient choice
for $|0 \rangle_i$ is the state with only the lowest level
occupied. One may choose a basis in the qudit Hilbert space involving
the vacuum state $|0 \rangle_i$ and the ``excited states'', such that
each basis state represents a joint eigenstate of all generators
$\lambda_{i}^{\kappa_i}$ in the Cartan subalgebra
($\kappa_i = 1,\ldots,d_{i}-1$).
 The eigenvalues of $\left\{
\lambda_{i}^{\kappa_i}\right\}$ thus provide good quantum numbers
labeling the qudit state.

Next, we choose a set of commuting nilpotent generators $\left\{
\nu_{i}^{\kappa_i}\right\} \subseteq {\cal E}_{i}$ that may be employed
as nilpotent variables. To be specific, let us enumerate these
variables such that $\left\vert 1\right\rangle _{i}=\nu_{i}%
^{1}\left\vert 0\right\rangle _{i}$ corresponds to the first excited
state of $i$-th element, $\left\vert 2\right\rangle _{i}=\nu_{i}%
^{2}\left\vert 0\right\rangle _{i}$ to the second, \textit{etc, }
A polynomial $F\left\{
\nu_{i}^{\kappa_i}\right\}$ of these nilpotent arguments characterizes a
generic quantum state of the assembly as the latter may be obtained
acting by the operator $F$ on the corresponding assembly reference
state $\left\vert \mathrm{O}\right\rangle =%
{\textstyle\prod\nolimits_{i}} \left\vert 0\right\rangle _{i}$.

In analogy with the procedure  for two qutrits described above, we can, using
local $SU(d_{i})$ operations, maximize the population of the vacuum
state and eliminate the terms linear in the nilpotent operators.  The
function $F$ acquires the form generalizing Eq.~(\ref{EQ28}),
 \begin{equation}
F=1+\alpha_{ij}^{\kappa_{i}\kappa_{j}}\nu_{i}^{\kappa_{i}}\nu_{j}^{\kappa
_{j}}+\alpha_{ijk}^{\kappa_{i}\kappa_{j}\kappa_{k}}\nu_{i}^{\kappa_{i}}%
\nu_{j}^{\kappa_{j}}\nu_{k}^{\kappa_{k}}\ldots ,
\label{canqutrit}
\end{equation}
where repeated indices imply summation. The state is normalized to
unit amplitude of the reference state, as earlier. As a next step, we
maximize the population of the symmetric state $|1, \ldots, 1 \rangle$
using the transformations of the subgroup $\otimes_i SU(d_{i}-1)$, where the 
tensor product is taken over all $i$ with $d_i > 2$,
which preserve the reference state.  At the third step,  we maximize the population
of the state $|\{k_i\}\rangle $, where $k_i = 2$ for the elements with $d_i > 2$ while for qubits the label $k_i$
is ``frozen'' at the value $k_i = 1$, and  maximization is done using the transformations of the
subgroup $\otimes_i SU(d_{i}-2)$ that affect neither the reference
state nor the first excited state, and the tensor product involves now the elements with $d_i > 3$. 
If the assembly involves 5-level elements,  we are allowed at the next  step to  
maximize the population of the state  $|\{k_i\}\rangle $, where $k_i = 1$ for qubits, $k_i = 2$ for qutrits 
and $k_i = 3$ for the elements with $d_i > 3$, etc.

 For example, for an assembly of a five-level system, a
four-level system, and a qutrit, one should consecutively maximize:
 
{\it (i)} the population of the state $|0,0,0\rangle$ by the
transformations from $SU(5)\otimes SU(4) \otimes SU(3)$;

 {\it (ii)} the population of the state $|1,1,1\rangle$ by the
transformations from $SU(4)\otimes SU(3) \otimes SU(2)$;

 {\it (iii)} the population of the state $|2,2,2\rangle$ by the
transformations from $SU(3)\otimes SU(2) $;

 {\it (iv)} the population of the state $|3,3,2\rangle$ by the
transformations from the remaining $SU(2)$ mixing the 3-d and the 4-th
excited states of the five-level system.
 
\begin{figure}[h]
{\centering{\includegraphics*[width=0.5\textwidth]{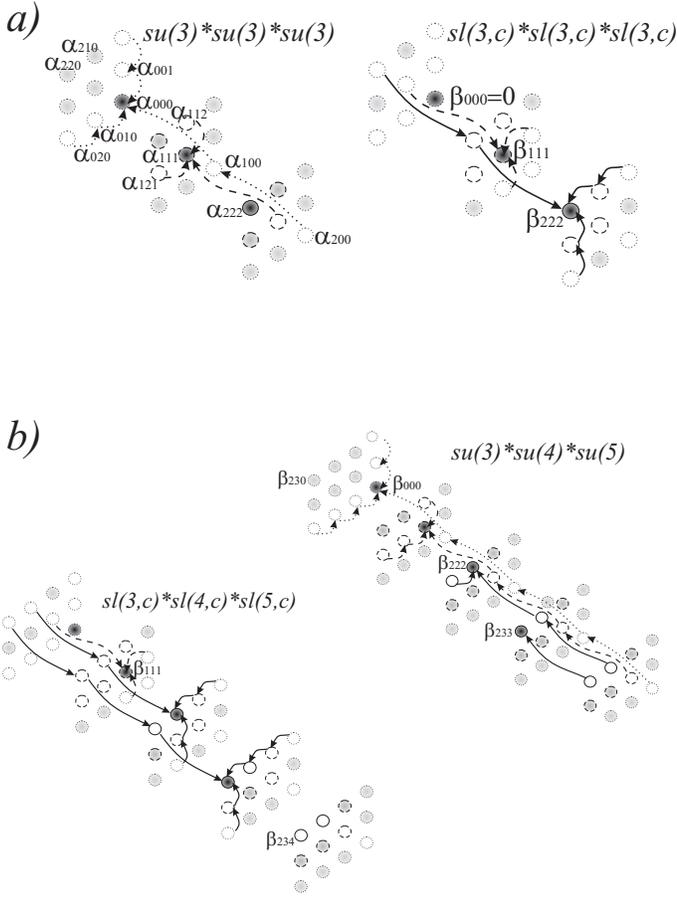}}} 
\caption{a)A possible strategy for identifying the tanglemeter $f_c$
for three qutrits. The choice of the parameters $P_{i}^{\kappa}$ of
local unitary transformations is done in such a way that the
transformations sequentially eliminate the terms in $F_c$ that are
coupled to the states $\left\vert k,k,k\right\rangle $ by just one
local operator. The $SU(3)$ transformations maximize the population of
the vacuum $|\mathrm{O}\rangle=\left\vert 0,0,0\right\rangle $. Next,
the $SU(2)$ transformations not acting on the states $|0_i\rangle$
shown by dotted circles, and only the states shown by dashed and solid
circle are mixed, maximize the population of the state $\left\vert
1,1,1\right\rangle $.  The population of the state $\left\vert
2,2,2\right\rangle $ cannot be further maximized by local unitary
operations. However, $SL(3)$ and $SL(2)$ dynamic transformations can
eliminate some of the coefficients $\beta$ directly connected to such
states.  b) The same diagrams for the case of different elements.}
\label{FIG5}
\end{figure}

This is all illustrated in Fig.\,\ref{FIG5}.

This procedure eventually reduces the nilpotent polynomial $F$ to the
canonic form of Eq.~(\ref{canqutrit}), where some coefficients
$\alpha_{ij}^{\kappa_{i}\kappa_{j}}$,
$\alpha_{ijk}^{\kappa_{i}\kappa_{j}\kappa_{k}}$ now vanish. In order
to better understand the pattern, consider two examples: an assembly
of three qutrits, and of two qutrits and a qubit.  In the first case,
the canonic form is
\begin{widetext}
\begin{align}
F_{c} &  =1+\alpha_{011} t_{2}^{+} t_{1}^{+} +\alpha_{012} t_{2}^{+}u_{1}^{+}  +\alpha_{021} u_{2}^{+} t_{1}^{+} 
+\alpha_{022} u_{2}^{+} u_{1}^{+}
+\alpha_{101}t_{3}^{+}t_{1}^{+}+\alpha_{102} t_{3}^{+} u_{1}^{+} +\alpha_{201} u_{3}^{+} t_{1}^{+} + 
\alpha_{202}u_{3}^{+}u_{1}^{+} +
\alpha_{110}t_{3}^{+}t_{2}^{+}\nonumber\\
&  +\alpha_{120}t_{3}^{+} u_{2}^{+} +\alpha_{210}u_{3}^{+} t_{2}^{+} +\alpha_{220}u_{3}^{+}u_{2}^{+}+
\alpha_{111}t_{3}^{+}t_{2}^{+}t_{1}^{+}+
\alpha_{122}t_{3}^{+} u_{2}^{+}u_{1}^{+} +\alpha_{212}u_{3}^{+}t_{2}^{+}u_{1}%
^{+}+\alpha_{221}u_{3}^{+}u_{2}^{+} t_{1}^{+} +\alpha_{222}u_{3}^{+}u_{2}%
^{+}u_{1}^{+}\,,\label{EQ30a}%
 \end{align}
whereas for two qutrits and a qubit (labeled by the index 3), it looks as follows%
\begin{equation}
F_{c}=1+\alpha_{4}t_{2}^{+}t_{1}^{+}+\alpha_{5} t_{2}^{+} u_{1}^{+} +\alpha_{7} u_{2}^{+} t_{1}^{+} +
\alpha_{8}u_{2}^{+}u_{1}^{+} + \alpha_{10}t_{3}^{+}%
t_{1}^{+}+\alpha_{11}t_{3}^{+} u_{1}^{+}  +\alpha_{12}t_{3}^{+}t_{2}^{+}%
+\alpha_{15}t_{3}^{+} u_{2}^{+} +\alpha_{13}t_{3}^{+}t_{2}^{+}t_{1}^{+}%
+\alpha_{17}t_{3}^{+} u_{2}^{+}u_{1}^{+} \,,\label{EQ30a1}%
\end{equation}%
\end{widetext}%
with the identification $\nu_{i}^{1}=t_{i}^{+}$ and $\nu_{i}^{2}%
=u_{i}^{+}$. In Eq.~(\ref{EQ30a}), the indices of $\alpha$ explicitly
label the individual states of the qutrits, while in
Eq.~(\ref{EQ30a1}) the notation relies on the decimal representation
of the base-$3$ numbers associated with these indices,
$112\rightarrow14$, \textit{etc}.  The canonic forms are defined, as
before, up to phase factors of the nilpotent variables
$\nu_{j}^{\kappa_{j}}$.

The forms of Eqs.~(\ref{EQ30a},\ref{EQ30a1}) do not involve linear
terms. Neither do they involve any term proportional to $t_3^+t_2^+u_1^+ ,\ t_3^+  u_2^+ t_1^+ $, 
and (for three qutrits) 
$ u_3^+ t_2^+ t_1^+$. In other words, the amplitudes of the states $|1,1,2\rangle,\
|1,2,1\rangle$ and, for three qutrits, $|2,1,1\rangle$ vanish. This
vanishing is achieved at the second stage of our procedure maximizing
the population of the state $|1,1,1\rangle$ by the transformations
from $\otimes_{i=1}^3 SU(2)_i$ for three qutrits, and from $SU_2(2)
\otimes SU_1(2)$ for two qutrits and a qubit. Indeed, from the
viewpoint of the remaining $SU(2)$ transformations, the state
$|1\rangle$ may be regarded as a reference state, and amplitudes like
$|1,1,2\rangle$ vanish for the same reason why the amplitudes
$|1,0,\ldots,0\rangle$ vanished when the population of $|{\Large
O}\rangle$ were maximized.  It is clear that, e.g., for an assembly of
three four-level systems, where the procedure involves 3 steps, the
canonic state has vanishing coefficients of the basis states $|2,1,1\rangle, 
|1,2,1\rangle, |1,1,2\rangle, |3,1,1\rangle, 
|1,3,1\rangle, |1,1,3\rangle, |3,2,2\rangle$, \\  $|2,3,2\rangle$ and
$|2,2,3\rangle$. Generally, if the procedure involves the maximization
of the population of the state $|\{k_i\}\rangle$, the vanishing amplitudes
are $|\{k_i'\}\rangle$, where the sets $\{k_i\}$ and $\{k_i'\}$ differ only in one position
$i_0$ and $k_{i_0}' > k_{i_0}$.

Going back to Eq.~(\ref{EQ30a}), we observe that the canonic polynomial
depends in this case on $17$ complex parameters, $6$ of which can be
set real by a proper choice of the phase factors of the nilpotent
variables $u_{i}^{+}$ and $t_{i}^{+}$. This coincides with the number
$N_{I}=28$ suggested by the asymptotic formula given in
Eq.~(\ref{EQ29aa}) valid for $n>2$.

The coefficients $\alpha_{i\ldots j}^{\kappa_i\ldots\kappa_{j}}$ can
be treated as invariants characterizing the qudit entanglement. But the
coefficients of the tanglemeter $f_{c}=\ln F_{c}$ provide, as we
already saw, a more direct and physical description.  In particular,
the criterion of Eq.~(\ref{EQ14}), indicating whether two groups, $A$
and $B$, of a bipartition are entangled, may be straightforwardly
generalized.  We have
\[
\frac{\partial^{2}f(\{\nu_{i}^{\kappa_{i}}\})}{\partial\nu_{i}^{\kappa_{i}%
}\partial\nu_{m}^{\kappa_{m}}}=0,\;\;\;i\in A,\quad m\in B \, .
\]
As before, this criterion holds even for a noncanonic nilpotential.

The construction of the $sl$-tanglemeter may be accomplished by
analogy to the qubit case. One can derive and directly employ the
dynamic equations for $f$, eliminate the states adjacent in the sense explained above
 to $\left\vert
k,\ldots,k\right\rangle $ by a proper choice of the transformation
parameters $P_{i}^{\kappa}$, thereby obtaining the $sl$-canonic form:
\begin{equation}
f_C =
\beta_{ij}^{\kappa_{i}\kappa_{j}}\nu_{i}^{\kappa_{i}}\nu_{j}^{\kappa_{j}} + \ldots 
\label{EQ30aa}%
\end{equation}
 Like it was the case for the $su$-tanglemeter, many of the coefficients $\beta_{ij}^{\kappa_{i}\kappa_{j}}$,
$ \beta_{ijk}^{\kappa_{i}\kappa_{j}\kappa_{k}}$, etc vanish, however.
Actually, as $SL$ transformations have more parameters than $SU$-transformations, we can now bring
to zero {\it more} coefficients than in the $SU$ case. Namely, the  
 sum on the right hand side of Eq.(\ref{EQ30aa})  contains {\em no} terms
corresponding to the states directly coupled to $\left\vert
k,\ldots,k\right\rangle $ by any single local transformation, and not
only those involving ``higher" states with $k^{\prime }\geqslant
k$ (for simplicity, we discuss here only the case when all elements have the same dimension).
Without explicitly presenting the corresponding dynamic equations,
we illustrate this idea in Fig.~\ref{FIG5} for the case of qutrits.
As one can see there, the generic $SL(3,\mathbb{C})$ canonic form of
$f$ contains $8$ complex parameters, $6$ of which may be specified by
a proper choice of the scaling factors. Thus,%
\begin{align*}
\label{tmr3qutrSL}
f_C  &  =\beta_{g}\left( u_{3}^{+}  t_{2}^{+} + u_{2}^{+} t_{1}^{+} + t_{3}^{+} u_{1}^{+}
\right)  +t_{3}^{+}t_{2}^{+}t_{1}^{+}\\
&  +\beta_{u}\left( t_{2}^{+} u_{1}^{+} + u_{3}^{+} t_{1}^{+} + t_{3}^{+} u_{2}^{+} \right) 
 + u_{3}^{+}u_{2}^{+}u_{1}^{+}\:.
\end{align*}

\subsection{Generalized entanglement and generating functions}

We now consider a situation where the local operations available for
qudits are\textit{ restricted} by some operational or fundamental
constraint, so that they cannot ensure universal local
transformations.  In particular, we focus on the case where the
restricted local operations form a subgroup $SU(m) \subset SU(d)$,
with $m<d$, of the full unitary transformation set.  As mentioned,
this is a special relevant instance of a more general,
subsystem-independent entanglement setting ({\em generalized
entanglement}) formalized in Refs.~\cite{Barnum1,Barnum2}. One of the
main implications of the latter approach is, in turn, to point to an
intimate connection between entanglement and so-called {\em
generalized coherent states} (GCSs)~\cite{Perelomov,Zhang}, which is
also useful to place our current analysis in a broader context.

Generalized coherent states may be constructed for quantum systems described by a
dynamic-symmetry Lie group, which typically is assumed to be reductive
or semisimple: a family of GCSs may be thought of as an orbit
resulting from application of the Lie group to a reference state,
which is identified with an extremal-weight state in the irreducible
representation of the underlying dynamical Lie algebra.  Within this
general setting, the canonical coherent states of a quantum
harmonic-oscillator~\cite{Glauber} may be recovered as resulting from
the application of the displacement operator $[\exp\left( az^{\ast}-
a^{\dagger }z\right)]$ to the ground state $\left\vert
0\right\rangle$.  Here, as usual, $a$ and$\ a^{\dagger}$ denote
annihilation and creation operators, respectively, and
$z=x+\mathrm{i}p \in {\mathbb C}$ stands for displacement in the phase
space $\left( x,p\right) $. The relevant dynamic groups are generated
by the Heisenberg-Weyl algebras, either
$\mathfrak{h}_{3}=\{\mathbb{I},a,a^{\dagger }\}$ that allows for
arbitrary displacements, or $\mathfrak{h}_{4}
=\{\mathbb{I},a,a^{\dagger},a^{\dagger}a\}$ that in addition involves
the phase space rotation $\exp\left( \mathrm{i}\varphi
a^{\dagger}a\right)$ by an angle $\varphi$~\cite{note4}. Physically,
the most representative manifestation of the coherent states is the
fact that external interventions realized via $a\dag$ and $a$ cannot
change the structure of the wave function expressed in terms of the
Wigner function, but just displace and rotate it in the phase
space. One can also include the operation $[\exp \mathrm{i}\left(s
(a^{\dagger})^{2}+s^{\ast}a^{2}\right)]$ of squeezing $s \in {\mathbb
C}$ by remaining within a finite-dimensional (so-called
``two-photon'') dynamic algebra.  Clearly, the algebra describing
arbitrary transformation on the oscillator Hilbert space is
infinite-dimensional, and destroys the manifestations typical of
coherent states.

Interestingly, the mathematical techniques employed to describe GCSs
are extremely close to the ones we use for nilpotent polynomials: any
GCS is constructed from a minimum-weight reference state of a
semisimple Lie algebra by application of a function of the appropriate
Cartan-Weyl raising operators which, for a finite dimensional Hilbert
space, are indeed nilpotent. One can always construct GCSs of a qudit,
relative to its {\em full} $su(d)$ algebra of transformations. The
resulting family is, of course, identical with the set of possible
pure states of the system.  For a qudit assembly with $n$ elements, it
was shown in~\cite{Barnum1} that conventional \textit{un}entangled
states are identical with the set of GCSs under \textit{arbitrary
local transformations}, that is, with respect to the algebra
$\oplus_{i=1}^n su(d_{i})$, with rank
$R={\textstyle\sum\nolimits_{i}}(d_{i}-1)$.  Geometrically, such GCSs
are {\em extremal} (in the convex sense) in the set of states which
may specified only through expectation values of arbitrary local
observables.  However, such extremality property may no longer be
fulfilled upon restricting the transformations on each element to a
proper subalgebra, $m_i < d_i$ -- implying the possibility of
generalized entanglement {\em relative to the restricted observable
set}.  Here, we further develop the connection between entanglement
and GCSs in terms of the nilpotent polynomials approach.  In a way, we
can say that we consider entanglement related to GCSs both within each
element as well as entanglement among GCSs of different elements of
the assembly, and construct in each case appropriate characteristics
based on nilpotent variables.

The first step toward accomplishing the above goal is to obtain a
proper description of the group of restricted local transformations,
by embedding it as a subgroup into the full-rank $r_{i}=d_{i}-1$ group
of local transformations. In other words, among the generators of full
group we have to specify the linear combinations that correspond to
the generators of the restricted local transformations. To this end,
it turns out that so-called \textit{generating function technique}
offers a convenient tool. Let $\left\vert \Psi\right\rangle $ be a
state of the assembly normalized to the unit amplitude of the vacuum
state, and apply the operator $\exp\left[ \sum_{i}x_{i}^{\kappa_{i}
}\left( \nu_{i}^{\kappa_{i}}\right) ^{\dagger}\right]$, where the
nilpotent operators $\nu_{i}^{\kappa_{i}}$ correspond to the
Cartan-Weyl decomposition of the full $su(d_{i})$ algebra of $i$-th
element, and $x_{i}^{\kappa_{i}} \in {\mathbb C}$. We finally project
the result onto the vacuum state and obtain
\begin{equation}
F\left(  \left\{  x_{i}^{\kappa_{i}}\right\}  \right)  =\left\langle
0\right\vert \exp\left[  \sum x_{i}^{\kappa_{i}}\left(  \nu_{i}^{\kappa_{i}%
}\right)  ^{\dagger}\right]  \left\vert \Psi\right\rangle \,, \label{EQ31}%
\end{equation}
which is the generating function for the set of variables $\left\{
x_{i}^{\kappa_{i}}\right\}$. This coincides with the function $F\left(
\left\{  \nu_{i}^{\kappa_{i}}\right\}  \right)  $ given by the nilpotent
polynomial $F$ where all the operators $\sigma^{+}$ are replaced by the
corresponding variables $x$.

As long as the set $\nu_{i}^{\kappa_{i}}$ corresponds to the full
algebra, introducing the generating function adds nothing new to the
characterization of entanglement. The situation changes when we
consider subalgebras of rank $m_{i}<r_{i}$. The generating function
takes then the form%
\begin{equation}
F\left(  \left\{  x_{i}^{\kappa_{i}}\right\}  \right)  =\left\langle
0\right\vert \exp\left[  \sum x_{i}^{\kappa_{i}}\left(  \mu_{i}^{\kappa_{i}%
}\right)  ^{\dagger}\right]  \left\vert \Psi\right\rangle \,, \label{EQ32}%
\end{equation}
where the commuting nilpotent operators $\left\{ \mu_{i}^{\kappa_{i}
}\right\}$ belong to an algebra of rank $m_{i}$. The elements $\mu
_{i}^{\kappa_{i}}$ are linear combinations $\left\{
\mu_{i}^{\kappa_{i} }\right\} \subset\mathrm{span}\left\{
\nu_{i}^{\kappa_{i}}\right\}$ of the nilpotent elements in the
Cartan-Weyl decomposition $L^{z}\oplus L^{+}\oplus L^{-}$. Among the
Cartan generators $\left\{ \lambda^{\kappa }\right\} $, we also have
to single out a subset $\left\{ \upsilon^{\kappa }\right\} \subset
L^{z}$ of the operators corresponding to the choice of $\left\{
\mu_{i}^{\kappa_{i}}\right\} $. Note that linear combinations
$\mu_{i}^{\kappa_{i}}$ of the nilpotent operators
$\nu_{i}^{\kappa_{i}}$ are also nilpotent, however of {\em higher
order}, that is, such that $\left( \mu_{i}^{\kappa_{i}}\right)
^{p_{i}^{\kappa_{i}}}=0$ for some integer $p_{i}^{\kappa_{i}}>1$. For
example, for $\mu=t^{+}+s^{+}$ of Eqs.~(\ref{EQ24})-(\ref{EQ26}), one
finds $\mu^{2}=u^{+}\neq0$, $\mu^{3}=0$.  Therefore, unlike all cases
discussed so far, {\em the generating function $F\left( \left\{
x_{i}^{\kappa_{i}}\right\} \right)$ may contain some higher powers of
the variables}\emph{ } $x_{i}^{\kappa_{i}}$, along with the terms
linear in $x_{i}^{\kappa_{i}}$.

The generating function $F\left(\left\{ x_{i}^{\kappa_{i}}\right\}
\right)$ may also be reduced to canonic form. This has to be
accomplished, however, only by resorting to local unitary operations
which belong to the subalgebra of restricted local transformations,
\begin{widetext}%
\begin{equation}
F_{c}\left(  \left\{  x_{i}^{\kappa_{i}}\right\}  \right)  =\left\langle
0\right\vert \exp\left[  \sum x_{i}^{\kappa_{i}}\left(  \mu_{i}^{\kappa_{i}%
}\right)  ^{\dagger}\right]  \exp\left[  \mathrm{i}\sum Z_{i}^{\kappa_{i}%
}\upsilon_{i}^{\kappa_{i}}+R_{i}^{\kappa_{i}}\left(  \mu_{i}^{\kappa_{i}%
}\right)  ^{\dagger}+\left(  R_{i}^{\kappa_{i}}\right)  ^{\ast}\mu_{i}%
^{\kappa_{i}}\right]  \left\vert \Psi\right\rangle . \label{EQ32aa}%
\end{equation}
\end{widetext}
Note that one may also generalize this approach to the case of
restricted $SL$ transformations, where in place of
$Z_{i}^{\kappa_{i}}$, $R_{i}^{\kappa_{i}}$, and $\left(
R_{i}^{\kappa_{i}}\right) ^{\ast}$, one has to substitute in
Eq.~(\ref{EQ32aa}) independent complex parameters. The corresponding
nilpotent polynomial $F_{c}\left( \left\{ \mu_{i}^{\kappa_{i}%
}\right\} \right) $ may now \textit{contain powers of the nilpotent
variables}, which is a signature of a very important fact:
Entanglement is no longer necessarily associated with different
subsystems, but may occur within each single element as well. 

\subsection{Examples: Generalized entanglement for one and two spin-$1$
systems}

We first consider generalized entanglement in the simplest example of
a three-level system (see Refs.~\cite{Barnum1,Somma}) and show how this
result may be interpreted in terms of nilpotent variables. Though a
three-level system corresponds to a full $su(3)$ algebra, we consider
it here as a spin-$1$ system that is, concentrate on a situation where
the physical observables are restricted to the subalgebra $su(2)$ of
spin operators, $S_{\pm}=S_{x}\pm\mathrm{i}S_{y}$ and $S_{z}$. Note
that the latter are equivalent to the $u^{\pm}+t^{\mp}$, and
$\lambda^{3}$ generators of $su(3)$, respectively.  The spin states
are characterized by the eigenvalues of $\upsilon=\lambda^{3}$, which
is the only Cartan generator of the $su(2)$ subalgebra. The spin-down,
lowest-weight state $\left\vert -1\right\rangle $ is chosen as the
reference state. The operators $s^{+}$ and $t^{-}$ form the commuting
nilpotent subalgebra $\left\{ \nu\right\} $ of $su(3)$ and give two
other \textquotedblleft excited" states $\left\vert 0\right\rangle
=t^{-}\left\vert -1\right\rangle $ and $\left\vert 1\right\rangle
=s^{+}\left\vert -1\right\rangle $.
They characterize the quantum states of the three-level system
according to the relation $\left\vert \Psi\right\rangle
=(1+\alpha_{s}s^{+}+\alpha_{t} t^{-})\left\vert -1\right\rangle
$. The operator $S_{+}=u^{+}+t^{-}$ is the only element of the
nilpotent $\left\{ \mu_{i}^{\kappa_{i}}\right\} $-subalgebra $su(2)
\subset su(3)$.

Now we show that {\em the state $|0\rangle$ is generalized entangled
with respect to $SU(2)$}. Indeed, by the unitary matrix
$\mathrm{e}^{-\mathrm{i}\pi S_{x}/\sqrt{2}}$ the state $|0\rangle$ may 
be transformed to the state with the maximum vacuum population
$\left\vert \psi_{C}\right\rangle =\left( \left\vert -1\right\rangle
+\left\vert 1\right\rangle \right) /\sqrt{2}$, which is evidently
different from the reference state $\left\vert -1\right\rangle $. The
corresponding canonic state normalized to unit reference state
amplitude reads%
\begin{equation}
\left\vert \Psi_{c}\right\rangle =(1+s^{+})\left\vert -1\right\rangle =\exp
s^{+}\left\vert -1\right\rangle \,, \label{EQ32b}%
\end{equation}
hence $f_{c}=s^{+}$ and $F_{c}=1+s^{+}$. We now construct the
generating function of Eq.~(\ref{EQ31}) by employing
$S_{+}=u^{+}+t^{-}$ as the only element of the nilpotent $\left\{
\mu_{i}^{\kappa_{i}}\right\}$-subalgebra of $su(2)$ for our
system. This yields%
\begin{align}
F\left(  x\right)   &  =\left\langle -1\right\vert \exp\left[  x\left(
u^{+}+t^{-}\right)  ^{\dagger}\right]  \exp s^{+}\left\vert -1\right\rangle
\nonumber\\
&  =\left\langle -1\right\vert \exp\left[  x\left(  u^{-}+t^{+}\right)
\right]  \exp s^{+}\left\vert -1\right\rangle \nonumber\\
&  =1+x^{2}/2\:. \label{EQ32c}%
\end{align}
The presence of the {\em quadratic term} $x^{2}/2$ is the signature of
generalized entanglement.

One way to understand the meaning of this \textquotedblleft
self-entanglement\textquotedblright\ in the state $\left\vert
0\right\rangle $ is to see it as a consequence of the fact that the
operators $S_{\pm,z}$ of $su(2)$ cannot lift the degeneracy of the two
eigenstates of the operator $\lambda_{8}$, which together with
$\lambda_{3}=S_{z}$ labels the states in the unrestricted $su(3)$
algebra. In other words, within the group of restricted local
transformations, the transition from the state $\left\vert
0\right\rangle $ can only access the state $\left( \left\vert
-1\right\rangle +\left\vert 1\right\rangle \right) /\sqrt{2}$ but {\em
not} $\left( \left\vert -1\right\rangle -\left\vert 1\right\rangle
\right) /\sqrt{2}$, hence the amplitudes of the reference state
$\left\vert -1\right\rangle $ and that of the state $\left\vert
1\right\rangle $ are fully correlated.  Alternatively, one can say
that no $SU(2)$ transformation is able to connect the state
$|-1\rangle$, which is a $SU(2)$-GCS and is unentangled, to the state
$|0\rangle$ which is a a $SU(3)$-GCS but not a $SU(2)$-GCS.

For a generic $SU(3)$ state $(1+\alpha_{s}s^{+}+\alpha_{t}
t^{-})\left\vert -1\right\rangle $, the nilpotent $SU(2)$ polynomial
\begin{align}
F\left( s^+\right) &  =\left.F\left( x\right)\right|_{x\rightarrow s^+}
\end{align}
is given by the generating function
\begin{align}
F\left( x\right) &  =\left\langle -1\right\vert \exp\left[  x\left(
u^{+}+t^{-}\right)  ^{\dagger}\right]  \exp(\alpha_{s}s^{+}+\alpha_{t}%
t^{-})\left\vert -1\right\rangle \nonumber\\
&  =1+\alpha_{t}x+\frac{\alpha_{s}x^{2}}{2}\:. \label{EQ32d}%
\end{align}
The corresponding canonic nilpotent polynomial of Eq.~(\ref{EQ32aa}) and the tanglemeter
take the form%
\begin{align}
F_{c}(S_{+})  &  =1+\alpha_{s}^{\prime}S_{+}^{2}/2\nonumber\\
f_{c}(S_{+})  &  =\alpha_{s}^{\prime}S_{+}^{2}/2\:, \label{EQ32e}%
\end{align}
respectively,
where the phase of $\alpha_{s}^{\prime}$ can be set to $0$. Therefore,
generalized entanglement of a single spin-$1$ is characterized by a
single real parameter.

Consider now a second example, that is, entanglement between two
spin-$1$~\cite{Barnum1,Somma}, which we describe as two three-level
systems subject to the action of $SU(2)\oplus SU(2)$ local
operations. By analogy to the single spin-$1$ case, we chose nilpotent
variables $\left\{ \mu_{i}^{\kappa_{i}}\right\} =\left\{
S_{1,2}\right\} $ in $su(2)\oplus su(2)$, where $S_{1,2}\equiv\left(
S_{+}\right) _{1,2}=u_{1,2}^{+}+t_{1,2}^{-}$, and $\left\{
\upsilon^{\kappa}\right\} =\left\{ \left( S_{z}\right) _{1,2}\right\}
$. The state $\left\vert -1,-1\right\rangle $ is chosen as the $SU(2)$
reference state. As before, the operators $s_{1}^{+}$ , $t_{1}^{-}$,
$s_{1,2}^{+}$, and $t_{1,2}^{-}$ are the nilpotent variables in the
full $su(3)\oplus su(3)$ algebra.

A generic quantum state for a two-qutrit assembly 
\begin{widetext}%
\begin{equation}
\left\vert \Psi\right\rangle =(1+\alpha_{s;1}s_{1}^{+}+\alpha_{t;1}t_{1}%
^{-}+\alpha_{s;2}s_{2}^{+}+\alpha_{t;2}t_{2}^{-}+\alpha_{s,s}s_{2}^{+}%
s_{1}^{+}+\alpha_{t,s} s_{2}^{+} t_{1}^{-} +\alpha_{t,s}  t_{2}^{-} s_{1}^{+}%
+\alpha_{t,t}t_{2}^{-}t_{1}^{-})\left\vert -1,-1\right\rangle \:,
\label{EQ33aa}
\end{equation}
is now characterized by the nilpotent polynomials
\begin{align}
F\left(  \left\{  s^{+},t^{-}\right\}  \right)   &  =1+\alpha_{s;1}s_{1}%
^{+}+\alpha_{t;1}t_{1}^{-}+\alpha_{s;2}s_{2}^{+}+\alpha_{t;2}t_{2}^{-}%
+\alpha_{s,s}s_{2}^{+}s_{1}^{+}+\alpha_{t,s}s_{2}^{+} t_{1}^{-} +\alpha_{t,s}t_{2}^{-} s_{1}^{+} +
\alpha_{t,t}t_{2}^{-}t_{1}^{-}~,\label{EQ33bb}\\
f\left(  \left\{  s^{+},t^{-}\right\}  \right)   &  =\alpha_{s;1}s_{1}%
^{+}+\alpha_{t;1}t_{1}^{-}+\alpha_{s;2}s_{2}^{+}+\alpha_{t;2}t_{2}^{-}+\left(
\alpha_{s,s}-\alpha_{s;2}\alpha_{s;1}\right)  s_{2}^{+}s_{1}^{+}+\left(
\alpha_{t,s}-\alpha_{s;2}\alpha_{t;1}\right)  s_{2}^{+} t_{1}^{-}\nonumber\\
&  +\left(  \alpha_{s,t}-\alpha_{t;2}\alpha_{s;1}\right) t_{2}^{-}  s_{1}^{+} + 
\left(  \alpha_{t,t}-\alpha_{t;2}\alpha_{t;1}\right)  t_{2}^{-}t_{1}%
^{-})\:, \label{EQ33cc}%
\end{align}
whereas for the $SU(2)$ characterization, Eq.~(\ref{EQ32}) gives the
generating function%
\begin{equation}
F\left(  x,y\right)  =\left\langle -1,-1\right\vert \exp\left[  x\left(
u_{1}^{+}+t_{1}^{-}\right)  ^{\dagger}+y\left(  u_{2}^{+}+t_{2}^{-}\right)
^{\dagger}\right]  F\left(  \left\{  s^{+},t^{-}\right\}  \right)  \left\vert
-1,-1\right\rangle \:. \label{EQ33c}%
\end{equation}
\end{widetext}Direct calculation yields%
\begin{align}
F\left(  x,y\right)  =1+  &  \alpha_{s;1}\frac{x^{2}}{2}+\alpha_{t;1}%
x+\alpha_{s;2}\frac{y^{2}}{2}+\alpha_{t;2}y+\alpha_{t,t}xy\nonumber\\
&  +\alpha_{t,s}x\frac{y^{2}}{2}+\alpha_{st}\frac{x^{2}}{2}y+\alpha_{s,s}%
\frac{x^{2}}{2}\frac{y^{2}}{2}\:. \label{EQ33d}%
\end{align}
This finally results in the nilpotent polynomial
\begin{align}
F\left(  S_{1},S_{2}\right)   &  =1+\alpha_{s;1}\frac{S_{1}^{2}}{2}%
+\alpha_{t;1}S_{1}+\alpha_{s;2}\frac{S_{2}^{2}}{2}+\alpha_{t;2}S_{2}%
\nonumber\\
&  +\alpha_{s,s}\frac{S_{2}^{2}}{2}\frac{S_{1}^{2}}{2}+\alpha_{t,s}
\frac{S_{2}^{2}}{2} S_{1} +\alpha_{st} S_{2} \frac{S_{1}^{2}}{2} +\alpha_{t,t}S_{2}%
S_{1}\:, \label{EQ33e}%
\end{align}
where the subscripts $+$ of the nilpotent variables $S$ are
implicit.  In the canonic form maximizing population of the
reference state, the population of the states $\left\vert
0,-1\right\rangle $ and $\left\vert -1,0\right\rangle $ vanish, thus 
one obtains%
\begin{align}
F_c\left(  S_{1},S_{2}\right)   &  =1+\alpha_{s;1}\frac{S_{1}^{2}}{2}%
+\alpha_{s;2}\frac{S_{2}^{2}}{2}+\alpha_{s,s}\frac{S_{2}^{2}}{2}\frac
{S_{1}^{2}}{2}\label{EQ33f}\\
&  +\alpha_{t,s} \frac{S_{2}^{2}}{2} S_{1} +\alpha_{st} S_{2} \frac{S_{1}^{2}}{2}%
 +\alpha_{t,t}S_{1}S_{2}~,\nonumber\\
f_c\left(  S_{1},S_{2}\right)   &  =\beta_{s;1}\frac{S_{1}^{2}}{2}+\beta
_{t,t}S_{1}S_{2}+\beta_{st}\frac{S_{1}^{2}}{2}S_{2}+\beta_{s;2}\frac{S_{2}%
^{2}}{2}\nonumber\\
&  +\beta_{t,s}S_{1}\frac{S_{2}^{2}}{2}+\beta_{s,s}\frac{S_{2}^{2}}{2}%
\frac{S_{1}^{2}}{2}\:,\nonumber
\end{align}
where $\beta_{s,s}=\alpha_{s,s}-2\alpha_{t,t}{}^{2}$, and all other
$\beta=\alpha$. As before, by exploiting the freedom of phase
transformations on the nilpotent variables, the parameters
$\alpha_{s;1}$ and $\alpha_{s;2}$ (or $\beta_{s;1}$ and $\beta_{s;2}$)
characterizing generalized entanglement within each of the three-level
systems can be set real, and we are left with four complex numbers
$\alpha_{s,s}$, $\alpha_{t,s}$, $\alpha_{st}$, and $\alpha_{t,t}$
characterizing the generalized inter-spin entanglement.

One may also ask the following question: How can we characterize
generalized entanglement under $SL$ transformations? For qubits, the
resulting classification is based on the dynamic
equation~(\ref{EQ49.11}) and the conditions
Eq.~(\ref{EQ62a})-(\ref{EQ62b}) of the exponential decrease of the
chosen coefficients in the course of controlled local
$SL(2,\mathbb{C})$ dynamics. In order to suggest a strategy for
characterizing an ensemble of spin-$1$ elements, we note that an
analog of Eq.~(\ref{EQ49.11}) may be derived for nilpotent polynomials
on $S_{i}$ with the help of the analogs of Eqs.~(\ref{EQ21aa}). A
requirement of eliminating certain terms in $f$ by a proper choice of
the local $SL(2,\mathbb{C})$ control parameters can also be imposed by
analogy. In particular, the part of $f$, which contains only
multi-linear terms in $S_{i}$ may be reduced to the $SL(2)$ canonic
form which, up to the replacement $\sigma_{i}^{+}\mapsto S_{i}$, is
identical to that of qubits. This specifies the canonic form of the
entire $f$, whereas the terms containing at least one $S_{i}^{2}$
factor comprise a generic polynomial with the coefficient specified in
the process of reducing the multi-linear part to the canonic form.

\subsection{Transformation of the nilpotential under change of
partition}

We next consider a situation which, in a sense, is the opposite to the
above-discussed scenario of generalized entanglement, whereby the
ranks of the algebras employed for the entanglement classification
\textit{exceed }the dimensions of the local Hilbert space of the
elements. This is the case of a partition of a composite system, where
each part is composite by itself and may contains multiple
elements. In other words, from an initial assembly we compose a new
one by considering groups of elements as new elements and by
describing the quantum state of each group by a single, collective
quantum number. One can say that the new assembly results from {\em
merging} of elements in the old one. 

We begin with the example of $n$-qubit system characterized by the
nilpotential of Eq.~(\ref{EQ12}),%
\begin{equation}
f(\{\sigma_{i}^{+}\})=\sum_{\{ k_i\}=0,1 }\beta_{k_{n},k_{n-1}\ldots k_{1}}
\prod_{i=1}^{n}\left(  \sigma_{i}^{+}\right)^{k_{i}}\:,\label{EQ34}
\end{equation}
partitioned in three parts $A$, $B$, and $C$, each of which contains
$n_{A}$, $n_{B}$, and $n_{C}$ qubits, respectively. The particular
case $C=\varnothing$ recovers the bipartite setting. The new assembly
thus consists of three elements with $d_{A}=2^{n_{A}}$,
$d_{B}=2^{n_{B}}$, and $d_{C}=2^{n_{C}}$.

By exploiting a standard Hubbard-Stratonovich procedure~\cite{HS}, we
can represent the polynomial $F=\mathrm{e}^{f}$ corresponding to
Eq.~(\ref{EQ34}) in the form of an integral%
\begin{equation}
F(\{\sigma_{i}^{+}\})=\int\mathrm{e}^{f(\{z_{i}\})}\prod_{i=1}^{n}%
\mathrm{e}^{-\left\vert z_{i}\right\vert ^{2}+z_{i}^{\ast}\sigma_{i}^{+}}%
\frac{\mathrm{d}^{n}z_{i}\mathrm{d}^{n}z_{i}^{\ast}}{\pi}~, \label{EQ35}%
\end{equation}
where the integration has to be performed independently over both the complex
variables $z_{i}$ and their complex conjugates $z_{i}^{\ast}$. This suggests a
straightforward separation of the system into three parts,
\begin{widetext}%
\begin{equation}
F\left\vert 0\right\rangle =\int\exp\left\{  \sum\limits_{i=n_{A}+n_{B}+1}%
^{n}z_{i}^{\ast}\sigma_{i}^{+}\right\}  \left\vert 0\right\rangle _{C}%
\exp\left\{  \sum\limits_{i=n_{A}+1}^{n_{A}+n_{B}}z_{i}^{\ast}\sigma_{i}%
^{+}\right\}  \left\vert 0\right\rangle _{B}\exp\left\{  \sum\limits_{i=0}%
^{n_{A}}z_{i}^{\ast}\sigma_{i}^{+}\right\}  \left\vert 0\right\rangle
_{A}\mathrm{e}^{f(\{z_{i}\})}\prod_{i=1}^{n}\mathrm{e}^{-\left\vert
z_{i}\right\vert ^{2}}\frac{\mathrm{d}^{n}z_{i}\mathrm{d}^{n}z_{i}^{\ast}}%
{\pi}\:, \label{EQ36}%
\end{equation}%
\end{widetext}%
where $\left\vert 0\right\rangle _{A;B;C}$ denote vacuum states of the
new elements, which we still can choose as product states of qubits
included within the new elements.

Let the Cartan subalgebras and the commuting nilpotent elements of the
$su(2^{n_{A}})$, $su(2^{n_{B}})$, and $su(2^{n_{C}})$ algebras be
denoted by $\left\{ \lambda_{A}^{\kappa_{A}}\right\} $ $\left\{
\nu_{A}^{\kappa_{A} }\right\} $, $\left\{
\lambda_{B}^{\kappa_{B}}\right\} $ $\left\{ \nu
_{B}^{\kappa_{B}}\right\}$, and $\left\{
\lambda_{C}^{\kappa_{C}}\right\} $ $\left\{
\nu_{C}^{\kappa_{C}}\right\}$, respectively, and let the state of the
assembly be characterized by a nilpotent polynomial on $\nu^{\kappa}$
upon noticing that $\left( \nu^{\kappa}\right)^{2}=0$. This yields%
\begin{widetext}
{%
\begin{align}
F(\{\nu_{A}^{\kappa_{A}},\nu_{B}^{\kappa_{B}},\nu_{C}^{\kappa_{C}}%
\})=\int\left(  1+\sum_{\kappa_{A}}\left\langle 0\right\vert _{A}\left(
\nu_{A}^{\kappa_{A}}\right)  ^{\dag}\mathrm{e}^{%
{\textstyle\sum\nolimits_{i=1}^{n_{A}}}
z_{i}^{\ast}\sigma_{i}^{+}}\left\vert 0\right\rangle _{A}\nu_{A}^{\kappa_{A}%
}\right)   &  \left(  1+\sum_{\kappa_{B}}\left\langle 0\right\vert _{B}\left(
\nu_{B}^{\kappa_{B}}\right)  ^{\dag}\mathrm{e}^{%
{\textstyle\sum\nolimits_{i=n_{A}+1}^{n_{A}+n_{B}}}
z_{i}^{\ast}\sigma_{i}^{+}}\left\vert 0\right\rangle _{B}\nu_{B}^{\kappa_{B}%
}\right)  \nonumber\\
\left(  1+\sum_{\kappa C}\left\langle 0\right\vert _{C}\left(  \nu_{C}%
^{\kappa_{C}}\right)  ^{\dag}\mathrm{e}^{%
{\textstyle\sum\nolimits_{i=n_{A}+n_{B}+1}^{n}}
z_{i}^{\ast}\sigma_{i}^{+}}\left\vert 0\right\rangle _{C}\nu_{C}^{\kappa_{C}%
}\right)   &  \mathrm{e}^{f(\{z_{i}\})}\prod_{i=1}^{n}\mathrm{e}^{-\left\vert
z_{i}\right\vert ^{2}}\frac{\mathrm{d}^{n}z_{i}\mathrm{d}^{n}z_{i}^{\ast}}%
{\pi}\:.\label{EQ37}%
\end{align}
}
One may rewrite this in the form%
\begin{equation}
F(\{\nu_{A}^{\kappa_{A}},\nu_{B}^{\kappa_{B}},\nu_{C}^{\kappa_{C}}%
\})=\sum_{\kappa_{A},\kappa_{B},\kappa_{C}=0}^{2^{n_{A}};2^{n_{B}};2^{n_{C}}%
}\widetilde{\alpha}_{\kappa_{A},\kappa_{B},\kappa_{C}}\nu_{A}^{\kappa_{A}}%
\nu_{B}^{\kappa_{B}}\nu_{C}^{\kappa_{C}}\:,\label{EQ38}%
\end{equation}
where the expression%
\begin{align}
\widetilde{\alpha}_{\kappa_{A},\kappa_{B},\kappa_{C}}  & =\int\left\langle
0\right\vert _{A}\left(  \nu_{A}^{\kappa_{A}}\right)  ^{\dag}\exp\left\{
\sum\limits_{i=0}^{n_{A}}z_{i}^{\ast}\sigma_{i}^{+}\right\}  \left\vert
0\right\rangle _{A}\left\langle 0\right\vert _{B}\left(  \nu_{B}^{\kappa_{B}%
}\right)  ^{\dag}\exp\left\{  \sum\limits_{i=n_{A}+1}^{n_{A}+n_{B}}z_{i}%
^{\ast}\sigma_{i}^{+}\right\}  \left\vert 0\right\rangle _{B}\nonumber\\
& \left\langle 0\right\vert _{C}\left(  \nu_{C}^{\kappa_{C}}\right)  ^{\dag
}\exp\left\{  \sum\limits_{i=n_{A}+n_{B}+1}^{n}z_{i}^{\ast}\sigma_{i}%
^{+}\right\}  \left\vert 0\right\rangle _{C}\mathrm{e}^{f(\{z_{i}\})}%
\prod_{i=1}^{n}\mathrm{e}^{-\left\vert z_{i}\right\vert ^{2}}\frac
{\mathrm{d}^{n}z_{i}\mathrm{d}^{n}z_{i}^{\ast}}{\pi}\:, \label{EQ39}%
\end{align}%
\end{widetext}%
explicitly gives the coefficients of the new nilpotent polynomials
characterizing entanglement in the new assembly. Here, the identity
operators $\nu_{A}^{0}{}$, $\nu_{B}^{0}$, and $\nu_{C}^{0}{}$ are
included in the sets $\left\{ \nu_{A}^{\kappa_{A}}\right\} ,\left\{
\nu_{B}^{\kappa_{B}}\right\} $, and $\left\{
\nu_{C}^{\kappa_{C}}\right\}$, respectively. Note that for bipartite
and tripartite entanglement, the expressions for $\widetilde{\alpha }$
and for $\widetilde{\beta}$ are identical, provided that one
eliminates terms of $F$ linear in $\nu$ by local transformations
$SU(2^{n_{A}})\otimes SU(2^{n_{B}})\otimes SU(2^{n_{C}})$, and
normalizes the reference state population to unity.  Also note that
for a bipartite case, the sum $\rho_{\kappa_{A},\kappa_{A}^{\prime}}=$
$\sum_{\kappa_{B}}\widetilde{\alpha
}_{\kappa_{A},\kappa_{B}}\widetilde{\alpha}_{\kappa_{A}^{\prime},\kappa_{B}%
}^{\ast}$ gives the density matrix of part $A$ normalized to unit
population of the reference state.

The generalization of expressions Eq.~(\ref{EQ38})-(\ref{EQ39}) to 
a larger number of new elements is straightforward:%
\begin{align}
F(\{\nu_{A}^{\kappa_{A}},\ldots,\nu_{W}^{\kappa_{W}}\})  & =\sum_{\kappa
_{A},\ldots,\kappa_{W}=0}^{2^{n_{A}};\ldots;2^{n_{W}}}\widetilde{\alpha
}_{\kappa_{A},\ldots,\kappa_{W}}\nu_{A}^{\kappa_{A}}\ldots\nu_{W}^{\kappa_{W}%
},\nonumber\\
f(\{\nu_{A}^{\kappa_{A}},\ldots,\nu_{W}^{\kappa_{W}}\})  & =\sum_{\kappa
_{A},\ldots,\kappa_{W}=0}^{2^{n_{A}};\ldots;2^{n_{W}}}\widetilde{\beta
}_{\kappa_{A},\ldots,\kappa_{W}}\nu_{A}^{\kappa_{A}}\ldots\nu_{W}^{\kappa_{W}%
}\:,
\end{align}
where the integrand of Eq.~(\ref{EQ39}) for
$\widetilde{\alpha}_{\kappa _{A},\ldots,\kappa_{W}}$ now contains more
factors 
$$\left\langle 0\right\vert _{K}\left(
\nu_{K}^{\kappa_{A}}\right) ^{\dag}\exp\left\{
{\textstyle\sum}_i
z_{i}^{\ast}\sigma_{i}^{+}\right\} \left\vert 0\right\rangle _{K}\:,$$
with $i$ running over all qubits included in the new element set $K$.
Once the reference state amplitude is normalized to $1$, the
nilpotential $f$ may be found by direct calculation of $\ln F.$ This
yields relations among $\widetilde{\alpha}$ and $\widetilde{\beta}$
that even for the canonic state do not coincide when the assembly
comprises more than three new elements.

One can finally raise the following question: What could be a
reasonable choice of a state marking an orbit in the case where the
dimensions of elements become large, and the exponential complexity
makes the identification of a global maximum of the reference state
population an intractable problem? One of the possibilities is to rely
on dynamics and manipulations exclusively with the nilpotential $f$,
aiming at the elimination of the coefficients $\beta$ of all the
states connected to the states $\left\vert k,\ldots,k\right\rangle $
by a single local transformation, as illustrated in
Fig.~\ref{FIG5}. However, we note that such a choice, though providing
a unique characterization of entanglement, may yield canonic forms not
corresponding to the maximum population of the reference state. Hence
it can be ambiguous, leading to different polynomials for the same
orbit, as it is the case of the example presented in Eq.~(25) of
Ref.~\cite{Gingrich}. Still, such a choice might have some advantages
in view of \textquotedblleft operational compatibility" important for
large ensembles, since it relies exclusively on the manipulations with
the extensive polynomial $f$ and does not invoke $F$, thereby avoiding
the need for the exponentially long procedure of $F\Leftrightarrow f$
conversion. Another advantage is that the generalization to the $SL$
case is straightforward.

\section{Summary and outlook}

We conclude by summarizing the general idea of our entanglement description
in composite quantum systems, based on the selection of a product reference
state and on the introduction of appropriate local nilpotent operators. We
also discuss the results obtained with the help of the nilpotent polynomials
and mention possible future applications of this technique to other physical
problems relevant to quantum entanglement and quantum information science.

Dealing with a physical system composed of distinguishable parts, we need
to specify among which subsets of the parts we wish to consider.
 In order to avoid confusion, we consider each of these subsets
as a single element and characterize it by a single quantum number. We call
an assemply, the collection of all the elements.

The main purpose for introducing the nilpotent variables technique is to
obtain a characterization of entanglement by \textit{extensive} quantities
that are sums of the characteristics of the unentangled parts of the system.
This approach critically relies on an important property of nilpotent
polynomials: any analytical function of a polynomial is also a polynomial,
whereas a key role is played by the logarithm function, which enables one to
relate the product state to a sum of independent terms, each of which
represents a part of the system unentangled with the rest. The extensive
characteristic, nilpotential $f$, that emerges, for a product quantum state
is indeed a sum of nilpotentials of the unentangled parts, whereas the
presence of entanglement among the elements of different parts is
represented by the corresponding cross-terms. Verification whether or not
such terms are present in the nilpotential thus serves as the entanglement
criterion.

By the very meaning of entanglement among elements of an assembly, these
characteristics should be insensitive to local transformations, which result
from arbitrary reversible (both unitary and not unitary) individual
manipulations of the elements. Therefore, the entire orbit of states, that
is the manifold of all states of the assembly that can be reached from a
given initial state by local operations, should correspond to the same
parameter values characterizing entanglement. The number of these
parameters, that is the dimension $D$ of the orbit coset, depends on the
type of local transformations allowed. For both unitary $SU$ and non-unitary 
$SL$ transformations, we propose a canonic form of the assembly state, which
depends on $D$ parameters and serves as the orbit marker.  For qubits, the
canonic form relies on the choice of a reference product state of the
assembly, the vacuum, and on maximization of the population of this state
via local transformations.

Given the canonic state of an assembly of qubits, we represent it in the
form of a polynomial $F_{c,C}(\left\{ \sigma _{i}^{+}\right\} )$ on the
raising operators $\sigma _{i}^{+}$ acting on the vacuum state. After
normalizing to unit vacuum state amplitude, we calculate another polynomial,
the \emph{tanglemeter} $f_{c,C}(\left\{ \sigma _{i}^{+}\right\} )=\ln
F_{c,C}(\left\{ \sigma _{i}^{+}\right\} )$, which depends on exactly $D$
parameters and contains all the information about entanglement in the form
of the coefficients standing in front of different nilpotent monoms $%
\prod\nolimits_{i}\sigma _{i}^{+}$. Subscripts $C$ and $c$ show that we
remain within $SU$ or $SL$ group of local transformations, respectively.  In
contrast to state given by the polynomial $F_{c,C}$, the tanglemeter $f_{c,C}
$ is an \emph{extensive} quantity. In statistical physics, extensive
quantities scale linearly with the size of the ensemble, like free energy
given by the logarithm of the partition function. The tanglemeter has
similar additive properties: for a composite system representing a set of
unentangled parts, the tanglemeter equals the sum of the tanglemeters of the
constituent parts. Straightforward inspection of the second derivatives $%
\partial ^{2}f_{c,C}/\partial \sigma _{i\in A}^{+}\partial \sigma _{j\in
B}^{+}$ allows one to check whether or not groups $A$ and $B$ of elements
are entangled.

We have presented several examples of tanglemeters for systems of a few
qubits. Still there may exist a certain number of states, for which the
tanglemeters, though dependent on the same number $D$ of parameters, cannot
have the chosen structure. These states comprise singular classes. In the
four-qubit example, we have shown in detail how a classification can be
constructed by considering local infinitesimal transformations sequentially
diminishing the coefficients of the monoms to be eliminated. This procedure
is described by a set of differential equations for the coefficients of $f$
and can be seen as control process, when we impose feedback conditions on
the parameters of continuous local transformations. By properly adjusting
these parameters to current values of the coefficients of the nilpotential,
we rapidly arrive at the tanglemeter $f_{c}$ in the limit $t\rightarrow
\infty $. The latter thus appears as a stationary stable point of the set of
equations with feedback. However, the proper choice of the parameters
required for elimination of the monoms cannot be made in certain domains in
the space of the nilpotential coefficients, where this procedure fails as a
result of the ``loss of complete controllability". In these domains, the
determinants of matrices relating the time derivatives of the coefficient
with the parameters of the local transformations vanish. The cases where
one, two, three, or all four eigenvalues of the determinant equal zero
correspond to different entanglement classes. More systematic exploration of
this classification in larger assemblies of qubits looks like an interesting
prospective task for immediate future research.

We have demonstrated that the nilpotent polynomial technique, initially
developed for qubits, can be extended to $d$-level elements, qudits, each of
which has different dimention $d_{i}$. In this case, a larger number of the
nilpotent variables per subsystem are required for the construction of the
polynomials. Let us spell out the salient features of the proposed
description of entanglement for the general case.

\begin{itemize}
\item[(i)] We start by choosing a reference state $|0\rangle _{i}$ for an
individual element and a corresponding state for the composite quantum
system, an assemply consisting of $d$-level systems, $|\mathrm{\ O}\rangle
=|0,\ldots ,0\rangle $. All other states are obtained by the action on $|%
\mathrm{\ O}\rangle $ by polynomials of nilpotent local operators. For
qubits, these nilpotent operators are simply the spin raising operators $%
\sigma _{i}^{+}$. For qudits, we invoke $d-1$ nilpotent commuting operators $%
\nu _{i}^{\kappa _{i}}$ that create $d_{i}-1$ excited states $\left\vert
\kappa _{i}\right\rangle =\nu _{i}^{\kappa _{i}}$ $|0\rangle _{i}$ out of
the vacuum state $|0\rangle _{i}$. This choice of basis is natural in the
framework of the Cartan-Weyl decomposition of the $su(d)$ algebra. Thereby
any state of the composite system may be represented by a polynomial $F(\nu
_{i}^{\kappa _{i}})$ acting on the reference state.

\item[(ii)] By applying local transformations to a given assembly state we
bring it into a certain \textit{canonic form}. For systems of qubits, the
latter is characterized by maximum population of the reference state $|%
\mathrm{\ O}\rangle $. In the case of qudits, we need to additionally
maximize the population of the maximum symmetric excited states. When the
dimensions $d_{i}$ of all elements are equal, these are the states $%
|1,\ldots ,1\rangle $, ..., $|\kappa ,\ldots ,\kappa \rangle $\ldots , $%
|d-1,\ldots ,d-1\rangle $, while for different $d_{i}$, in this sequence,we
replace $\kappa $ by $d_{i}-1$ for the elements of the dimention $%
d_{i}<\kappa $.

\item[(iii)] Any analytical function of a polynomial of nilpotent arguments $%
\nu _{i}^{\kappa _{i}}$ is also a polynomial. The logarithm of the canonic
polynomial $F_{c}$, \textit{tanglemeter,} has coefficients that by
construction are invariant with respect to local transformations. This
offers a simple way to systematically list all the invariant characteristics
of entanglement for a system of an arbitrary number of qudits. 
\end{itemize}

Another extension of the technique applies to a case relevant to generalized
entanglement, where the algebra of local transformations has a rank strictly
less than that suggested by the dimension of the elements. In such a
situation, the number of nilpotent variables per element is less than $d-1$,
and moreover, the form of the nilpotent polynomial is sensitive to the
choice of reference state, thus being \textquotedblleft reference" and
\textquotedblleft observable" dependent. Identification of classes for large
elements and for generalized entanglement based on the dynamic equations are
two immediate open questions concerning the proposed technique. It would be
interesting to identify the classes in the case of generalized entanglement
and to derive the dynamic equations describing the evolution of the
corresponding nilpotential. This constitutes yet another direction for
future studies.

Manipulation of quantum assemblies and, in particular, quantum computation,
implies application of both local and nonlocal two-particle gate 
transformations. This is also the case for system composed of naturally
interacting elements, such as spin chains, cold Rydberg gases, and arrays of
cold two-level atoms trapped in a standing electromagnetic wave. The Schr%
\"{o}dinger equation yields a dynamic equation for the nilpotent polynomials
-- a linear one for the polynomial $F$, which is in turn linear in the state
amplitudes, and a nonlinear one for $f=\ln F$. For an ensemble of $n$
qubits, both dynamic equations involve $\sim 2^{n}$ variables, the
polynomial coefficients and their time derivatives. At first glance, linear
equations are always simpler. However, this is not necessarily the case: we
have shown an example of universal evolution that yields an equation for $f$%
, which is equivalent to the classical Hamilton-Jacobi equation in just $n$
dimensions. How far does this analogy go? What are the consequence for
entanglement dynamics and, in particular, for quantum algorithms? These are
intriguing questions to be further addressed.

Finally, one could also conceive applications of the nilpotent polynomial
technique for analytical investigations of entanglement in correlated
many-body systems -- including the problem of better characterizing quantum
phase transitions~\cite{Somma} -- and extend the present consideration to
the case where non-unitary decoherence and dissipation effect are included
in the dynamics. In an even broader context, one can think of employing this
technique as a tool for establishing relations between entanglement dynamics~%
\cite{Amico,Cubitt} and the known exactly solvable problems of statistical
mechanics and complex quantum systems \cite{Akulin,Takahashi}.

\section*{Acknowledgments}

A.~M. would like to thank her advisor J.~W. Clark for supporting her
during this work. She also acknowledges the support by the U.~S.\
National Science Foundation under Grant No.~PHY-0140316 and by the
Nipher Fund. V.M.A. gratefully acknowledges the support of QUACS EC
network grant. A.~M. and V.M.A. are grateful to J.~W. Clark, I. Dumer,
J. Siewert, V. Ka\c{c}, V.  Kravtsov I. Lerner and V. Youdson for
useful discussions and communication, as well as to the organizers of
\textsl{Madeira Math Encounters XXIX} for their
hospitality. L.~V. gratefully acknowledges partial support from
Constance and Walter Burke through their Special Project Fund in
Quantum Information Science.

\section{Appendices}

\subsection{Dimension of cosets}

\paragraph{\textsl{Two qubits}}
The fact that the expression Eq.~(\ref{EQ2}) is invariant implies that
the counting $2^{n+1}-3n-2$ of the number of invariants breaks for
$n=2$.  That means that out of $3n=6$ local transformations, only $5$
act faithfully (non-trivially). In other words, there is a certain
local transformation which leaves a generic wave function intact. On
can amuse oneself and find it explicitly. A generic infinitesimal
local transformation is
\begin{equation}
\delta\psi_{ij}\ =\ iP_{2}^{\kappa}(\sigma_{2}^{\kappa})_{ii\prime}%
\psi_{i\prime j} + iP_{1}^{\kappa}(\sigma_{1}^{\kappa})_{jj\prime}
\psi_{ij\prime}  \, . \label{EQA1}
\end{equation}
Leaving only the components $\psi_{00}$ and $\psi_{11}$ (as dictated by
the canonic form), and writing $P_{r}^{\kappa}\sigma_{r}^{\kappa}=P_{r}%
^{z}\sigma^{z}+P_{r}^{-}\sigma^{+}+P_{r}^{+}\sigma^{-}$ with $r=1,2$, we
arrive at the system of equations 
\begin{align}
-i\delta\psi_{11}  &  = (P_{1}^{z}+P_{2}^{z})\psi_{11}=0\nonumber\\
-i\delta\psi_{00}  &  = -(P_{1}^{z}+P_{2}^{z})\psi_{00}=0\nonumber\\
-i\delta\psi_{01}  &  = P_{1}^{-}\psi_{00}+P_{2}^{+}\psi_{11}=0\nonumber\\
-i\delta\psi_{10}  &  = P_{2}^{-}\psi_{00}+P_{1}^{+}\psi_{11}=0 \label{EQA2}%
\end{align}
From the last two equations, it follows that $P_{1}^{+}=P_{2}^{+}=0$
(for generic $\psi_{00}$, $\psi_{11}$), while the first and the second
ones give the same condition $P_{1}^{z}+P_{2}^{z}=0$ (or $\phi_1 +
\phi_2 + \varphi_1 + \varphi_2 = 0$ in the notation of
Eq.~(\ref{n2trans})), i.e. we have, indeed, a one-parametric set of
local transformations that leave $\psi_{ij}$ intact.
\vspace{1cm}

\paragraph{\textsl{Three qubits}}

Take a special canonic wave function with only four nonzero
components: $\psi_{011},\psi_{101},\psi_{110}$ and $\psi_{000} = 1$.
The infinitesimal local transformations are
\begin{align}
\delta\psi_{ijk}\  &  =\ iP_{3}^{\kappa}(\sigma_{3}^{\kappa})_{ii^{\prime}%
}\psi_{i^{\prime}jk}+iP_{2}^{\kappa}(\sigma_{2}^{\kappa})_{jj^{\prime}}%
\psi_{ij^{\prime}k}\nonumber\\
&  +iP_{1}^{\kappa}(\sigma_{1}^{\kappa})_{kk^{\prime}}\psi_{ijk^{\prime}\ }.
\label{EQA3}%
\end{align}
Let us show that $\delta\psi_{ijk}=0$ implies $P_{r}^{\kappa}=0$ for
all $\kappa=\pm,z$, and $r=1,2,3$. The corresponding system of
equations is
\begin{align}
-i\delta\psi_{000}  & = -(P_3^z + P_2^z + P_1^z) = 0 \, ,\label{EQA4.0}\\
-i\delta\psi_{100}  &  =P_{2}^{+}\psi_{110}+P_{1}^{+}\psi_{101} + P_3^- 
=0 \, , \label{EQA4.1}\\
-i\delta\psi_{010}  &  =P_{3}^{+}\psi_{110}+P_{1}^{+}\psi_{011} + P_2^-
=0 \, , \label{EQA4.2}\\
-i\delta\psi_{001}  &  =P_{3}^{+}\psi_{101}+P_{2}^{+}\psi_{011} + P_1^-
=0 \, , \label{EQA4.3}\\
-i\delta\psi_{011}  &  =(P_{2}^{z}+P_{1}^{z}-P_{3}^{z})\psi_{011}%
=0 \, , \label{EQA4.4}\\
-i\delta\psi_{101}  &  =(P_{1}^{z}+P_{3}^{z}-P_{2}^{z})\psi_{101}%
=0 \, , \label{EQA4.5}\\
-i\delta\psi_{110}  &  =(P_{3}^{z}+P_{2}^{z}-P_{1}^{z})\psi_{110}%
=0 \, , \label{EQA4.6}\\
-i\delta\psi_{111}  &  =P_{3}^{-}\psi_{011}+P_{2}^{-}\psi_{101}+P_{1}^{-}%
\psi_{110}=0 \, .\label{EQA4.7}%
\end{align}

Eqs.~(\ref{EQA4.1})-(\ref{EQA4.3}) have the form of a homogeneous
linear system of six equations for the real and imaginary parts of
$P_{r}^{+}$. One can be convinced that the determinant of this system
does not vanish for generic $\psi_{011} ,\psi_{101},\psi_{110}$ and
this implies that the only solution is $P_{r}^{+}=0$.
Eqs.~(\ref{EQA4.4})-(\ref{EQA4.6}) give in turn a homogeneous system
for the three parameters $P_{r}^{z}=0$. Its determinant does not
vanish for nonzero $\psi_{011} ,\psi_{101},\psi_{110}$ and this
implies $P_{r}^{z}=0$. \hfill\textit{Q.E.D.}

Note that this statement would not be correct for \textit{any} pure
state. The state having $\psi_{000}$ as the only non-vanishing
component is annihilated by two linearly independent generators (with
$P_{1}^{z}+P_{2}^{z}=0$ and $P_{1}^{z}+P_{3}^{z}=0$). For some other
choice, there is only one trivially acting generator. But for a
generic state there is none.
\vspace{1cm}

\paragraph{\textsl{$n>3$ qubits}}

The local transformations involve $3n$ parameters
$P_{j}^{\kappa}$. Introduce, as before, complex
$P_{j}^{\pm}=P_{j}^{x}\pm i P_{j}^{y}$. Take the wave function with
$\psi_{000} = 1$ and $n$ other nonzero components $\psi_{01\ldots1}$,
$\psi_{101\ldots1}$, \ldots, $\psi_{1\ldots10}$. Consider
$\delta\psi_{01\ldots1}$, etc. We obtain a system of $n$ linear
homogeneous equations for $n$ real variables $P_{i}^{z}$. The matrix
of the system
\begin{equation}
M\ =\ \left(
\begin{array}
[c]{cccc}%
-1 & 1 & \ldots & 1\\
1 & -1 & \ldots & 1\\
\ldots & \ldots & \ldots & \ldots\\
1 & \ldots & \ldots & -1
\end{array}
\right)  \label{matr}%
\end{equation}
(cf. Eqs.~(\ref{EQA4.4})-(\ref{EQA4.6})) has a non-vanishing
determinant for $n>2$. Indeed, $M=A-2I$, where $A$ is the matrix with
all components equal to $1$ and $I$ is the unit matrix. But $A$ is a
matrix of rank $1$. It has $n-1$ degenerate eigenvalues $\lambda=0$
and one eigenvalue $\lambda=n$. And $\lambda=2$ is not an
eigenvalue. It follows from this that the matrix $M$ does not have
zero eigenvalues and the only solution to the equation system is
$P_{j}^{z}=0$.

Consider now $\delta\psi_{\{k_{n}\}}$, where the set $\{k_{n}\}$
involves $2$ zeros and $n-2$ unities. We obtain a system of
$n(n-1)/2$ linear homogeneous equations for $n$ complex variables
$P_{j}^{+}$. Let us prove by induction that it has only zero solutions
for generic $\psi_{01\ldots1}$, etc. Consider first
$\delta\psi_{\{k_{n}\}}$, with the unity at the leftmost position
($\{k_{n}\}= \{ 1, \{k_{n-1}\}\} $). The equation system for
$P_{2}^{+} ,\ldots,P_{n}^{+}$ derived from this has the same form as
the equation system $\delta\psi_{\{k_{n-1}\}}$ in the case of $n-1$
qubits. By the inductive assumption, it has only zero
solutions. Knowing that $P_{2,\ldots, n}^{-}=0$ and considering, say,
the equation
\[
\delta\psi_{001\ldots1}=P_{1}^{+}\psi_{10\ldots1}+P_{2}^{+}\psi_{01\ldots
1}=0\,,
\]
we derive that in addition that $P_{1}^{+}=0$. 

This proves that no
infinitesimal transformation leaving the generic $n$-qubit state
exists for $n>2$.
\vspace{1cm}

\paragraph{\textsl{Remark on the invariance of the canonic form}}

Let us consider the canonic state suggested by Eq.~(\ref{EQ6}) and try to
prove that this form is unique. Consider infinitesimal local transformations
$1+\mathrm{i}\sum_{i}P_{i}^{+}\sigma_{i}^{-}+\mathrm{i}\sum_{i}P_{i}^{-}%
\sigma_{i}^{+}$ of a canonic state vector and require that they do not bring
about linear in $\sigma_{j}^{+}$ terms. The requirement means that
\[
\delta\psi_{10\ldots0}=\ldots=\delta\psi_{0\ldots01}=0
\]
and implies
\begin{align}
P_{1}^{-}\psi_{0\ldots0}+P_{2}^{+}\psi_{110\ldots0}+\ldots+P_{n}^{+}%
\psi_{10\ldots01}\  &  =\ 0\newline\label{otklon}\\
&  \ldots\nonumber\\
\newline P_{1}^{+}\psi_{10\ldots01}+P_{2}^{+}\psi_{010\ldots01}+\ldots
+P_{n}^{-}\psi_{0\ldots0}\  &  =\ 0\ .\nonumber
\end{align}
This is a system of $n$ complex equations for $n$ complex parameters
$P_{j}^{-}=\left(  P_{j}^{+}\right)  ^{\ast}$, which we have encountered in
Eq.~(\ref{EQ62a}). It can be given in the form
\begin{equation}
P_{j}^{-}+\sum_{i=1}^{n}P_{i}^{+}M_{i,j}=0, \label{ur}%
\end{equation}
which with the allowance for Eq.~(\ref{EQ60}) coincides with the right hand
side of Eq.~(\ref{EQ62}), which we have encountered discussing the dynamic
reduction of nilpotentials to the canonic form. There we have seen that all
eigenvalues of $M_{i,j}$ are never equal $1$ for generic states in the canonic
form, since it would contradict to the requirement of the maximum vacuum state
population. Therefore the determinant of the system Eq.~(\ref{ur}) does not
vanish in the generic case, \ and hence the system Eq.~(\ref{otklon}) has only
the trivial solution $\alpha_{j}=0$. The canonic form of the generic state is
thus unique and can only experience phase transformations, unless the phases
are specified by additional requirements.

\subsection{Graph approach. $4$-qubit $sl$-classification\label{B}}

In this Appendix we attribute a physical meaning to the coefficients of the
tanglemeter by relating these to concurrence and
 $3$-tangle. Based on these results, in the first section of this Appendix we
construct graphs that illustrate the entanglement topology. In the second
section, we summarize the $sl$-entanglement classes for four qubits as they
emerge from the dynamic equations for the tanglemeter.\vspace{1cm}

\textsl{a.~~Graphical interpretation of the tanglemeter coefficients}
Starting with two qubits in a pure state, the tanglemeter
involves only one coefficient, $\beta_3$,  in terms of which
we can express the von Neumann entropy $S_{vN}=-\mathrm{Tr}[\rho_{A}\log\rho_{A}]$ 
or the concurrence, by 
\begin{equation}
C_{12}=2\sqrt{\mathrm{{det}}\rho}=2\beta_{11}/(1+\beta_{11}^{2})~~.
\label{con}%
\end{equation}
Since the tanglemeter in this case is nothing but the Schmidt decomposition,
nothing new is introduced; a nonzero $\beta$ coefficient implies the presence of
bipartite entanglement among the two qubits. 

For three qubits in a pure state, both $C$ and $\tau$ can be
expressed in terms of the amplitudes $\psi$ of the $su$-canonical state Eqs.~(\ref{concurrence}-
\ref{threetangle}).
 These expressions are shorter than for a
non-canonical state vector, since they do not contain the amplitudes $\psi
_{100}$, $\psi_{010}$, and $\psi_{001}$.

For more than three qubits one can evaluate the concurrence.
Since the expressions are rather complicated, we shall instead employ
A. Peres' separability criterion of partial transposition \cite{Peres}:
the qubits are not entangled iff the eigenvalues of the partially 
transposed reduced density matrix  of qubits 1 and 2
are non negative.  
The first case we  considered is that in which  only bilinear terms
are present in the canonical state of four qubits.
The eigenvalues of the partial transposed reduced density matrix are found to
 satisfy the following relationships:
\begin{align}
\label{4bi}
 \kappa_1 \kappa_2= & \left|\psi_{0101}\psi_{1001}+\psi_{1010}\psi_{0110}\right|^2 
-\left|\psi_{0011}\psi_{0000}\right|^2, \nonumber\\
\kappa_1 + \kappa_2= & 2 \mathrm{Re} \left(\psi_{1001}^*\psi_{1010}+\psi_{0101}^*\psi_{0110}\right), \nonumber\\
 \kappa_3 \kappa_4 = & - \left|\psi_{0101}\psi_{1001}+\psi_{1010}\psi_{0110}\right|^2 \nonumber\\
 &+ \left|\psi_{0011}\psi_{0000}\right|^2+\left|\psi_{0011}\psi_{1100}\right|^2, \nonumber\\
 \kappa_3 + \kappa_4=& \left|\psi_{0000}\right|^2+\left|\psi_{0011}\right|^2+\left|\psi_{1100}\right|^2
 \end{align}
In the second case  we consider,  only trilinear terms are present, and we find
\begin{align}
\label{4tri}
 \kappa_1 \kappa_2= & , \left|\psi_{0000}\psi_{0111}\right|^2+ \left|\psi_{0000}\psi_{1011}\right|^2
 -\left|\psi_{1110}\psi_{1101}\right|^2,\nonumber\\
\kappa_1 + \kappa_2= & \left|\psi_{0000}\right|^2+\left|\psi_{1011}\right|^2+\left|\psi_{0111}\right|^2, \nonumber\\
 \kappa_3 \kappa_4 = &  \left|\psi_{1110}\psi_{1101}\right|^2, \nonumber\\
 \kappa_3 + \kappa_4=& \left|\psi_{1110}\right|^2+\left|\psi_{1101}\right|^2.
 \end{align}
 
The considerable simplification achieved for the formulas of familiar measures can help us
to construct a topological picture of entanglement based on the invariant coefficients
of the canonical form. This analysis also suggests an alternative derivation of the $3$-tangle measure.

The rules that we prescribe for constructing the graphs to depict entanglement based on the coefficients
of the tanglemeter  are as follows:
\begin{itemize}
	\item Assign for each bilinear term $\beta_{ij}$ a line connecting the  qubits $i$ and $j$.
	\item For each trilinear term $\beta_{ijk}$, a surface on the plane confined by the $i,j,k$ qubits.
	\item For higher order terms, volumes, among the  qubits involved.
\end{itemize}

\begin{figure}[h]
{\centering{\includegraphics*[width=0.5\textwidth]{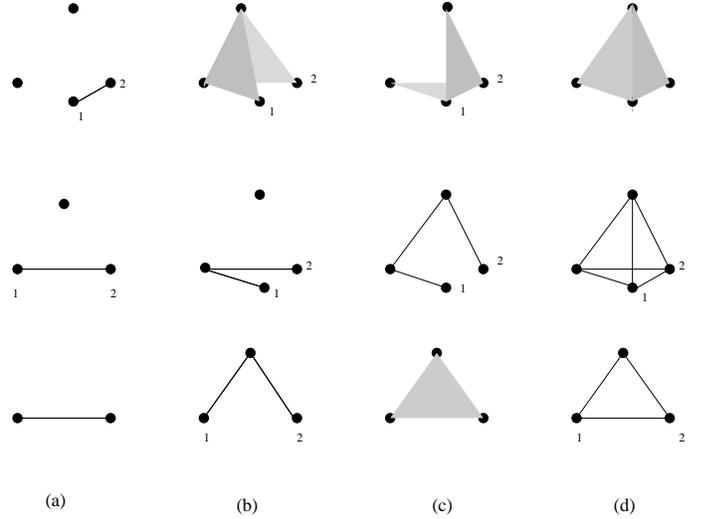}}} \vspace
{0.1cm}\caption{Bipartite entanglement among the qubits 1 and 2. 
(a) direct, (b) indirect, (c)
configurations without  entanglement, and (d) 
configurations that can result in cancellations of entanglement.
 }
\label{Bi}%
\end{figure}

From  Eqs.~(\ref{con}),~(\ref{concurrence}),~(\ref{4bi}), one sees that the existence
of a bilinear term $\beta_{ij}$   in the general case implies the presence
of bipartite entanglement between the qubits $i$ and $j$;  we call
this type of bipartite entanglement {\it direct} and represent it
graphically as in Fig.~\ref{Bi}~(a). 
Bipartite entanglement is also present when
 the   two qubits are {\it indirectly} connected  by a line
that passes through a third qubit. We can see from  Eq.~(\ref{concurrence})
that the concurrence $C_{12}$
 is nonzero  also when both $\beta_5,\beta_6$ are present, Fig.~\ref{Bi}~(b).
On the other hand, there is no indirect bipartite entanglement
if the line connecting
the two qubits involves more than two edges, as in Fig.~\ref{Bi}~(c).
In the general case, in which both direct and indirect contributions are present,
 there are cancellation effects instead of addition, since
 the closed loops contribute to tripartite entanglement (see below). 
There are also cases in which bipartite entanglement is due to higher-order
 terms, as in  Fig.~\ref{Bi}~(b), where the terms $\beta_{1110}, \beta_{1101}$
are present and the eigenvalues $\kappa_1$ and $\kappa_2$  in Eq.~(\ref{4tri}) become negative.
 Another configuration of surfaces,
like that in Fig.~\ref{Bi}.(c), does not create bipartite entanglement.

\begin{figure}[h]
{\centering{\includegraphics*[width=0.3\textwidth]{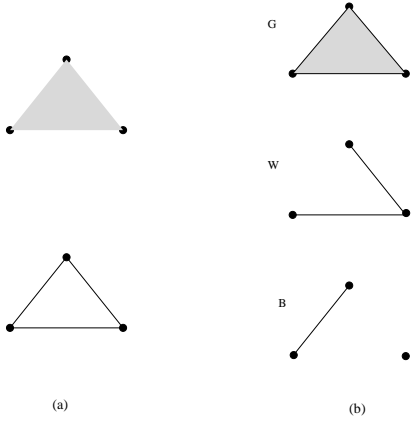}}} \vspace
{0.1cm}\caption{(a) Tripartite entanglement in two different configurations. 
(b) The three classes of entanglement for three qubits. 
 $B$ stands for biseparable,
$W$ for the singular class containing the Werner state and $G$ for the  the general  orbit.
 }
\label{Ci}%
\end{figure}

Trilinear terms correspond to  surfaces and by the Eq.~(\ref{concurrence})
we  see that their presence in the canonical form of three qubits
results in nonvanishing $3$-tangle, as in Fig.~\ref{Ci}.(a).  
 This is not the only configuration that permits  tripartite entanglement;
a loop consisting of three lines is also a configuration with genuine
tripartite entanglement, as in  Eq.~(\ref{concurrence}). The inverse statement is not true
for three qubits, since the term $\beta_{7}$ is not present in the expression
of concurrence. 

As an application, in Fig.~\ref{Ci}.(b) we represent the three classes of entanglement
for three qubits  diagrammatically way. It is important to note that if a diagrams is connected,
i.e., there is a closed loop or surface,
then this property cannot change under general local transformations.

We anticipate that the diagrammatic technique can be generalized
to more qubits and  improved with directed
lines, surfaces and volumes that would aid in visualizing
of cancellations or joint contributions. It appears that $N$-partite
entanglement can be attributed not only to the coefficients in front
of the $N$-th order monomials,
 but also to products  of  lower-order coefficients. \vspace{1cm}

\textsl{b.~~Summarizing entanglement classes under $SL(2,\mathbb{C})$
transformations}
In Sect. III.D.4, we have shown how do the $sl$-entanglement classes emerge
from the consideration of the tanglemeter dynamics under controlled action of
continuous local transformations when proper feedback requirements are
imposed. The general class has been identified along with four different
singular classes, corresponding to one, two, three, and four vanishing
eigenvectors of the determinant (\ref{EQ62e}). All these classes were
characterized by three complex parameters. Here, we further refine this
analysis and add some singular classes that depend on less parameters in order
to compare this classification with that given in Ref.~\cite{Verstraete}. We
illustrate the correspondence, when normalize the states of Ref.~\cite{Verstraete} 
to unit vacuum amplitude and compare these states with ones
suggested by $sl$-tanglemeters.

Let us remind that the $sl$-tanglemeter for four qubits we obtain in two
steps. We first employ eight out of twelve complex parameters of local $SL$
transformations and put the generic state depending initially on fifteen
complex parameters to the form of Eq.~(\ref{EQ14b}), which depends only on
seven complex parameters. Then we chose the rest of the available
transformations, four scaling operators $e^{B_{i}\sigma_{i}^{z}}$, in order to
reduce the $sl$-tanglemeter to the form of Eq.~(\ref{EQ14bb}). However, the
second step not always possible to perform: when the tanglemeter is singular
and one or more of the coefficients $\beta$ equal zero, the scalings can
simplify the nilpotential further, although not to the form of Eq.~
(\ref{EQ14bb}). For example, when in Eq.~(\ref{EQ14b}) $\beta_{3}=0$, the
tanglemeter can be set in the form
\begin{align*}
f_{C}  &  =\sigma_{3}^{+}\sigma_{4}^{+}+\beta_{5}(\sigma_{1}^{+}\sigma_{3}%
^{+}+\sigma_{2}^{+}\sigma_{4}^{+})\\
&  +\beta_{6}(\sigma_{1}^{+}\sigma_{4}^{+}+\sigma_{2}^{+}\sigma_{3}%
^{+})+\sigma_{1}^{+}\sigma_{2}^{+}\sigma_{3}^{+}\sigma_{4}^{+}~,
\end{align*}
characterized by only two parameters. For $\beta_{3}=\beta_{10}=0$  it reads
\[
f_{C}=\sigma_{3}^{+}\sigma_{4}^{+}+\sigma_{1}^{+}\sigma_{3}^{+}+\beta
_{6}(\sigma_{1}^{+}\sigma_{4}^{+}+\sigma_{2}^{+}\sigma_{3}^{+})+\sigma_{1}%
^{+}\sigma_{2}^{+}\sigma_{3}^{+}\sigma_{4}^{+}~,
\]
and $\beta_{3}=\beta_{10}=\beta_{9}=0$ results in
\[
f_{C}=\sigma_{3}^{+}\sigma_{4}^{+}+\sigma_{1}^{+}\sigma_{3}^{+}+\sigma_{2}%
^{+}\sigma_{3}^{+}+\sigma_{1}^{+}\sigma_{2}^{+}\sigma_{3}^{+}\sigma_{4}^{+}~.
\]
When the four-linear coefficient $\beta_{15}=0$ and one or more of the
quadratic coefficients are also zero, the singular classes of states without
fourpartite entanglement emerge: the $sl$-tanglemeter of four-qubit $W$ state
\[
f_{C}=\sigma_{3}^{+}\sigma_{4}^{+}+\sigma_{1}^{+}\sigma_{3}^{+}+\sigma_{2}%
^{+}\sigma_{3}^{+},
\]
belongs to one of these classes, and separable states with tanglemeters of the
type
\[
f_{C}=\sigma_{3}^{+}\sigma_{4}^{+}+\sigma_{2}^{+}\sigma_{3}^{+}+\sigma_{2}%
^{+}\sigma_{4}^{+}~,
\]
and similar, belong to other.

Consider now several special cases of the singular class Eq.~(\ref{EQ14bbbb}),
where all four of the eigenvalues of the determinant (\ref{EQ62e}) are zero.
When, in addition, one or several of the trilinear coefficients equal zero, it
becomes possible to rescale one or more of the bilinear coefficients to unity.
Less general classes with $sl$-tanglemeters like
\begin{align*}
f_{C}  &  =\sigma_{1}^{+}\sigma_{3}^{+}\sigma_{4}^{+}+\sigma_{1}^{+}\sigma
_{2}^{+}\sigma_{4}^{+}+\sigma_{2}^{+}\sigma_{3}^{+}\sigma_{4}^{+}+\sigma
_{1}^{+}\sigma_{2}^{+}\\
&  +\beta_{5}\sigma_{1}^{+}\sigma_{3}^{+}+\beta_{6}\sigma_{2}^{+}\sigma
_{3}^{+}~,\\
f_{C}  &  =\sigma_{1}^{+}\sigma_{2}^{+}\sigma_{4}^{+}+\sigma_{2}^{+}\sigma
_{3}^{+}\sigma_{4}^{+}+\sigma_{1}^{+}\sigma_{2}^{+}+\sigma_{1}^{+}\sigma
_{3}^{+}+\beta_{6}\sigma_{2}^{+}\sigma_{3}^{+}~,\\
f_{C}  &  =\sigma_{2}^{+}\sigma_{3}^{+}\sigma_{4}^{+}+\sigma_{1}^{+}\sigma
_{2}^{+}+\sigma_{1}^{+}\sigma_{3}^{+}+\sigma_{2}^{+}\sigma_{3}^{+}%
\end{align*}
emerge as a result. One reveals more singular classes, when besides of several
zero cubic coefficients, two or more quadratic terms vanish. Two examples
\begin{align*}
f_{C}  &  =\sigma_{1}^{+}\sigma_{3}^{+}\sigma_{4}^{+}+\sigma_{1}^{+}\sigma
_{2}^{+}\sigma_{4}^{+}+\sigma_{2}^{+}\sigma_{3}^{+}\sigma_{4}^{+}~,\\
f_{C}  &  =\sigma_{1}^{+}\sigma_{3}^{+}\sigma_{4}^{+}+\sigma_{1}^{+}\sigma
_{2}^{+}\sigma_{4}^{+}+\sigma_{2}^{+}\sigma_{3}^{+}~,
\end{align*}
illustrate this case.

Different four-qubit classes emerging from this classification are presented
in Table \ref{tables} along with the results of Ref.~\cite{Verstraete}. We
include the general and main singular classes and omit a number of separable
singular classes corresponding to product states. For the classes not
symmetric under cyclic permutation of qubit indexes, this permutations is
implicit. We note that our classes result from consideration of dynamic
evolution that implies a series of sequential infinitesimal local operations
preserving $su$-canonic form of the nilpotential. Therefore the situation,
where some of the obtained classes turn out to be equivalent under a finite
local $sl$-transformation, yet cannot be excluded with certainty. Keeping this
in mind, it is easy to see that $G_{a}$ is identical to $G_{abcd}$ class that
is also suggested in Ref.~\cite{Akimasa}. The class $L_{abc_{2}}$ corresponds
to $LG2_{a}$, while $L_{a_{2}b_{2}}$ coincides with $LG1_{a}$. After applying
$SL$ transformations on qubits 3 and 4 of the $L_{ab_{3}}$ class, the latter
reduces to a form that is a singular case of the general class. The state
$L_{a_{4}}$ can be set in the canonic form by flipping the second and third
qubit, and then it is a special case of $G_{c}$ for $\beta_{6}=0$ and
$\beta_{7}=i$. Continuing, $L_{a_{2}0_{3\oplus\bar{3}}}$ coincides with
$S_{f}$, $L_{0_{7\oplus\bar{1}}}$ with $S_{b}$, and $L_{0_{3\oplus\bar{1}%
}0_{3\oplus\bar{1}}}$ with $S_{d}$. The singular class $L_{0_{5\oplus\bar{3}}%
}$ is of the $S_{a}$ form. Thus, we can conclude that the two classifications
do overlap and complement each other.

\begin{widetext}%

\begin{table}
\label{tables}
\begin{tabular}
[c]{cc}%
\emph{General class} & 3 complex parameters\\\hline
$G_{a}$ & $f=\beta_{3}(\sigma_{1}^{+}\sigma_{2}^{+}+\sigma_{3}^{+}\sigma
_{4}^{+})+\beta_{5}(\sigma_{1}^{+}\sigma_{3}^{+}+\sigma_{2}^{+}\sigma_{4}%
^{+})$\\
& $+\beta_{6}(\sigma_{1}^{+}\sigma_{4}^{+}+\sigma_{2}^{+}\sigma_{3}%
^{+})+\sigma_{1}^{+}\sigma_{2}^{+}\sigma_{3}^{+}\sigma_{4}^{+}$\\
\emph{Singular $3D$ classes} & 3 complex parameters\\\hline
$G_{b}$ & $f=\beta_{3}\left(  \sigma_{1}^{+}\sigma_{2}^{+}+\sigma_{3}%
^{+}\sigma_{4}^{+}\right)  +\beta_{5}\left(  \sigma_{1}^{+}\sigma_{3}%
^{+}+\sigma_{2}^{+}\sigma_{4}^{+}\right)  +\beta_{6}\left(  \sigma_{1}%
^{+}\sigma_{4}^{+}+\sigma_{2}^{+}\sigma_{3}^{+}\right)  $\\
& $+\sigma_{1}^{+}\sigma_{2}^{+}\sigma_{3}^{+}-\sigma_{1}^{+}\sigma_{2}%
^{+}\sigma_{4}^{+}+\sigma_{1}^{+}\sigma_{3}^{+}\sigma_{4}^{+}-\sigma_{2}%
^{+}\sigma_{3}^{+}\sigma_{4}^{+}$\\
& $+2\left(  \beta_{5}\beta_{6}-\beta_{3}\beta_{6}+\beta_{3}\beta_{5}\right)
\sigma_{1}^{+}\sigma_{2}^{+}\sigma_{3}^{+}\sigma_{4}^{+}$\\
& \\
$G_{c}$ & $f=\sigma_{1}^{+}\sigma_{2}^{+}+\sigma_{1}^{+}\sigma_{3}^{+}%
+\sigma_{2}^{+}\sigma_{4}^{+}+\sigma_{3}^{+}\sigma_{4}^{+}+\beta_{6}\left(
\sigma_{1}^{+}\sigma_{4}^{+}+\sigma_{2}^{+}\sigma_{3}^{+}\right)  $\\
& $+\beta_{7}\left(  \sigma_{1}^{+}\sigma_{2}^{+}\sigma_{3}^{+}-\sigma_{2}%
^{+}\sigma_{3}^{+}\sigma_{4}^{+}\right)  +\beta_{11}\left(  \sigma_{1}%
^{+}\sigma_{2}^{+}\sigma_{4}^{+}-\sigma_{1}^{+}\sigma_{3}^{+}\sigma_{4}%
^{+}\right)  $\\
& $+2\sigma_{1}^{+}\sigma_{2}^{+}\sigma_{3}^{+}\sigma_{4}^{+}$\\
& \\
$G_{d}$ & $f=\sigma_{1}^{+}\sigma_{2}^{+}+\sigma_{1}^{+}\sigma_{3}^{+}%
+\sigma_{2}^{+}\sigma_{4}^{+}+\sigma_{3}^{+}\sigma_{4}^{+}+\sigma_{1}%
^{+}\sigma_{4}^{+}+\sigma_{2}^{+}\sigma_{3}^{+}$\\
& $+\beta_{14}\left(  \sigma_{1}^{+}\sigma_{2}^{+}\sigma_{3}^{+}-\sigma
_{1}^{+}\sigma_{2}^{+}\sigma_{4}^{+}+\sigma_{1}^{+}\sigma_{3}^{+}\sigma
_{4}^{+}-\sigma_{2}^{+}\sigma_{3}^{+}\sigma_{4}^{+}\right)  $\\
& $+\beta_{13}\left(  \sigma_{1}^{+}\sigma_{2}^{+}\sigma_{3}^{+}+\sigma
_{1}^{+}\sigma_{2}^{+}\sigma_{4}^{+}-\sigma_{1}^{+}\sigma_{3}^{+}\sigma
_{4}^{+}-\sigma_{2}^{+}\sigma_{3}^{+}\sigma_{4}^{+}\right)  $\\
& $+\beta_{11}\left(  \sigma_{1}^{+}\sigma_{2}^{+}\sigma_{3}^{+}-\sigma
_{1}^{+}\sigma_{2}^{+}\sigma_{4}^{+}-\sigma_{1}^{+}\sigma_{3}^{+}\sigma
_{4}^{+}+\sigma_{2}^{+}\sigma_{3}^{+}\sigma_{4}^{+}\right)  $\\
& $+2\sigma_{1}^{+}\sigma_{2}^{+}\sigma_{3}^{+}\sigma_{4}^{+}$\\
& \\
$G_{e}$ & $f=\sigma_{1}^{+}\sigma_{2}^{+}\sigma_{3}^{+}+\sigma_{1}^{+}%
\sigma_{2}^{+}\sigma_{4}^{+}+\sigma_{1}^{+}\sigma_{3}^{+}\sigma_{4}^{+}%
+\sigma_{2}^{+}\sigma_{3}^{+}\sigma_{4}^{+}+$\\
& $\beta_{3}\sigma_{1}^{+}\sigma_{2}^{+}+\beta_{6}\sigma_{2}^{+}\sigma_{3}%
^{+}+\beta_{5}\sigma_{2}^{+}\sigma_{3}^{+}$\\
& \\
\emph{Singular $2D$ classes} & 2 complex parameters\\\hline
& \\
$LG2_{a}$ & $f=\sigma_{3}^{+}\sigma_{4}^{+}+\beta_{5}(\sigma_{1}^{+}\sigma
_{3}^{+}+\sigma_{2}^{+}\sigma_{4}^{+})+\beta_{6}(\sigma_{1}^{+}\sigma_{4}%
^{+}+\sigma_{2}^{+}\sigma_{3}^{+})+\sigma_{1}^{+}\sigma_{2}^{+}\sigma_{3}%
^{+}\sigma_{4}^{+}$\\
& \\
$LG2_{b}$ & $f=\sigma_{1}^{+}\sigma_{2}^{+}+\sigma_{3}^{+}\sigma_{4}^{+}%
+\beta_{5}(\sigma_{1}^{+}\sigma_{3}^{+}+\sigma_{2}^{+}\sigma_{4}^{+}%
)+\beta_{6}(\sigma_{1}^{+}\sigma_{4}^{+}+\sigma_{2}^{+}\sigma_{3}^{+})$\\
& \\
$LG2_{c}$ & $f=\sigma_{1}^{+}\sigma_{3}^{+}\sigma_{4}^{+}+\sigma_{1}^{+}%
\sigma_{2}^{+}\sigma_{4}^{+}+\sigma_{2}^{+}\sigma_{3}^{+}\sigma_{4}^{+}%
+\sigma_{1}^{+}\sigma_{2}^{+}+\beta_{5}\sigma_{1}^{+}\sigma_{3}^{+}+\beta
_{6}\sigma_{2}^{+}\sigma_{3}^{+}$\\
& \\
\emph{Singular $1D$ classes} & 1 complex parameters\\\hline
& \\
$LG1_{a}$ & $f=\sigma_{1}^{+}\sigma_{2}^{+}+\sigma_{1}^{+}\sigma_{3}^{+}%
+\beta_{6}(\sigma_{1}^{+}\sigma_{4}^{+}+\sigma_{2}^{+}\sigma_{3}^{+}%
)+\sigma_{1}^{+}\sigma_{2}^{+}\sigma_{3}^{+}\sigma_{4}$\\
& \\
$LG1_{b}$ & $f=\sigma_{1}^{+}\sigma_{2}^{+}\sigma_{4}^{+}+\sigma_{2}^{+}%
\sigma_{3}^{+}\sigma_{4}^{+}+\sigma_{1}^{+}\sigma_{2}^{+}+\sigma_{1}^{+}%
\sigma_{3}^{+}+\beta_{6}\sigma_{2}^{+}\sigma_{3}^{+}$\\
& \\
\emph{Singular point classes} & no parameters\\\hline
$S_{a}$ & $f=\sigma_{1}^{+}\sigma_{2}^{+}\sigma_{3}^{+}+\sigma_{1}^{+}%
\sigma_{3}^{+}\sigma_{4}^{+}+\sigma_{2}^{+}\sigma_{4}^{+}$\\
$S_{b}$ & $f=\sigma_{1}^{+}\sigma_{2}^{+}\sigma_{3}^{+}+\sigma_{1}^{+}%
\sigma_{3}^{+}\sigma_{4}^{+}+\sigma_{1}^{+}\sigma_{2}^{+}\sigma_{4}^{+}$\\
$S_{c}$ & $f=\sigma_{1}^{+}\sigma_{2}^{+}\sigma_{3}^{+}+\sigma_{1}^{+}%
\sigma_{3}^{+}\sigma_{4}^{+}$\\
$S_{d}$ & $f=\sigma_{1}^{+}\sigma_{2}^{+}\sigma_{3}^{+}$\\
$S_{e}$ & $f=\sigma_{3}^{+}\sigma_{4}^{+}+\sigma_{1}^{+}\sigma_{3}^{+}%
+\sigma_{2}^{+}\sigma_{3}^{+}+\sigma_{1}^{+}\sigma_{2}^{+}\sigma_{3}^{+}%
\sigma_{4}^{+}$\\
$S_{f}$ & $f=\sigma_{1}^{+}\sigma_{2}^{+}+\sigma_{2}^{+}\sigma_{3}^{+}%
+\sigma_{3}^{+}\sigma_{1}^{+}+\sigma_{1}^{+}\sigma_{2}^{+}\sigma_{3}^{+}%
\sigma_{4}^{+}$\\
$\dots$ & $\dots$\\
&
\end{tabular}
\caption{Classification of four-qubit entanglement classes following from
$SL(2,\mathbb{C})$ transformation properties of the canonic form.}%
\end{table}%
\end{widetext}%

\subsection{Equation for local and two-body interaction}

We give here some more details on the derivation of the dynamic
equation for the nilpotential $f$.  Consider a single gate operation applied to qubits
$i$ and $j$. We cast the function $f$ in the form
$f_{00}+\sigma_{i}^{+}f_{01}+\sigma_{j}^{+}
f_{10}+\sigma_{i}^{+}\sigma_{j}^{+}f_{11}$, and find%
\begin{align}
\mathrm{e}^{-f}  &  =\mathrm{e}^{-f_{00}}\left[  1-\sigma_{i}^{+}f_{01}
-\sigma_{j}^{+}f_{10}-\sigma_{i}^{+}\sigma_{j}^{+}(f_{11}-f_{01}f_{10})\right]
\nonumber\\
\mathrm{e}^{f}  &  =\mathrm{e}^{f_{00}}\left[  1+\sigma_{i}^{+}f_{01}
+\sigma_{j}^{+}f_{10}+\sigma_{i}^{+}\sigma_{j}^{+}(f_{11}+f_{01}
f_{10})\right]  ~, \label{EQ53}%
\end{align}
where the coefficients 
\begin{widetext}
\begin{align}
f_{00}  &  =f-\sigma_{i}^{+}\frac{\partial f}{\partial\sigma_{i}^{+}}
-\sigma_{j}^{+}\frac{\partial f}{\partial\sigma_{j}^{+}}+\sigma_{i}^{+}
\sigma_{j}^{+}\frac{\partial^{2}f}{\partial\sigma_{i}^{+}\partial\sigma
_{j}^{+}}\,, \nonumber\\
f_{01}   & =\frac{\partial f}{\partial\sigma_{i}^{+}}-\sigma_{j}^{+}
\frac{\partial^{2}f}{\partial\sigma_{i}^{+}\partial\sigma_{j}^{+}}\,,
\hspace{5mm}
f_{10}   =\frac{\partial f}{\partial\sigma_{j}^{+}}-\sigma_{i}^{+}
\frac{\partial^{2}f}{\partial\sigma_{i}^{+}\partial\sigma_{j}^{+}}\,,
\hspace{5mm}
f_{11}    =\frac{\partial^{2}f}{\partial\sigma_{i}^{+}\partial\sigma_{j}^{+}}\,,
\label{EQ54}%
\end{align}
\end{widetext}
are independent of $\sigma_{i}^{+},\sigma_{j}^{+}$. Upon substituting
these equations into Eq.~(\ref{EQ21af}) for the Hamiltonian in
Eq.~(\ref{EQ52}), bearing in mind Eq.~(\ref{EQ49.11}), we obtain
\begin{widetext}
\begin{align}
\mathrm{i}\frac{\partial f}{\partial t}  &  =P_{i}^{-}\sigma_{i}^{+}+P_{j}%
^{-}\sigma_{j}^{+}+P_{i}^{+}\frac{\partial f}{\partial\sigma_{i}^{+}}%
+P_{j}^{+}\frac{\partial f}{\partial\sigma_{j}^{+}}-P_{i}^{+}\sigma_{i}%
^{+}\left(  \frac{\partial f}{\partial\sigma_{i}^{+}}\right)  ^{2}-P_{j}%
^{+}\sigma_{j}^{+}\left(  \frac{\partial f}{\partial\sigma_{j}^{+}}\right)
^{2}\nonumber\\
&  +G_{ij}\bigg[  \sigma_{j}^{+}\frac{\partial f}{\partial\sigma_{i}^{+}%
}+\sigma_{i}^{+}\frac{\partial f}{\partial\sigma_{j}^{+}}-\sigma_{i}^{+}%
\sigma_{j}^{+}\left(  \frac{\partial f}{\partial\sigma_{i}^{+}}\right)
^{2}-\sigma_{i}^{+}\sigma_{j}^{+}\left(  \frac{\partial f}{\partial\sigma
_{j}^{+}}\right)  ^{2}\bigg]  \label{EQ55}\:.
\end{align}
\end{widetext}
Summing over $i,j$ yields Eq.~(\ref{EQ56}), where the condition
$G_{ij}=G_{ji}$ is taken into account.\

\subsection{Derivation of $3$-tangle based on the tanglemeter\label{F}}

Here we present  a derivation of $3$-tangle
that  is  based directly
on the transformations of amplitudes under local operations.
The  explicit
expression for $3$-tangle has the rather simple form of Eq.~(\ref{threetangle}),
when written for the $su$-canonical state. When considering $SL$ transformations, we
 look for a quantity that takes different values within
the generic $sl$ orbit of three qubits, and vanishes outside this orbit.
Performing $SL$-transformations,  we can still set appropriate
conditions so as to preserve  the $su$-canonical form of states.

We start with the $sl$-canonical state, i.e., is the maximum entangled state \\
$\left(  \left\vert 000\right\rangle +\left\vert 111\right\rangle \right)
/\sqrt{2}$, written in terms of nilpotent variables and normalizing to unit
reference state amplitude:
\[
|\psi\rangle=\frac{1}{\sqrt{2}}(1+\sigma_{1}^{+}\sigma_{2}^{+}\sigma_{3}%
^{+})|000\rangle ~~.
\]
Then we apply to each qubit the general $SL(2,\mathbb{C})$ transformation, which
reads
\[
e^{A_{i}\sigma_{i}^{-}}e^{B_{i}\sigma_{i}^{z}}e^{C_{i}\sigma_{i}^{+}}%
,~~A_{i},B_{i},C_{i}\in{\mathbb{C}},\,i=1,2,3.
\]
This transformation results in the most general admissible $su$-canonical form
of the $su$-canonical state,
\begin{align*}
|\psi^{\prime}\rangle &  =(\psi_{000}^{\prime}+\psi_{011}^{\prime}\sigma
_{1}^{+}\sigma_{2}^{+}+\psi_{110}^{\prime}\sigma_{2}^{+}\sigma_{3}^{+}\\
&  +\psi_{101}^{\prime}\sigma_{1}^{+}\sigma_{3}^{+}+\psi_{111}^{\prime}%
\sigma_{1}^{+}\sigma_{2}^{+}\sigma_{3}^{+})|000\rangle~,
\end{align*}
in which the coefficients take the forms
\begin{align}
\psi_{000}^{\prime} &  =\frac{(1+z)^{2}}{\sqrt{2}}\mathrm{e}^{-B_{1}%
-B_{2}-B_{3}},\label{cof}\\
\psi_{011}^{\prime} &  =-\frac{1+z}{z\sqrt{2}}C_{1}C_{2}\mathrm{e}%
^{B_{1}+B_{2}-B_{3}}\nonumber\\
\psi_{101}^{\prime} &  =-\frac{1+z}{z\sqrt{2}}C_{1}C_{3}\mathrm{e}%
^{B_{1}+B_{3}-B_{2}}\nonumber\\
\psi_{110}^{\prime} &  =-\frac{1+z}{z\sqrt{2}}C_{2}C_{3}\mathrm{e}%
^{B_{2}+B_{3}-B_{1}}\nonumber\\
\psi_{111}^{\prime} &  =-\frac{1+2z}{z^{2}\sqrt{2}}C_{1}C_{2}C_{3}%
\mathrm{e}^{B_{1}+B_{2}+B_{3}}~~.\nonumber
\end{align}
Here we have employed a complex variable $z$ that naturally emerges from the
requirements of $su$-canonical form, having the relations $z=A_{1}%
e^{2B_{1}}C_{1}=A_{2}e^{2B_{2}}C_{2}=A_{3}e^{2B_{3}}C_{3}$ and $(1+z)^{2}%
C_{2}C_{3}=-z^{2}$. One sees that the amplitudes of the $su$-canonical state
depend on seven real parameters, since in addition to five parameters of the
tanglemeter, we now allow for two more parameters characterizing the vacuum
state amplitude.

Being in the general orbit,
states of less general orbits can be 
reached if irreversible transformations are performed
\cite{Akimasa} corresponding to the limit $|z|\rightarrow\infty$.
At this point we want to construct a polynomial measure on the amplitudes $\psi'$,
such  a way that it is a function of $z$ and it vanishes outside the general
$sl$-orbit. 
Direct inspection shows that we can construct a function that 
depends only on the parameter $z$ 
\begin{equation}
\zeta(z)=\frac{\psi_{111}^{^{\prime}2}\psi_{000}^{^{\prime}2}}{\psi
_{000}^{\prime}\psi_{110}^{\prime}\psi_{011}^{\prime}\psi_{101}^{\prime}%
}=-\frac{(1+2z)^{2}}{z(1+z)}~,\label{zeta}%
\end{equation}
\begin{figure}[h]
{\centering{\includegraphics*[width=0.4\textwidth]{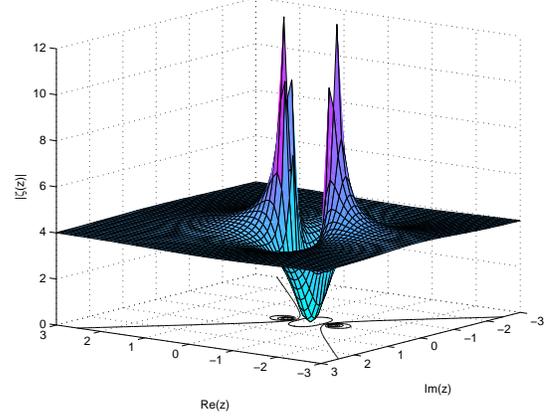}}} \vspace
{0.1cm}\caption{ Function $|\zeta(z)|$ as in Eq.~(\ref{zeta}). In the limit
$|z|\rightarrow\infty$, $|\zeta(z)|\rightarrow4$.}\vspace{6mm}
\label{Lorz}%
\end{figure}
and  is independent of the normalization,.
The function $\zeta(z)$ shown in Fig.~\ref{Lorz} can take any
value, but $-4$ that corresponds to the limit $|z|\rightarrow\infty$.
Therefore, the polynomial 
\[ 
\left|\psi_{111}^{^{\prime}2}\psi_{000}^{^{\prime}2}+4\psi
_{000}^{\prime}\psi_{110}^{\prime}\psi_{011}^{\prime}\psi_{101}^{\prime}\right|
\]
will be identically zero outside the general orbit, and  is otherwise nothing
but the $3$-tangle expressed in terms of the coefficients
of the canonical nilpotent polynomial,  Eq.~(\ref{threetangle}).

This procedure for constructing a measure can be in principle
extended to more qubits. However, the extension will require
some care. The general $sl$-orbit of four qubits is characterized by six parameters parameters, in contrast
to the three-qubits case. Consequently,
   irreversible transformations may
connect different $sl$-orbits that both contain $N$-partite entanglement.
Therefore the desired limits need to be clearly specified.

Finally, we would like to mention that \textit{Theorem 3} in Ref.~\cite{Verstraete2} 
suggests another useful application of the $\zeta(z)$
function. For an arbitrary $3$-qubit state expressed in canonical form,
one can calculate the value of $\zeta(z)$ by
direct substitution of the numerical values of amplitudes to Eq.~(\ref{zeta}),
 and then solve a binomial equation
 to find the root $z=z_{r}$. One can choose the coefficients
$A_{i},~B_{i}$, and $C_{i}$ such that they satisfy $A_{i}\mathrm{e}^{2B_{i}}%
C_{i}=z_{r}$. In this way a determinant-$1$ transformation that brings the
maximum entangled state to the chosen state can be identified explicitly. The
inverse transformation can be used for an optimal filtering procedure called purification,
a probabilistic procedure that transforms a state of the general orbit
to the one with the maximal entanglement,i.e. to the $GHZ$ state.



\begin{thebibliography}{99}
\bibitem {note0}Capital letters $SU$ and $SL$ will normally denote Lie
groups, while the small $su$ and $sl$ denote the corresponding Lie
algebras. However, we do not always follow this convention, unless it
creates a confusion.
\bibitem{expn} Still there is a strong indication that the 
maximum population is unique. Indeed, there exists a unique local
unitary operation which being applied to a single-qubit $2\times2$ density 
matrices maximizes its ground state polpulation. Such transformations
applied to all qubits maximize the assembly ground state population.
\bibitem{purification}We therefore do not address the purification
protocols.
\bibitem{xx} Actually, a finite number of different canonic states may exist,
that can be transformed one to another by $SL$-transformations consisting
of qubit-``flips'' and scaling.
In order to be specific we can choose one which has minimum $\left|\mathrm{ln}\sum\left|\psi\right|^2\right|$.
\bibitem{footnote0}
A mathematician would say that $\epsilon ^{ii^{\prime }}$ represents an
invariant tensor of both $SU(2)$ and $SL_{i}(2,\mathbb{C})$ transformations.
On the other hand, the tensor $\delta _{i}^{j}$ is invariant under $SU$, but
not under $SL$.
\bibitem {integration}This operation may also be defined as
integration over the nilpotent variable, by the analogy to well-known
anticommuting Grassman variables \cite{Berezin}.
\end{thebibliography}
\end{document}